\documentclass[12pt,preprint]{aastex}
\usepackage{apjfonts}
\usepackage{amsmath}
\usepackage{color}
\shorttitle{Effect of Interacting Rarefaction Waves on Relativistically Hot Jets}
\shortauthors{Matsumoto et al.}
\begin{document}
%%%%%%%%%%%%%%%%%%%%%%%%%%%%%%-------------------------- TITLE  & AUTHORS ---------------------- %%%%%%%%%%%%%%%%%%%%%%%%%%%%%%%%%
\title{Effect of Interacting Rarefaction Waves on Relativistically Hot Jets}
\author{Jin Matsumoto\altaffilmark{1,2,3}, Youhei Masada\altaffilmark{4}, and Kazunari Shibata\altaffilmark{1}}
\altaffiltext{1}{Kwasan and Hida Observatories, Kyoto University, Kyoto, Japan; jin@kusastro.kyoto-u.ac.jp}
\altaffiltext{2}{Department of Astronomy, Kyoto University, Kyoto, Japan}
\altaffiltext{3}{National Astronomical Observatory of Japan, Tokyo, Japan}
\altaffiltext{4}{Graduate School of System Informatics, Department of Computational Science, Kobe University, Kobe, Japan}
%%%%%%%%%%%%%%%%%%%%%%%%%%%%%%------------------------  ABSTRACT ------------------------------ %%%%%%%%%%%%%%%%%%%%%%%%%%%%%%%%%%
\begin{abstract}
The effect of rarefaction acceleration on the propagation dynamics and structure of relativistically hot jets is 
studied through relativistic hydrodynamic simulations. We emphasize the nonlinear interaction 
of rarefaction waves excited at the interface between a cylindrical jet and the surrounding medium. From 
simplified one-dimensional models with radial jet structure, we find that a decrease in the relativistic pressure 
due to the interacting rarefaction waves in the central zone of the jet transiently yields a more powerful boost of 
the bulk jet than that expected from single rarefaction acceleration. This leads to a cyclic in-situ energy conversion 
between thermal and bulk kinetic energies which induces radial oscillating motion of the jet. The oscillation timescale 
is characterized by the initial pressure ratio of the jet to the ambient medium, and follows a simple scaling 
relation $\tau_{\rm oscillation} \propto (P_{\rm jet,0}/P_{\rm amb,0})^{1/2}$. It is confirmed from extended 
two-dimensional simulations that this radial oscillating motion in the one-dimensional system manifests as modulation 
of the structure of the jet in a more realistic situation where a relativistically hot jet propagates through an ambient 
medium. It is found that when the ambient medium has a power law pressure distribution, the size of the reconfinement 
region along the propagation direction of the jet in the modulation structure $\lambda$ evolves according to a 
self-similar relation $\lambda \propto t^{\alpha/2}$ where $\alpha$ is the power-law index of the pressure distribution. 
\end{abstract}
\keywords{galaxies: jets --- hydrodynamics --- methods: numerical --- relativistic processes --- shock waves}
%%%%%%%%%%%%%%%%%%%%%%%%%%%%%%%---------------------- BODY OF THE PAPER ------------------------ %%%%%%%%%%%%%%%%%%%%%%%%%%%%%%%%%
%%%%%%%%%%%%%%%%%%%%%%%%%%%%%%%----------------------- S1 INTRODUCTION ------------------------- %%%%%%%%%%%%%%%%%%%%%%%%%%%%%%%%%
\section{Introduction}

Relativistic jets are collimated bipolar outflows that have a velocity almost equal to light 
speed. They are ubiquitous among astrophysical systems consisting of a compact object 
surrounded by an accretion disk, e.g., active galactic nuclei \citep[AGNs;][]{Ferrari98}, 
microquasars \citep{Mirabel99} and the central engine of gamma-ray bursts \citep[GRBs;][]{Piran04, Meszaros06}. 
Although there are many works that try to determine the mechanism by which relativistic 
jets are accelerated and collimated, the process is still not understood. 

\citet{Aloy06} recently reported that relativistic jets can be powerfully boosted 
along the interface between the jet and ambient medium, if the jet has sufficiently large
velocity and specific enthalpy, and is overpressured \citep[see also][2005]{Aloy03}. 
A rarefaction wave excited at the interface converts the relativistic thermal energy of the 
plasma into kinetic energy and yields an amplification of the Lorentz factor in the jet-ambient 
medium interface. This type of boost, which we label rarefaction acceleration, is not possible 
in Newtonian dynamics, but is an inherent process in relativistic hydrodynamics. 

The relativistic magnetohydrodynamic effects on the rarefaction acceleration were studied 
by \citet{Mizuno08} and \cite{Aloy08}. \citet{Mizuno08} showed that, especially 
in the case of a magnetic field perpendicular to the jet direction, the boost becomes more 
powerful than that expected from pure hydrodynamic models. \cite{Aloy08} studied 
the radiative output of magnetized jets bounded by anomalous shear layers in which jets are 
boosted due to the rarefaction wave. It was indicated theoretically that the boosted Lorentz 
factor is well described, as a function of the initial pressure ratio between the jet and the 
ambient medium, by a simple scaling law derived by \citet{Zenitani10} \citep[see also][]{Komissarov10}.

Rarefaction acceleration of jets is a process that is expected to occur under realistic conditions. 
Since a large amount of energy is stored in the jet at its launch site, the jet would be initially 
overpressured. Actually, the jet from AGNs often appear to be overpressured with relativistic 
velocity even as it propagates through the interstellar medium \citep{Bicknell96}.
In some cases it has two components: a faster, lighter component surrounded by a slower, denser 
component \citep{Giroletti04, Meliani09}. The jet that emerges from the central engine of GRBs 
is also expected to be hot and overpressured when it propagates inside the progenitor star. 
Furthermore, the fluid of the boundary layer of the jet would be boosted by rarefaction acceleration 
when the relativistic jet breaks out on the surface of the progenitor star and the external pressure support drops \citep{Tchekhovskoy10}.
The Lorentz factor of the jet can be boosted when it enters the low pressure region outside of the thick accretion torus 
which is a possible candidate of the central engine of short GRBs \citep{Aloy05}.

In this paper, we study in detail how the rarefaction acceleration affects the propagation dynamics 
of relativistically hot jets through one-dimensional (1D) and two-dimensional (2D) relativistic 
hydrodynamic simulations. The nonlinear interaction of rarefaction waves excited at the interface 
between the jet and ambient medium is especially focused upon because it might have a potential 
impact on the boosting process and even alter the dynamics and structure of the jet. 
This paper is organized as follows. In Section 2, the numerical results of our 1D relativistic hydrodynamic 
simulations are presented. We perform 2D simulations in Section 3. 
Finally, we discuss and summarize our findings in Section 4 and 5.
%%%%%%%%%%%%%%%%%%%%%%%%%%%%%%%%%%%%%%%%%%%%%%%%%%%%%%%%%%%%%%%%%%%%%%%%%%%%%%%%%%%%%%%%%%%%%%%
%%%%%%%%%%%%%%%%%%%%%%%%%%%%%%%---------------------- S2 1D Simulations ------------------------ %%%%%%%%%%%%%%%%%%%%%%%%%%%%%%%
%%%%%%%%%%%%%%%%%%%%%%%%%%%%%%%%%%%%%%%%%%%%%%%%%%%%%%%%%%%%%%%%%%%%%%%%%%%%%%%%%%%%%%%%%%%%%%%
\section{One-dimensional Simulations} 
\subsection{Numerical Models and Setup}
\begin{figure}[!htbp]
\begin{center}\scalebox{1.2}{\rotatebox{0}{\includegraphics{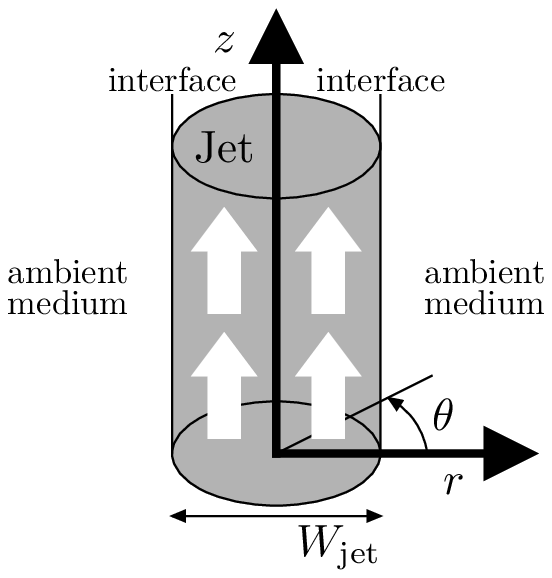}}}
\caption{The numerical setting for our 1D simulations. A relativistically hot jet region is bounded by ambient gas. 
Derivatives of the physical variables in the $z$-direction and $\theta$-direction are assumed to be zero. We calculate the 
evolution of the jet only in the radial direction.}
\label{fig1}
\end{center}
\end{figure}

In order to investigate the interaction of rarefaction waves excited at the jet-ambient medium interface, we initially set a jet 
surround by ambient gas in the calculation domain, as is schematically shown in Figure~\ref{fig1}. We solve the special 
relativistic hydrodynamic (RHD) equations in a one-dimensional axisymmetric cylindrical coordinate system. The fluid velocity 
has two components: the normal velocity $v_{r}$ and tangential velocity $v_{z}$ to the interface which separates the jet and ambient 
medium. Derivatives of the physical variables in the $z$-direction are assumed to be zero. We adopt, as the equation of state for 
the relativistic gas, the ideal gas law with a ratio of specific heats $\Gamma = 4/3$. 
The basic equations are then,
\begin{eqnarray}
\frac{\partial}{\partial t}(\gamma \rho)                   &+& \frac{1}{r}\frac{\partial}{\partial r}(r \gamma \rho v_{r}) = 0 \;, \label{eq: mass conservation} \\ 
\frac{\partial}{\partial t}(\gamma^2 \rho h v_{r}) &+& \frac{1}{r}\frac{\partial}{\partial r} \biggl [ r (\gamma^2 \rho h v_{r}^2 + Pc^2) \biggr ] = \frac{Pc^2}{r} \;, \label{eq: momentum conservation r} \\
\frac{\partial}{\partial t}(\gamma^2 \rho h v_{z}) &+& \frac{1}{r}\frac{\partial}{\partial r} \biggl [ r (\gamma^2 \rho h v_{z} v_{r}) \biggr ] = 0 \;, \label{eq: momentum conservation z} \\
\frac{\partial}{\partial t}(\gamma^2 \rho h - P)      &+& \frac{1}{r}\frac{\partial}{\partial r} \biggl [ r (\gamma^2 \rho h v_{r}) \biggr ] = 0 \;. \label{eq: energy conservation}
%
%\frac{1}{c}\frac{\partial}{\partial t} \Bigl (\gamma \rho \Bigr )                   &+& \frac{1}{r}\frac{\partial}{\partial r} \Bigl (r \gamma \rho \frac{v_{r}}{c} \Bigr ) = 0 \;, \label{eq: mass conservation} \\ 
%\frac{1}{c}\frac{\partial}{\partial t} \Bigl ( \gamma^2 \rho h \frac{v_{r}}{c} \Bigr ) &+& \frac{1}{r}\frac{\partial}{\partial r} \biggl [ r \Bigl ( \gamma^2 \rho h \frac{v_{r}^2}{c^2} + P \Bigr ) \biggr ] = 0 \;, \label{eq: momentum conservation r} \\
%\frac{1}{c}\frac{\partial}{\partial t} \Bigl ( \gamma^2 \rho h \frac{v_{z}}{c} \Bigr ) &+& \frac{1}{r}\frac{\partial}{\partial r} \biggl [ r \Bigl ( \gamma^2 \rho h \frac{v_{z} v_{r}}{c^2} \Bigr ) \biggr ] = 0 \;, \label{eq: momentum conservation z} \\
%\frac{1}{c}\frac{\partial}{\partial t} \Bigl ( \gamma^2 \rho h - P \Bigr )      &+& \frac{1}{r}\frac{\partial}{\partial r} \biggl [ r \Bigl ( \gamma^2 \rho h \frac{v_{r}}{c} \Bigr ) \biggr ] = 0 \;. \label{eq: energy conservation}
\end{eqnarray}
where $\gamma \equiv 1/\sqrt{1-(v_{x}/c)^2 - (v_{z}/c)^2}$ is the Lorentz factor and $h = c^2 + \Gamma P/ (\Gamma -1)\rho$ is 
the specific enthalpy. The other symbols have their usual meanings.

For our initial conditions, we assume a relativistically hot jet with larger pressure and smaller rest-mass density than the ambient medium. 
The initial density and pressure in the jet are chosen as $\rho_{\rm jet, 0} = 0.1$ and $P_{\rm jet, 0} = 1$, respectively. Those of the 
ambient medium are $\rho_{\rm amb, 0} = 1$ and $P_{\rm amb, 0} = 0.1$. In addition, the jet velocity to the z-direction is relativistic 
$v_{\rm jet, 0} = 0.99c$, with a Lorentz factor of $\gamma_{\rm jet,0} \sim 7$. The ambient medium does not move and the normal 
velocity $v_{r}$ is set to be zero initially in the calculation domain.  
In this paper, we set a simple model for the jet-ambient medium system in order to investigate the basic physics of the interaction of 
rarefaction waves excited at the jet-ambient medium interface. 

The normalization units in length, velocity, time, and energy density are chosen as the initial jet width $W_{\rm jet,0}$, light 
speed $c$, light crossing time over the initial jet width $W_{\rm jet,0}/c$, and rest mass energy density in the ambient medium 
$\rho_{\rm amb,0}c^2$. The computational domain spans $0 < r < 2$. 
A uniform grid with a grid size $\Delta r=10^{-3}$ is 
adopted for our calculations. We use a reflecting boundary condition on the axis $r = 0$. 
The outer boundary of the grid uses the outflow (zero gradient) boundary condition.

We use a relativistic HLLC scheme for hydrodynamics to solve the RHD equations \citep{Mignone05}.
The primitive variables are calculated from the conservative variables following the method of \citet{Mignone07}. 
We use a MUSCL-type interpolation method to attain second-order accuracy in space 
while the temporal accuracy obtains second-order by using Runge-Kutta time integration.
%%%%%%%%%%%%%%%%%%%%%%%%%%%%%%%%%%%%%%%%%%%%%%%%%%%%%%%%%%%%%%%%%%%%%%%%%%%%%%%%%%%%%%%%%%%%%
%%%%%%%%%%%%%%%%%%%%%%%%%%%---------------  Result ------------------%%%%%%%%%%%%%%%%%%%%%%%%%%%%
%%%%%%%%%%%%%%%%%%%%%%%%%%%%%%%%%%%%%%%%%%%%%%%%%%%%%%%%%%%%%%%%%%%%%%%%%%%%%%%%%%%%%%%%%%%%%
\subsection{Results of One-dimensional Simulations}
\subsubsection{The Jet Boosting Mechanism of Aloy \& Rezzola 2006}
Since the jet is initially hotter and has a higher pressure than the denser, colder ambient medium, 
three types of hydrodynamic wave are excited at the jet-ambient medium interface. A shock wave 
propagates outward through the ambient medium, and a rarefaction wave starts to travel toward 
the center of the jet. Behind the shock wave, there exists a contact discontinuity, that is an entropy 
wave, corresponding to the edge of the jet. The pressure decrease in the jet due to the rarefaction 
wave accelerates the gas in the jet-ambient medium interface in the tangential direction as a natural 
outcome of relativistic hydrodynamics, as was reported by \cite{Aloy06}. 

Following the analysis of \citet{Zenitani10}, by combining Equations~(\ref{eq: momentum conservation z}) 
and~(\ref{eq: energy conservation}) we can obtain
\begin{eqnarray}
\gamma^2 \rho h \biggl ( \frac{\partial }{\partial t} + v_{r} \frac{\partial}{\partial r} \biggr ) \frac{v_{z}}{c} \notag
&=& -\frac{v_{z}}{c}\frac{\partial P}{\partial t} \\
&\sim& -\frac{\partial P}{\partial t}\; \; \; \; (\because v_{z} \sim c) \; . \label{eq: EoM}
\end{eqnarray}
This indicates that a time-decreasing pressure is responsible for the acceleration of the gas in relativistic 
hydrodynamics, unlike the non-relativistic case. The conversion from the thermal energy to bulk kinetic 
energy of the jet is then constrained by the relativistic Bernoulli equation which provides enthalpy conservation, 
\begin{eqnarray} 
\gamma h \sim {\rm const.} \label{eq: gh}
\end{eqnarray}
The relativistic jet is accelerated by this mechanism, found by \citet{Aloy06}, in our numerical model. 

\subsubsection{Temporal Evolution of the System}

\begin{figure}[!htbp]
\begin{center}
\scalebox{1.2}{\rotatebox{0}{\includegraphics{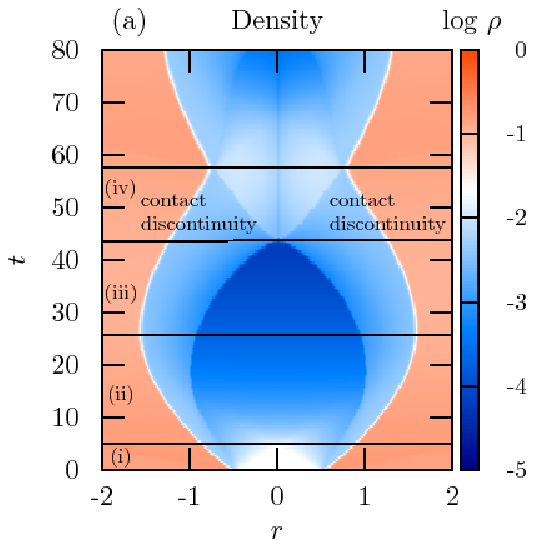}}}
\scalebox{1.2}{\rotatebox{0}{\includegraphics{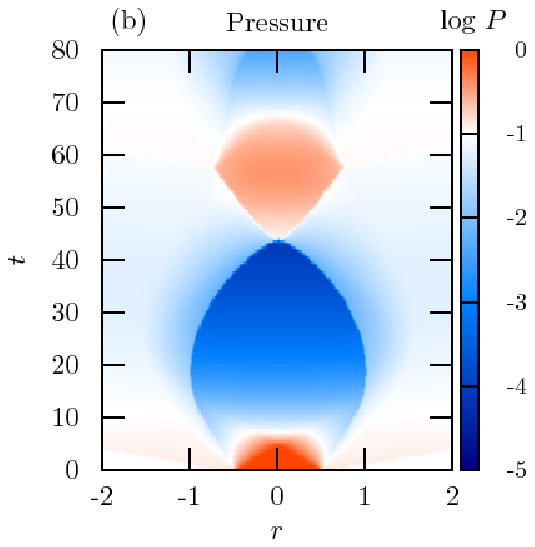}}}\\
\scalebox{1.2}{\rotatebox{0}{\includegraphics{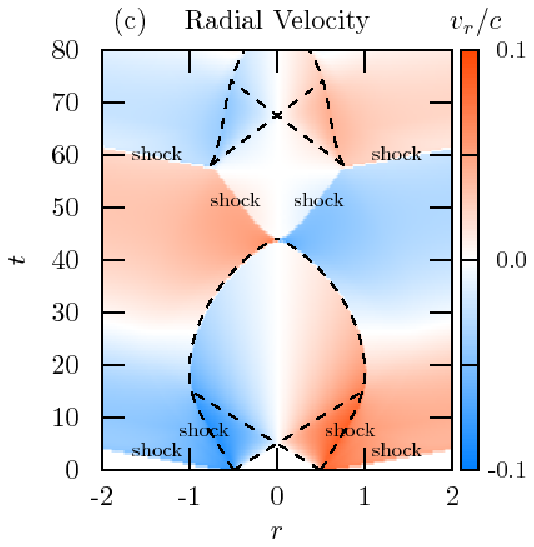}}}
\scalebox{1.2}{\rotatebox{0}{\includegraphics{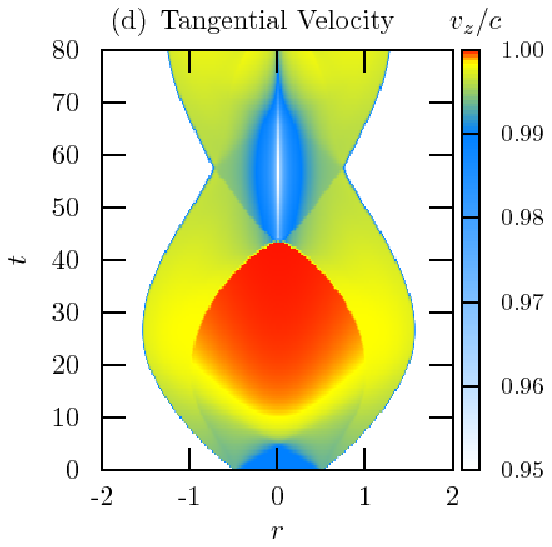}}}\\
\scalebox{1.2}{\rotatebox{0}{\includegraphics{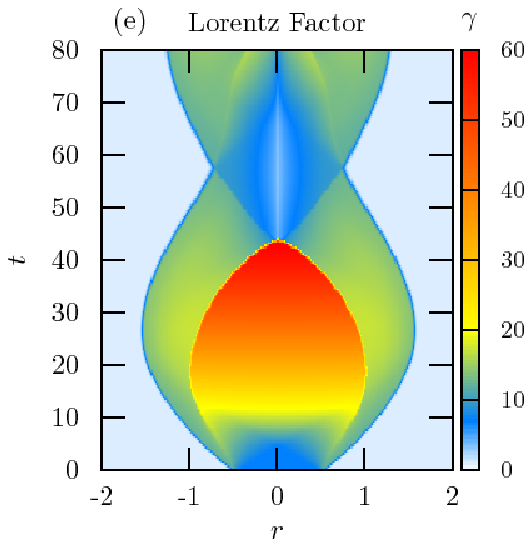}}}
% or
%\scalebox{1}{\rotatebox{0}{\includegraphics{f2a.eps}}}
%\scalebox{1}{\rotatebox{0}{\includegraphics{f2b.eps}}}
%\scalebox{1}{\rotatebox{0}{\includegraphics{f2c.eps}}}\\
%\scalebox{1}{\rotatebox{0}{\includegraphics{f2d.eps}}}
%\scalebox{1}{\rotatebox{0}{\includegraphics{f2e.eps}}}
\caption{Temporal evolution of the jet-ambient medium system: (a) the density, (b) the pressure, (c) the radial velocity, 
(d) the tangential velocity and (e) the Lorentz factor. In panel (c), the rarefaction region is enclosed by dashed-lines. 
The tangential velocity of the ambient medium is zero in the panel (d). 
(A color version of this figure is available in the online journal.)}
\label{fig2}
\end{center}
\end{figure}

\begin{figure}[!htbp]
\begin{center}
\scalebox{0.8}{\rotatebox{0}{\includegraphics{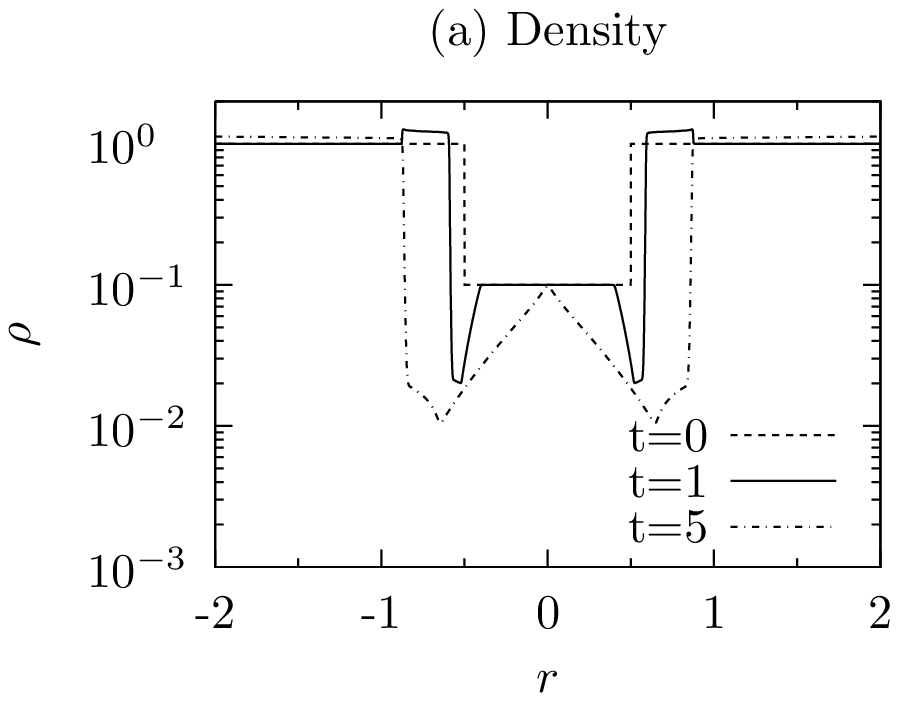}}}
\scalebox{0.8}{\rotatebox{0}{\includegraphics{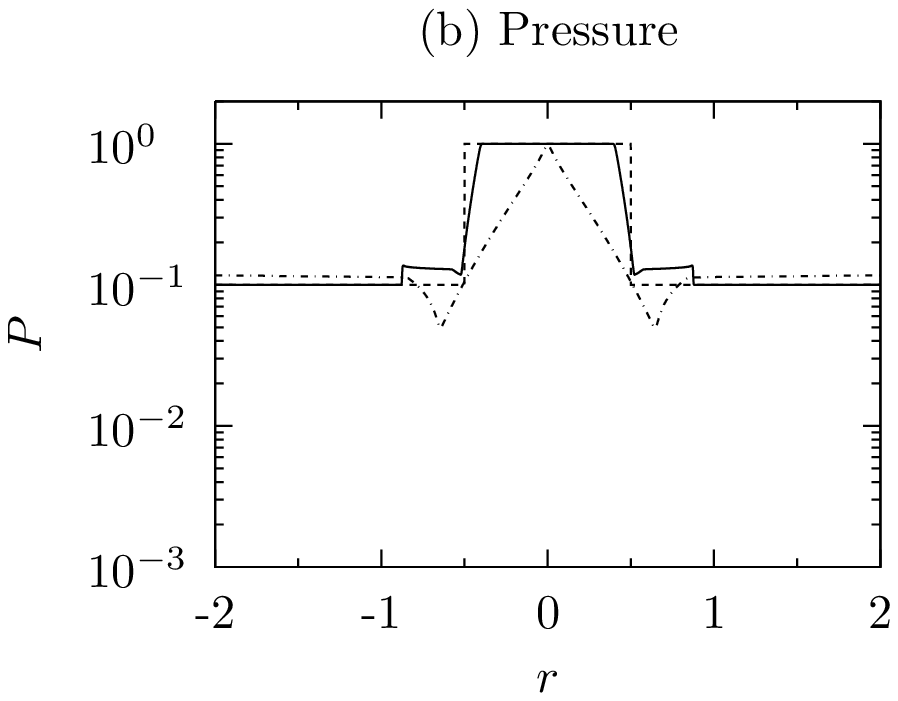}}}\\
\scalebox{0.8}{\rotatebox{0}{\includegraphics{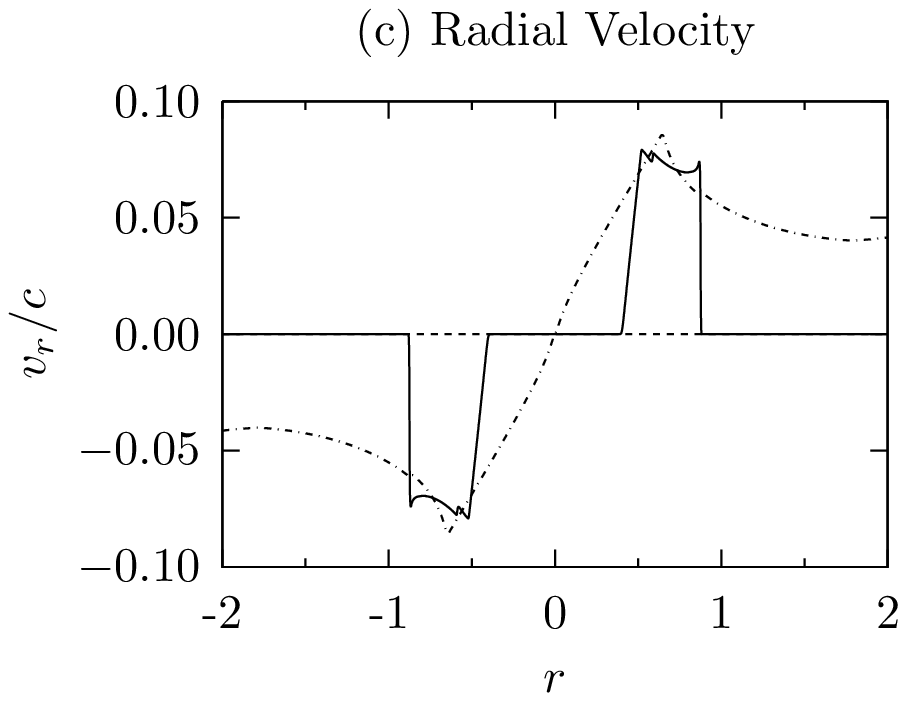}}}
\scalebox{0.8}{\rotatebox{0}{\includegraphics{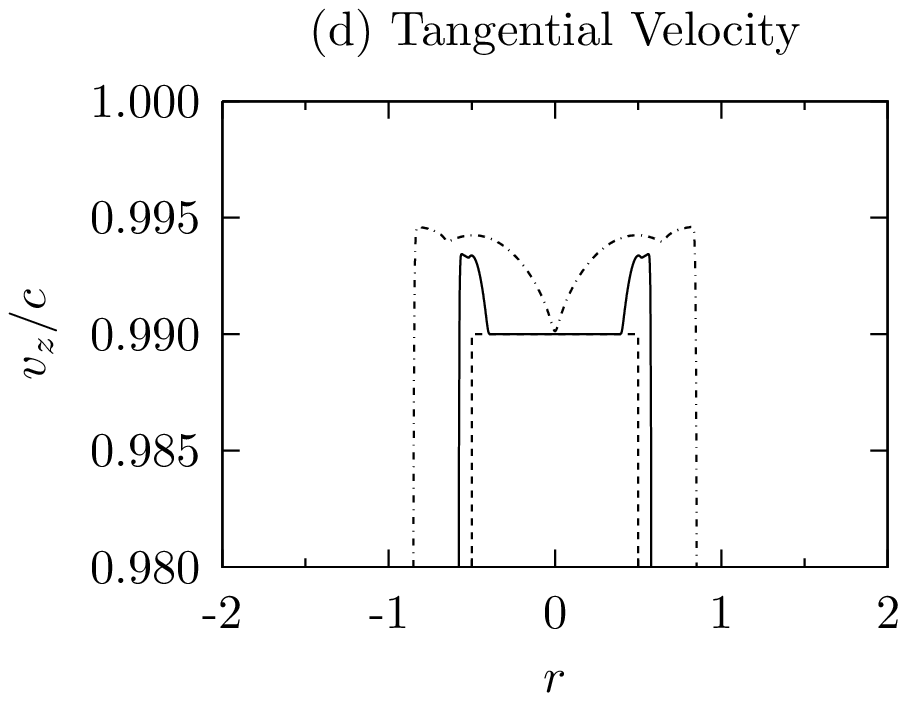}}}\\
\scalebox{0.8}{\rotatebox{0}{\includegraphics{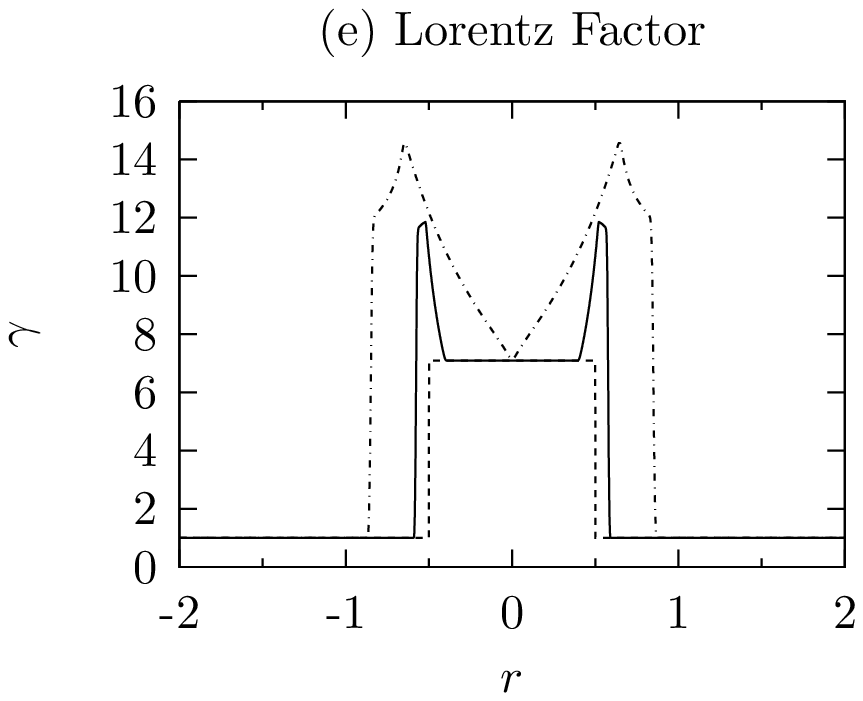}}}
% or 
%\scalebox{0.58}{\rotatebox{0}{\includegraphics{f3a.eps}}}
%\scalebox{0.58}{\rotatebox{0}{\includegraphics{f3b.eps}}}
%\scalebox{0.58}{\rotatebox{0}{\includegraphics{f3c.eps}}}\\
%\scalebox{0.58}{\rotatebox{0}{\includegraphics{f3d.eps}}}
%\scalebox{0.58}{\rotatebox{0}{\includegraphics{f3e.eps}}}
\caption{
The spatial distribution of the jet-ambient medium system at $t=0$ (dashed lines), $t=1$ (solid lines) and $t=5$ (dash-dotted lines): (a) the density, (b) the pressure, (c) the radial velocity, (d) the tangential velocity and (d) the Lorentz factor.
}
\label{fig3}
\end{center}
\end{figure}

\begin{figure}[!htbp]
\begin{center}
\scalebox{0.8}{\rotatebox{0}{\includegraphics{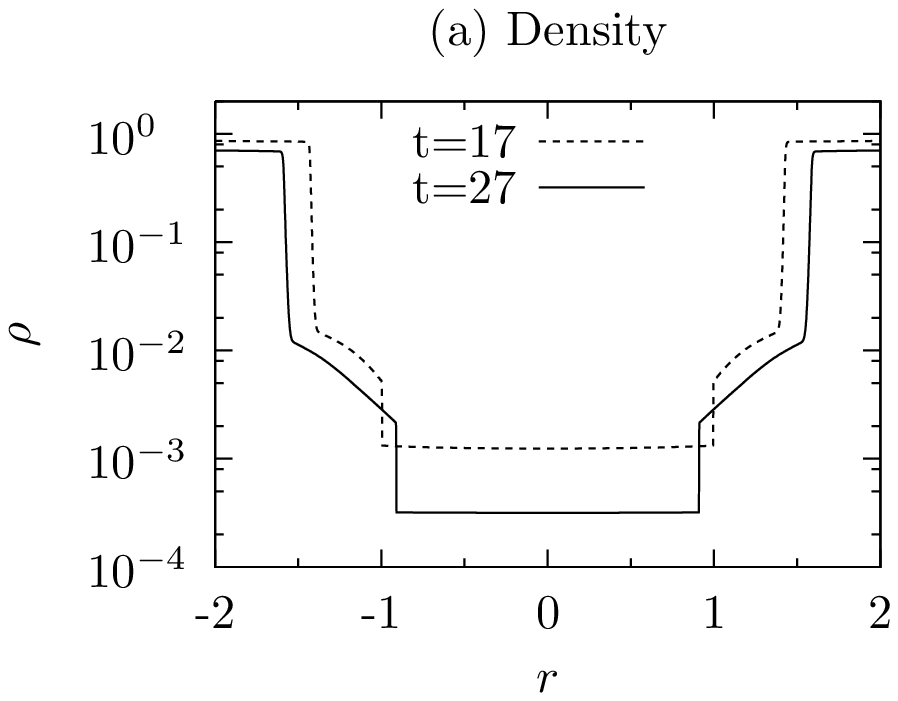}}}
\scalebox{0.8}{\rotatebox{0}{\includegraphics{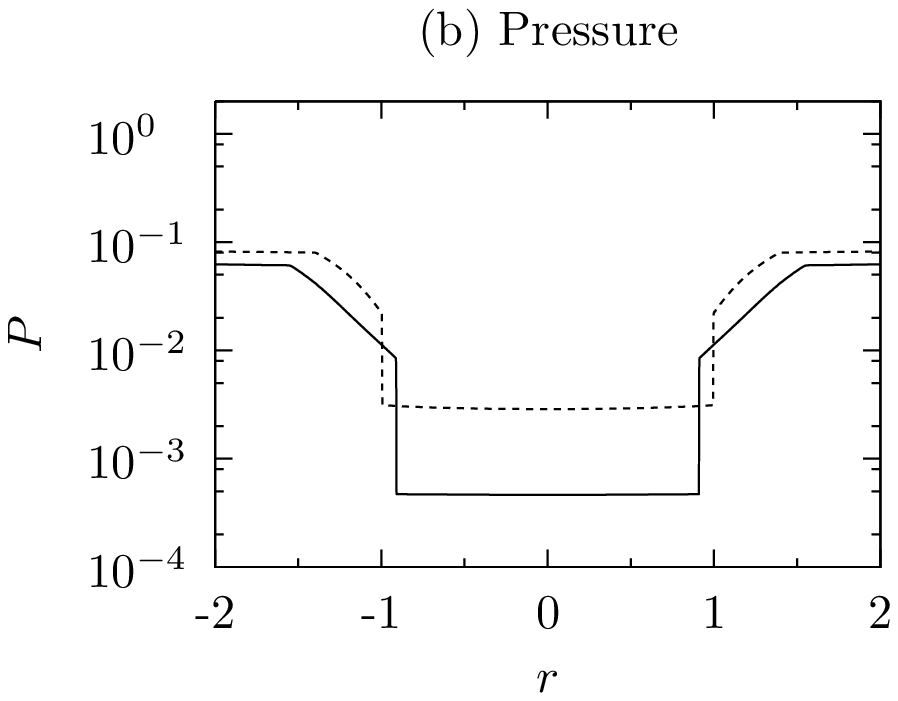}}}\\
\scalebox{0.8}{\rotatebox{0}{\includegraphics{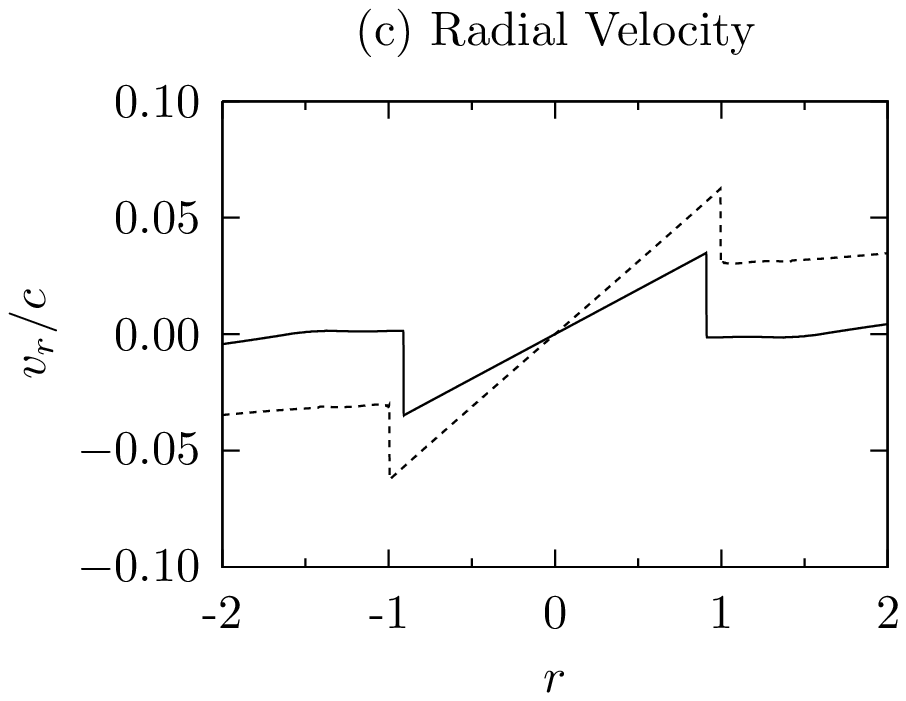}}}
\scalebox{0.8}{\rotatebox{0}{\includegraphics{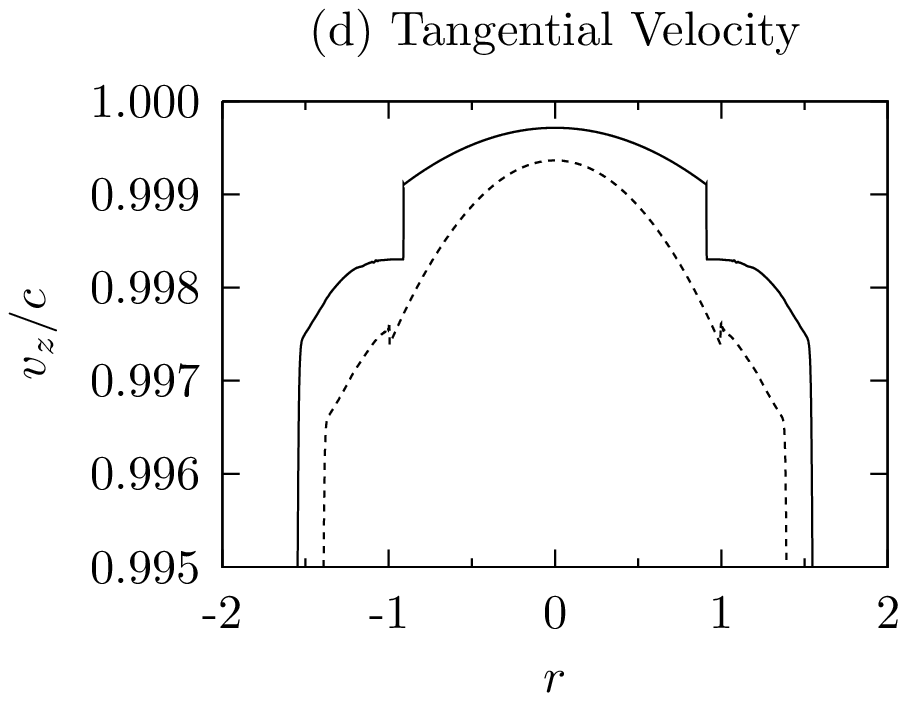}}}\\
\scalebox{0.8}{\rotatebox{0}{\includegraphics{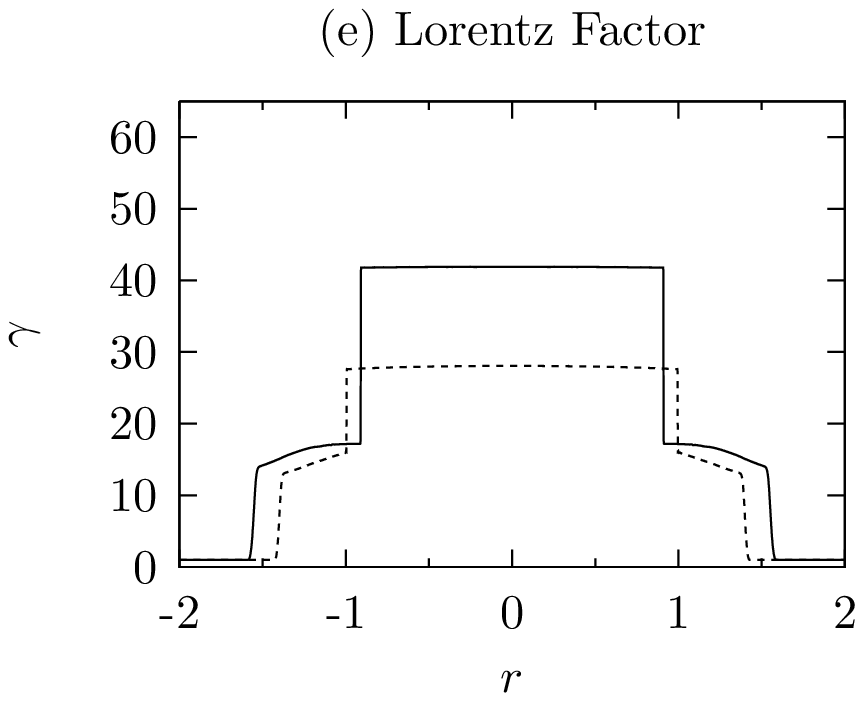}}}
% or
%\scalebox{0.58}{\rotatebox{0}{\includegraphics{f4a.eps}}}
%\scalebox{0.58}{\rotatebox{0}{\includegraphics{f4b.eps}}}
%\scalebox{0.58}{\rotatebox{0}{\includegraphics{f4c.eps}}}\\
%\scalebox{0.58}{\rotatebox{0}{\includegraphics{f4d.eps}}}
%\scalebox{0.58}{\rotatebox{0}{\includegraphics{f4e.eps}}}
\caption{
Same as Figure~\ref{fig3}, but at $t=17$ (dashed lines) and $27$ (solid lines).
}
\label{fig4}
\end{center}
\end{figure}

\begin{figure}[!htbp]
\begin{center}
\scalebox{0.8}{\rotatebox{0}{\includegraphics{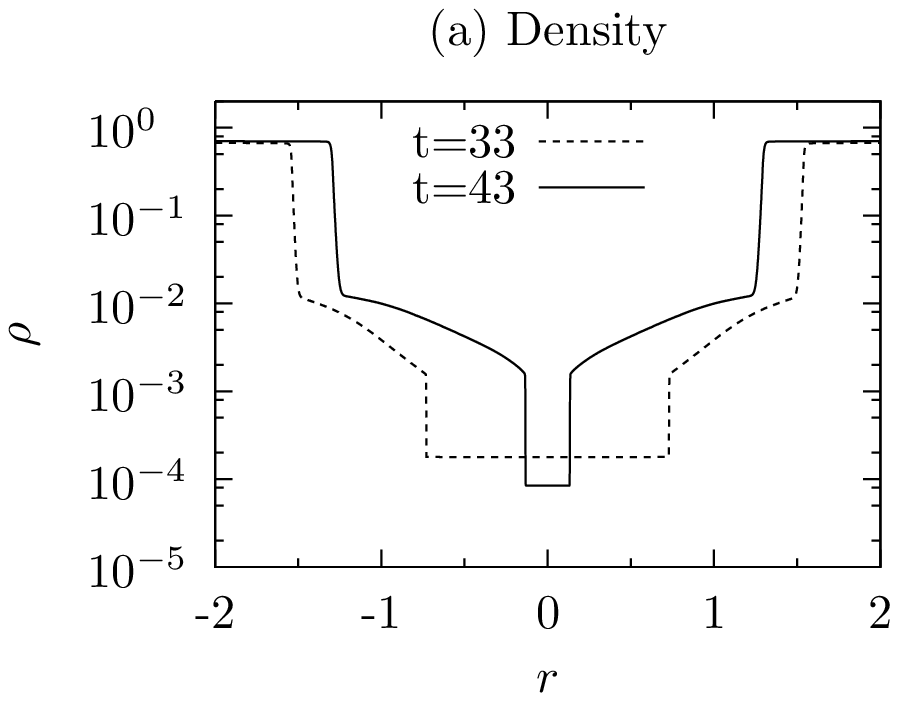}}}
\scalebox{0.8}{\rotatebox{0}{\includegraphics{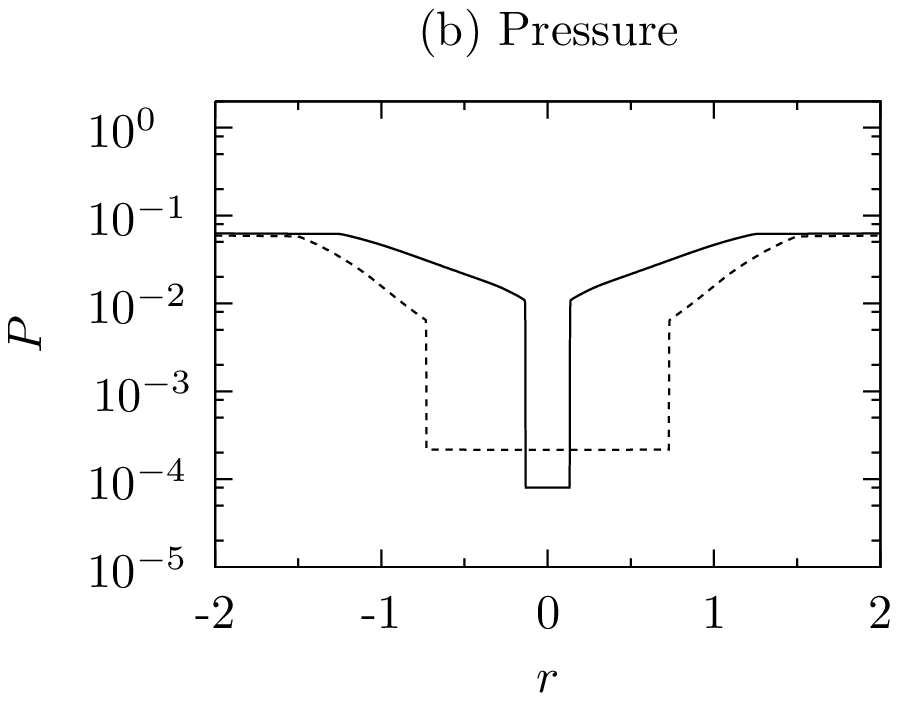}}}\\
\scalebox{0.8}{\rotatebox{0}{\includegraphics{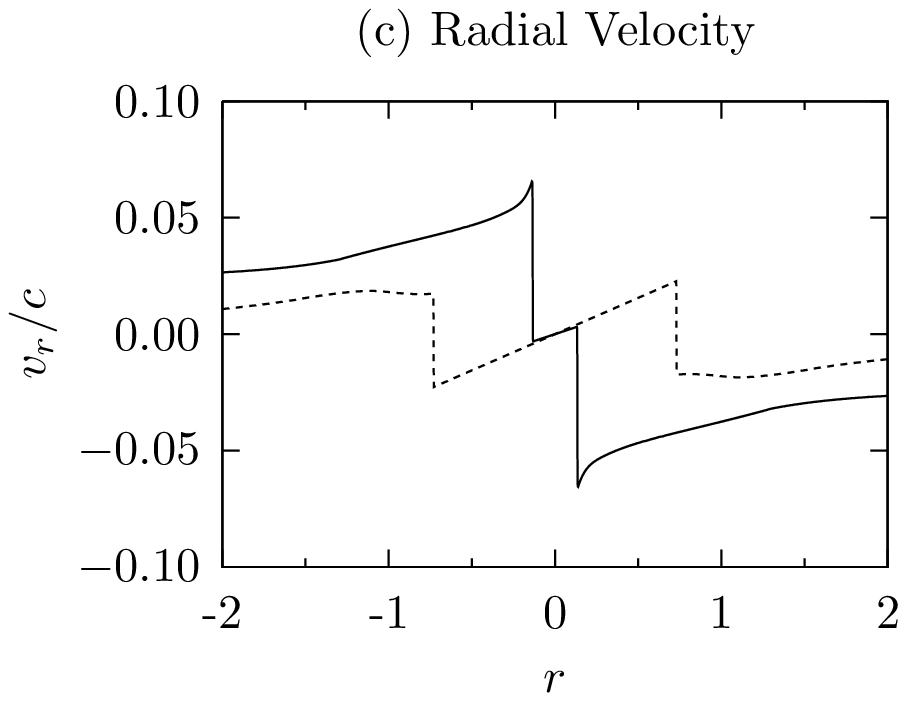}}}
\scalebox{0.8}{\rotatebox{0}{\includegraphics{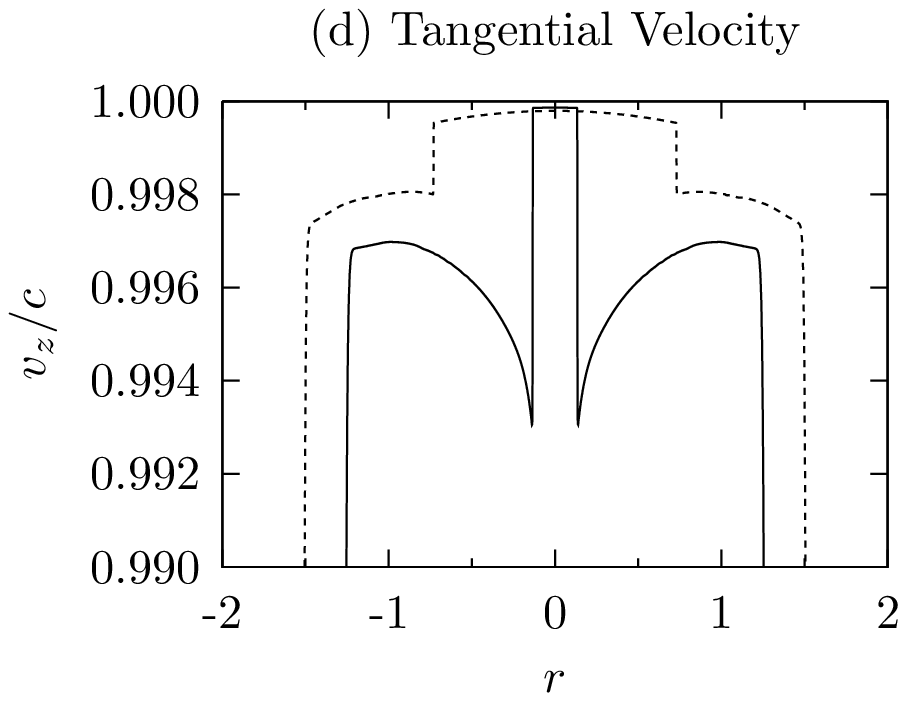}}}\\
\scalebox{0.8}{\rotatebox{0}{\includegraphics{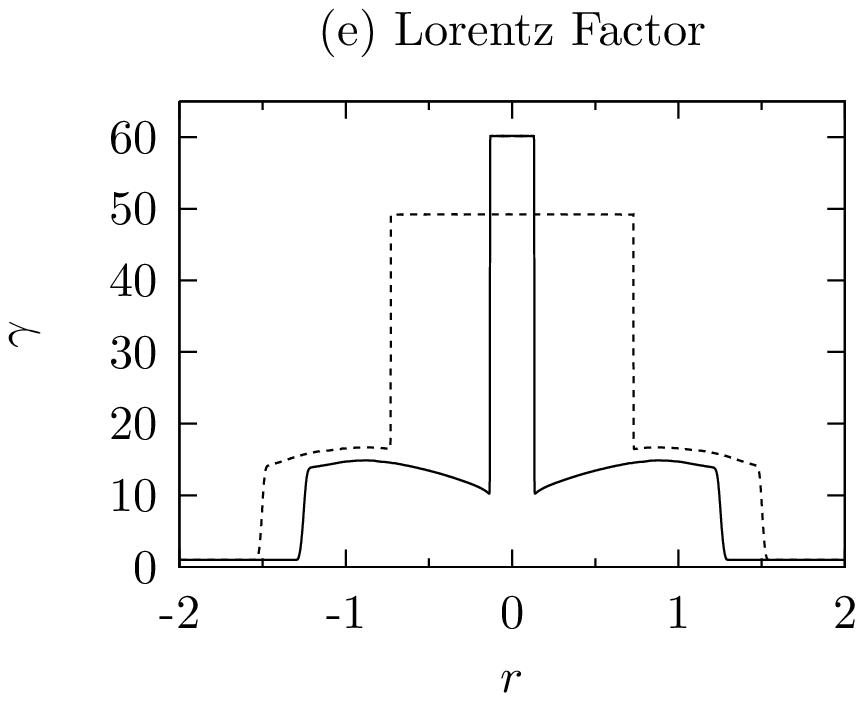}}}
% or
%\scalebox{0.58}{\rotatebox{0}{\includegraphics{f5a.eps}}}
%\scalebox{0.58}{\rotatebox{0}{\includegraphics{f5b.eps}}}
%\scalebox{0.58}{\rotatebox{0}{\includegraphics{f5c.eps}}}\\
%\scalebox{0.58}{\rotatebox{0}{\includegraphics{f5d.eps}}}
%\scalebox{0.58}{\rotatebox{0}{\includegraphics{f5e.eps}}}
\caption{
Same as Figure~\ref{fig3}, but at $t=33$ (dashed lines) and $43$ (solid lines).
}
\label{fig5}
\end{center}
\end{figure}

\begin{figure}[!htbp]
\begin{center}
\scalebox{0.8}{\rotatebox{0}{\includegraphics{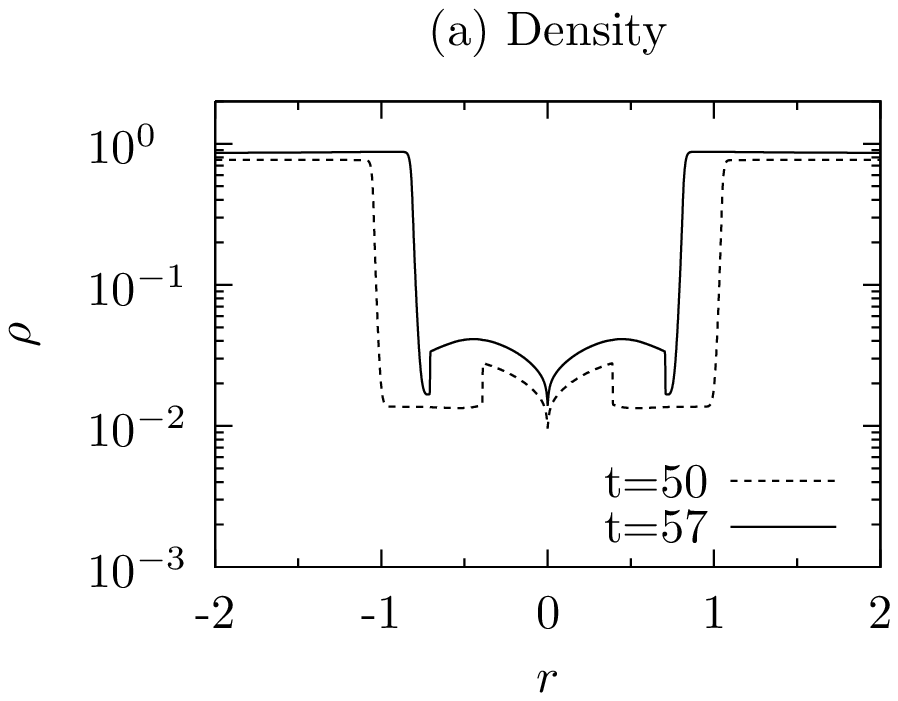}}}
\scalebox{0.8}{\rotatebox{0}{\includegraphics{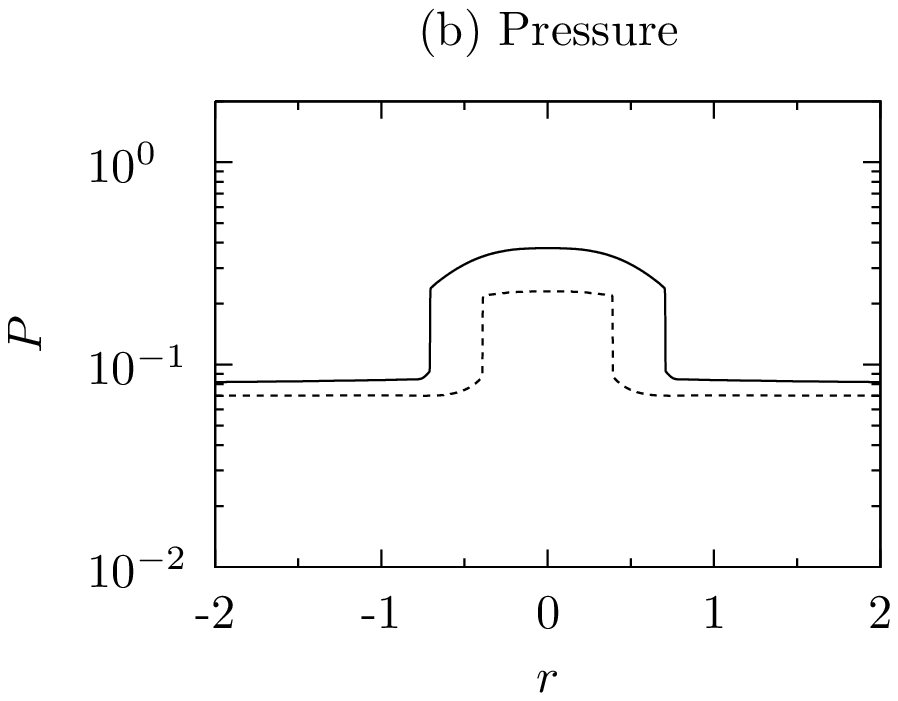}}}\\
\scalebox{0.8}{\rotatebox{0}{\includegraphics{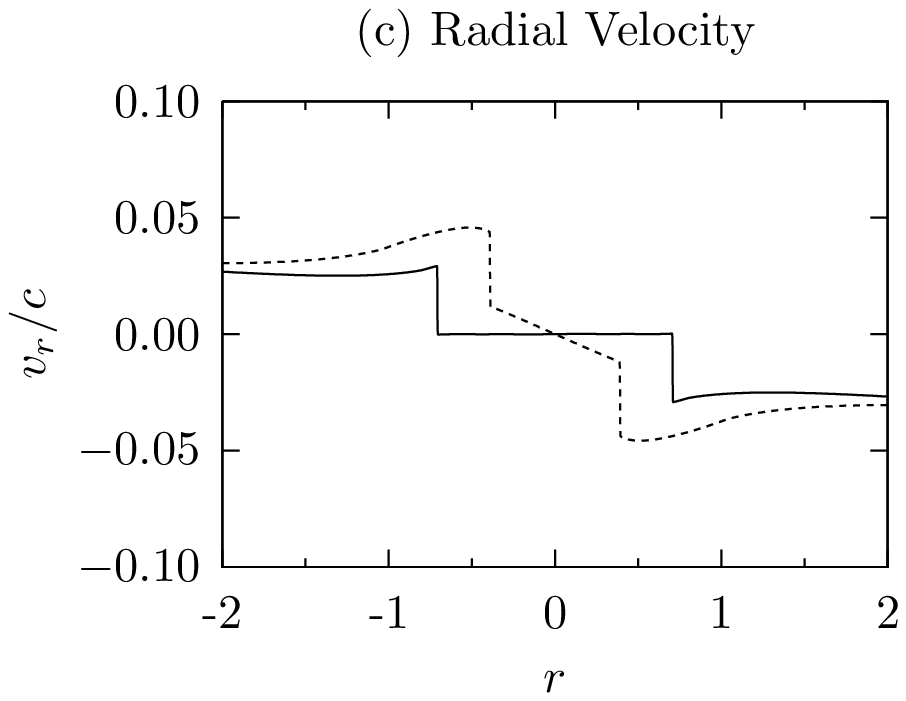}}}
\scalebox{0.8}{\rotatebox{0}{\includegraphics{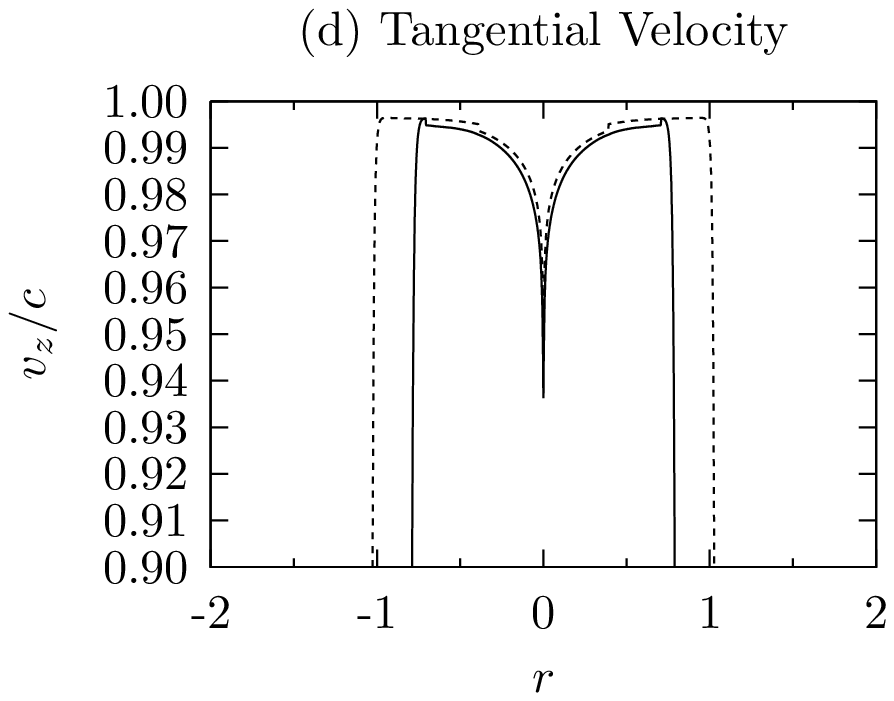}}}\\
\scalebox{0.8}{\rotatebox{0}{\includegraphics{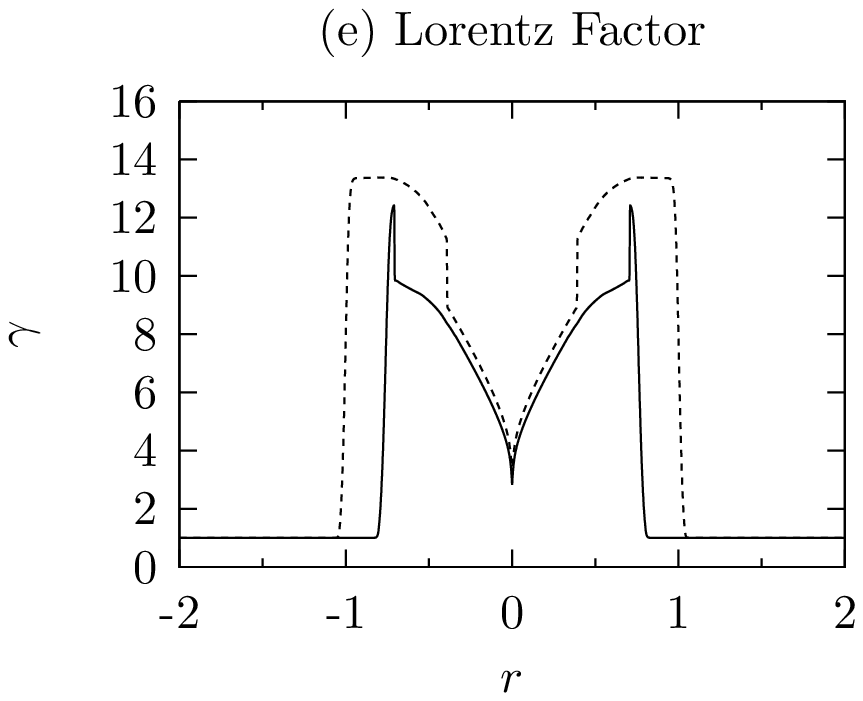}}}
% or
%\scalebox{0.58}{\rotatebox{0}{\includegraphics{f6a.eps}}}
%\scalebox{0.58}{\rotatebox{0}{\includegraphics{f6b.eps}}}
%\scalebox{0.58}{\rotatebox{0}{\includegraphics{f6c.eps}}}\\
%\scalebox{0.58}{\rotatebox{0}{\includegraphics{f6d.eps}}}
%\scalebox{0.58}{\rotatebox{0}{\includegraphics{f6e.eps}}}
\caption{
Same as Figure~\ref{fig3}, but at $t=50$ (dashed lines) and $57$ (solid lines).
}
\label{fig6}
\end{center}
\end{figure}

Figure~\ref{fig2} shows the temporal evolution of our jet-ambient medium system. The color contour 
represents the spatial distribution of (a) the density $\rho$, (b) the pressure $P$, (c) the radial velocity $v_r$,
(d) the tangential velocity $v_z$, and (e) the Lorentz factor $\gamma$. 
Snapshots of the spatial distribution of the hydrodynamics variables at the time t=5,
27, 43 and 57 are illustrated in Figure~\ref{fig3}, ~\ref{fig4}, ~\ref{fig5} and ~\ref{fig6}, respectively.
The dashed line in these figures denote the profiles of hydrodynamic variables at the time t = 0, 17, 33, 
and 50 for the better demonstration of the propagation of the rarefaction and shock waves.
In the early evolutionary stage ($0 < t < 5$: phase (i)), described in section 3.3.1, the single rarefaction acceleration of 
the gas is observed in the jet-ambient medium interface. Figure~\ref{fig3} gives the spatial distribution of the jet-ambient 
medium system at $t=0$, $1$ and $5$. The Lorentz factor of the fluid at the jet-ambient interface is boosted to 12 
due to the rarefaction acceleration when t=1 and 5. One can find that the contact 
discontinuity moves outward, that is, the jet expands since the contact discontinuity 
corresponds to the edge of the jet. Note that the head of the rarefaction wave intersects the jet axis while the Lorentz 
factor of the fluid inside the jet reaches its maximum value at the tail of the rarefaction wave when $t=5$.

Subsequently at $t \simeq 5$ the rarefaction waves converge on the central region of the jet and an incident shock wave 
is excited at the tail of the rarefaction wave in the cylindrical jet, bringing a substantial change in the dynamics (phase (ii) 
in Figure~\ref{fig2}). The shock wave is excited at the tail of the rarefaction wave as a natural result of the cylindrical 
shock tube problem in which the characteristic lines intersect. The shock wave is located at $r=1$ when $t=17$ as 
shown in Figure 4. The gas pressure in the interacting region of rarefaction 
waves, confined by the surface of the shock wave, is then further reduced and becomes lower than that of the ambient 
gas (see Figure~\ref{fig4}). Since, according to Equation~(\ref{eq: EoM}), the thermal energy is converted to the bulk 
kinetic energy of the gas, the Lorentz factor of the gas in the interacting region is further boosted. The peak Lorentz factor 
of the gas inside the rarefaction region reaches $\gamma \simeq 60$ at $t = 44$ (see Figure~\ref{fig5}), which is a factor 
$\simeq 5$ higher than that due to the single rarefaction acceleration at the jet-ambient medium interface. The boosted 
Lorentz factor of the fluid at the jet-ambient medium interface due to the single rarefaction acceleration is roughly 12 as 
shown in Figure 3, 4, 5, and 6.

The interaction of rarefaction waves generates a strong inward pressure gradient behind the jet-ambient medium 
interface which acts to decelerate the radial expansion of the jet, turning expansion of the jet into contraction around 
time $t=27$ (see Figure~\ref{fig2}(a), ~\ref{fig2}(c)). The contraction of the contact discontinuity results in converging 
flows inside the jet in the phase (iii). 

When $t=44$, the inward shock waves proceeding the converging flow collide with each other at the center of the jet and  
propagate outward. The gas bounded by the shock is compressed and heated. Since according to Equation (\ref{eq: EoM}) 
a time-increasing pressure decelerates the tangential velocity of the jet, the Lorentz factor of the jet reduces and the 
specific enthalpy increases (see Equation~(\ref{eq: gh})). Thus, in phase (iv), the Lorentz factor of the shock heated 
gas drops. The spatial distribution of the jet-ambient medium system at the end of the phase (iv) ($t=57$) is demonstrated 
in Figure~\ref{fig6}. 

When the shock wave encounters the contact discontinuity around $t=58$, the system has almost returned to its initial 
state. The jet still has a sufficiently larger tangential velocity and specific enthalpy than the ambient medium, but the gas 
pressure in the jet, however, has become smaller than that of the initial state. This is the result of the energy conversion 
from thermal energy to bulk kinetic energy of the jet. Since the system is restored to a state that is almost the same as 
the initial conditions, the three types of hydrodynamic waves, an outward propagating shock wave, a contact discontinuity 
(the edge of the jet), and a converging rarefaction wave, appear at the jet-ambient medium interface. Therefore the contracting 
radial motion of the jet becomes an expanding motion.

\subsubsection{Oscillation of the Relativistic Jet}
After phase (iv), the system has returned to conditions similar to the initial conditions, and the system repeats the cycle of 
phases (i)--(iv) until the pressure of the jet becomes equal to that of the ambient gas. During the cycle, the radial motion of 
the jet oscillates between expansion and contraction, and the system undergoes successive exchanges between thermal energy 
and kinetic energy. Figure~\ref{fig7} shows the temporal evolution of (a) the jet width, (b) the maximum and average tangential 
velocity, (c) the maximum and average Lorentz factor, and (d) the average of the specific enthalpy in the jet. The oscillations of the 
Lorentz factor and the averaged specific enthalpy are in anti-phase in accordance with the relativistic Bernoulli relation~(\ref{eq: gh}). 

When the pressure inside the jet becomes almost equal to the ambient gas around time $t=1000$, the oscillation of the jet ends. 
The spatial distributions of the density, pressure, radial and tangential velocity and Lorentz factor at $t=2000$ are depicted in Figure~\ref{fig8}. 
At the quasi-steady state, there exists an accelerated region localized to the boundary layer of the jet, as is shown in Figure~\ref{fig8}(d) and~(e). 
The physical parameters in the interface of the jet are comparable to those expected from single rarefaction acceleration 
\citep{Aloy06}. This is because the oscillating motion during the relaxation stage, shown in Figure~\ref{fig7}, has little 
influence on the physical parameters in the boundary layer. At the final quasi-steady state, the pressure in the jet is the almost 
same as that of the ambient medium. The initial relativistic thermal energy of the jet is converted into the bulk kinetic energy of the jet. 
%%%%%%%%%%%%%%%%%%%%%%%%%%%%%%%%%%%%%%%%%%%%%%%%%%%%%%%%%%%%%%%%%%%%%%%%%%%%%%%%%%%%%%%%%%%%%
%%%%%%%%%%%%%%%%%%%%%%%%%%%--------------- Scaling law for the oscillation timescale ------------------%%%%%%%%%%%%%%%%%%%%%%%%%%%%
%%%%%%%%%%%%%%%%%%%%%%%%%%%%%%%%%%%%%%%%%%%%%%%%%%%%%%%%%%%%%%%%%%%%%%%%%%%%%%%%%%%%%%%%%%%%%
\subsection{Scaling Law for the Oscillation Timescale}
As shown in the section 2.2.3, the initial non-equilibrium system evolves, through a transition stage, toward a quasi-steady state 
where a hydrostatic balance is established in the radial direction. During the transition stage, the oscillation timescale is 
almost constant while the oscillation amplitude gradually decreases. In this section, we derive a scaling relation which can 
reproduce the oscillation timescale of our jet-ambient medium system.

Figure~\ref{fig7}(a) indicates that, during the transition stage, the radial size of the jet oscillates around the jet width of 
the final quasi-steady state. We then approximate the typical oscillating width of the jet by the jet width of the saturated 
state illustrated in Figure~\ref{fig8}. The typical oscillation time of the jet would be determined by the propagation time of 
the sound waves over the typical oscillating width of the jet. When using the physical parameters at the quasi-steady state, 
the oscillating time is evaluated as 
\begin{eqnarray}
\tau = \bar{\gamma}_{\rm jet,s} W_{\rm jet,s}/C_{\rm s} \label{eq: tau}
\end{eqnarray}
in the laboratory frame. Here, the subscript s stands for the physical values in the final steady state. Since the thermal energy 
inside the jet is relativistic, the sound speed is well approximated by $C_{\rm s} \simeq c/\sqrt{3}$. 

In order to derive the typical oscillation time of the jet-ambient medium system, we need to estimate the average Lorentz factor in the 
jet $\bar{\gamma}_{\rm jet,s}$ in the quasi-steady state. Since the total energy in the jet is almost conserved during the transition 
phase\footnote{Our models indicate that the loss of the total energy is less than $0.4$\% in the transition phase}, we can obtain 
the following relation neglecting the rest mass energy, which is smaller than the relativistic thermal energy in the jet,
\begin{eqnarray}
{W_{\rm jet, s}}^2 {\bar{\gamma}_{\rm jet, s}}^2 P_{\rm amb, 0} = {W_{\rm jet, 0}}^2 {\gamma_{\rm jet, 0}}^2 P_{\rm jet, 0} \;. \label{eq: average}
\end{eqnarray}
Note that we replace the pressure in the jet in the steady state $P_{\rm jet, s}$ with that of the initial ambient 
medium $P_{\rm amb,0}$ because a pressure balance is established between inside and outside the jet in the 
final quasi-steady state.

From Equations~(\ref{eq: tau}) and~(\ref{eq: average}), we can give the scaling law for the oscillation time
\begin{eqnarray}
\tau  = \sqrt{3} \gamma_{\rm jet,0} \biggl ( \frac{W_{\rm jet,0}}{c} \biggr ) \biggl ( \frac{P_{\rm jet,0}}{P_{\rm amb,0}} \biggr )^{1/2} \;. \label{eq: t_oscillation}
\end{eqnarray}
In Figure~\ref{fig9}, we plot the oscillation time averaged over ten cycles for numerical runs with different initial pressure 
ratios. The solid line represents the analytic scaling we derived.  This indicates that our numerical results are well captured 
by our simple scaling law shown in Equation~(\ref{eq: t_oscillation}). 

Magnetic fields, which are expected to exist in relativistic jets from AGNs and GRBs \citep{Blandford82, Uchida85, Shibata86, 
Blandford00, Lyutikov03}, may alter the oscillation mechanism we have presented. In the poynting flux jet, the magnetic pressure is 
dominant over the gas pressure. Since the sum of the magnetic pressure and the gas pressure contributes to the rarefaction acceleration 
\citep{Mizuno08, Aloy08, Zenitani10, Tchekhovskoy10, Komissarov10}, it is expected that the magnetic pressure would play much 
the same role as the relativistic thermal pressure in the jet oscillation.
%%%%%%%%%%%%%%%%%%%%%%%%%%%%%%%%%%%%%%%%%%%%%%%%%%%%%%%%%%%%%%%%%%%%%%%%%%%%%%%%%%%%%%%%%%%%%
%%%%%%%%%%%%%%%%%%%%%%%%%%%--------------- S3 2D simulations ------------------%%%%%%%%%%%%%%%%%%%%%%%%%%%%
%%%%%%%%%%%%%%%%%%%%%%%%%%%%%%%%%%%%%%%%%%%%%%%%%%%%%%%%%%%%%%%%%%%%%%%%%%%%%%%%%%%%%%%%%%%%%
\section{Two-dimensional Simulations}

\subsection{Numerical Models and Setup}
We investigate how the oscillating motion found in the 1D model affects the propagation dynamics and the structure of 
the jet in a more realistic two dimensional axisymmetric system. Our axisymmetric simulation of the jet propagation has 
been carried out in cylindrical coordinates $(r,z)$, where the z-axis coincides with the symmetric axis (see Figure~\ref{fig1}). 
Relativistically hot flow is continuously injected into the ambient medium from the lower boundary of the computational 
domain, $z=z_{\rm low}$. The governing equations we solved are 
\begin{eqnarray}
\frac{\partial}{\partial t}(\gamma \rho)                   &+& \frac{1}{r}\frac{\partial}{\partial r}(r \gamma \rho v_{r})  
+ \frac{\partial}{\partial z}(\gamma \rho v_{z}) = 0 \;,  \\ 
\frac{\partial}{\partial t}(\gamma^2 \rho h v_{r}) &+& \frac{1}{r}\frac{\partial}{\partial r} \biggl [ r (\gamma^2 \rho h v_{r}^2 + Pc^2) \biggr ] 
+ \frac{\partial}{\partial z} \biggl [\gamma^2 \rho h v_{r} v_{z} \biggr ] = \frac{Pc^2}{r} \;,  \\
\frac{\partial}{\partial t}(\gamma^2 \rho h v_{z}) &+& \frac{1}{r}\frac{\partial}{\partial r} \biggl [ r (\gamma^2 \rho h v_{z} v_{r}) \biggr ] 
+ \frac{\partial}{\partial z} \biggl [ \gamma^2 \rho h v_{z}^2 + Pc^2 \biggr ]= 0 \;,  \\
\frac{\partial}{\partial t}(\gamma^2 \rho h - P)      &+& \frac{1}{r}\frac{\partial}{\partial r} \biggl [ r (\gamma^2 \rho h v_{r}) \biggr ] 
+ \frac{\partial}{\partial z} \biggl [\gamma^2 \rho h v_{z} \biggr ] = 0 \;, 
\end{eqnarray}
where we use the same normalization units and symbols defined in section 2 
in order to compare the results of 1D and 2D calculations. The injection flow has 
the same hydrodynamic parameters as the 1D model, that is $\rho_{\rm jet,0 } = 0.1$, 
$P_{\rm jet,0} = 1.0$, $v_{r,0} = 0$, $v_{z,0} = 0.99c$, and $\gamma_{\rm jet,0} \sim 7$. 

For the simulation presented here, we use simple power law models for the ambient medium. When assuming a polytropic 
atmosphere, the ambient medium has pressure and density distributions
\begin{eqnarray}
P_{\rm amb} &=& P_{\rm amb,0} \biggl ( \frac{\hat{r}}{W_{\rm jet,0}} \biggr )^{-\alpha} \label{eq: P_amb} \\
\rho_{\rm amb} &=& \rho_{\rm amb,0} \biggl ( \frac{\hat{r}}{W_{\rm jet,0}} \biggr )^{-3\alpha/4}\;, \label{eq: rho_amb}
\end{eqnarray} 
where $\hat{r}=\sqrt{r^2+z^2}$ is the spherical radius in the cylindrical coordinate system and $P_{\rm amb,0}=0.1$
and $\rho_{\rm amb,0}=1$. Note that the model with $\alpha = 0$ corresponds to a uniform ambient medium model. 
For simplicity we neglect the effect of the gravity and the ambient medium does not move initially in the calculation domain. 
We set the power law index of the ambient medium $\alpha$ as $0.0$, $0.4$, and $0.8$ in the following simulations. 

The calculation domain spans ($100 W_{\rm jet,0} \times 1500 W_{\rm jet,0}$) in the ($r \times z$)-plane which 
corresponds to a $1600 \times 24000$ grid. A uniform resolution of 8 numerical cells over the radius of the injection jet is used.
At the lower boundary ($z_{\rm low}=1$) hydrodynamic variables are fixed inside the jet injection region ($0 < r < 0.5$) 
while the boundary conditions are reflective outside the jet injection region. An outflow boundary condition is imposed 
on the outer boundaries of the grid and the symmetry axis is reflecting.

\subsection{Analytic Estimation: The Size of Cusp-Shaped Boosted Region}
Our one-dimensional models suggest that the relativistically hot jet alternates between 
acceleration and deceleration phases with the oscillation period derived in Equation~(\ref{eq: t_oscillation}) 
when it propagates through a uniform ambient medium. This is a consequence of the 
in-situ conversion between thermal and bulk kinetic energies inside the jet. When the 
relativistically hot jet is continuously injected into the ambient medium in more realistic 
multi-dimensional situations, we can expect that the radial oscillating motion that appeared 
in the one-dimensional system manifests as the periodically modulated structure of the jet
along the jet propagation direction. From the scaling law for the oscillation period of the 
jet obtained in section 2.3, we estimate here the typical size of the region where the 
relativistic jet is boosted.
%, that is a cusp-shaped structure confined by the shock wave.  

The space-time diagram of Figure~\ref{fig2} shows the boosted region of the jet has a cusp shape
which corresponds to the interacting region of rarefaction waves. The multi-dimensional 
extension of this jet oscillation in the 1D system would provide the periodic formation 
of the cusp-shaped boosted region confined by oblique shocks \citep{Norman82, Sanders83}
along the propagation direction of the jet. In previous numerical works, shock waves confining 
the flow to a narrow region are observed inside relativistic jets, that is called the reconfinemet shock  
\citep[e.g., ][]{Marti97, Gomez97, Komissarov97, Komissarov98, Aloy00a, Aloy00b, Agudo01, Zhang03, Zhang04, Mizuta06, Mizuta09, Perucho07, Morsony07, Morsony10, Lazzati09, Mimica09, Nagakura11}.
The cusp-shaped boosted region, which is result from the jet oscillation due to the interacting 
rarefaction waves, would appear in the multi-dimensional system as the reconfinemet region found 
in these previous works.

We can approximate the size of the cusp-shaped boosted region, that is the reconfinemet 
region, by the propagation distance of the jet fluid during the typical oscillation time $\tau$
written in equation~(\ref{eq: t_oscillation}). Since the fluid velocity of the relativistic jet is almost equal to the 
speed of light, the propagation distance $\lambda$ is given by 
\begin{equation}
\lambda = c\tau = \sqrt{3} \gamma_{\rm jet,0} W_{\rm jet,0} \biggl ( \frac{P_{\rm jet,0}}{P_{\rm amb,0}} \biggr )^{1/2} \;. \label{eq: lambda1}
\end{equation}
The size of the reconfinement region is proportional to the Lorentz factor of the injected jet and the
square root of the initial pressure ratio between the jet and ambient medium.
When the relativistically hot jet is injected continuously into the uniform ambient medium, 
the cusp-shaped reconfinement region with the size $\lambda$ will be formed periodically 
inside the jet. The scaling law which relates the typical size of the reconfinement region 
to the pressure of the ambient medium is also analytically derived by \citet{Daly88}. 
We compare our scaling to it later in the discussion section.

In a gravitationally bounded atmosphere, which is plausible in the central engine of relativistic 
jets, the ambient gas should be stratified and have a pressure distribution along the propagation 
direction of the jet. Since the radial oscillation of the relativistic jet is controlled by the pressure 
ratio between the jet and ambient medium, the size of the reconfinement region $\lambda$ evolves 
with time when the jet propagates through the stratified medium. Adopting the pressure profile for 
the stratified ambient medium described by equation~(\ref{eq: P_amb}), the size of the reconfinement 
region along the $z$-axis is evaluated as
\begin{equation}
\lambda = \sqrt{3} \gamma_{\rm jet,0} W_{\rm jet,0} \biggl ( \frac{P_{\rm jet,0}}{P_{\rm amb,0}} \biggr )^{1/2} \biggl ( \frac{z}{W_{\rm jet,0}} \biggr )^{\alpha/2} \;. \label{eq: lambda_z}
\end{equation}
Since the fluid velocity inside the reconfinement region should be relativistic, we can 
provide the following relation by replacing $z$ in equation~(\ref{eq: lambda_z}) with $ct$, 
\begin{equation}
\lambda \propto t^{\alpha/2} \;. \label{eq: lambda2}
\end{equation} 
This implies that the cusp-shaped reconfinement region evolves self-similarly in the multi-dimensional 
system \citep[see also][]{Komissarov98, Bromberg07, Nalewajko09, Kohler11}.

\subsection{Self-Similar Evolution of Reconfinement Region}

\subsubsection{Uniform Ambient Medium Model}

Figure~\ref{fig10} shows the temporal evolution of the injected relativistically hot jet for a uniform ambient model (power-law 
index $\alpha = 0$ in equations~(\ref{eq: P_amb}) and~(\ref{eq: rho_amb})) in the early evolutionary stage. The color contour represents the spatial 
distribution of (a) the density $\rho$, (b) the pressure $P$, and (c) the Lorentz factor $\gamma$ when $t = 100$ (left), 
$200$ (middle), and $300$ (right) respectively. As expected from one-dimensional models, the cusp-shaped rarefaction 
regions confined by oblique shocks are formed inside the jet. In those regions, the fluid of the jet is accelerated due to the 
interaction of rarefaction waves and subsequent in-situ energy conversion from relativistic thermal energy to bulk 
kinetic energy (see Section 2.2). Since the rarefaction waves are repeatedly excited behind the jet head, the reconfinement 
regions are periodically formed along the jet-propagation direction and create the modulatied structure of the jet in the early
evolutionary stage. 

Figure~\ref{fig11} gives the spatial distribution of (a) the density $\rho$, (b) the pressure $P$ and (c) the Lorentz factor $\gamma$ 
when $t = 2000$. As the jet propagates, the modulated structure of the jet has a loss in coherency except at the region near the injection 
point. This is because there are vortices that are episodically created by a Kelvin-Helmholtz like instability that develops between 
the jet and the backflow near the jet head \citep{Mizuta09}. Since the continuously induced vortices make the ambient medium 
inhomogeneous, the modulation structure of the jet becomes incoherent and then the size of the reconfinement region violently 
fluctuates, except the region near the injection point where the ambient inhomogeneity remains relatively weak. 

The spatial distribution of the Lorentz factor near the jet injection point is shown in Figure~\ref{fig12}(b). In order to
compare it to the result of the 1D calculation, the temporal evolution of the Lorentz factor in the jet-ambient system of 
1D simulation is shown in Figure~\ref{fig12}(a). The size of the cusp-shaped rarefaction region in both cases seems to be
almost the same, although the vertical axis is the time in Figure~\ref{fig12}(a). The typical size of the reconfinement region 
in the jet-propagation direction can be estimated theoretically from relation~(\ref{eq: lambda1}).  With using the parameters 
$\gamma_{\rm jet,0} \simeq 7$, $W_{\rm jet,0}=1$, and $P_{\rm jet,0}/P_{\rm amb,0} = 10$, we can obtain $\lambda \simeq 40$, 
indicating our scaling relation captures the result of 2D simulation. The width of the contact layer of 2D simulation is 
larger than that of the 1D calculation due to the lower resolution of the simulation. 

The pressure on the $z$-axis in the 1D and 2D simulations is shown in Figure~\ref{fig13}(a) and (b), respectively. Note that 
the horizontal axis represents the time $t$ in Figure~\ref{fig13}(a) and (b) depicts the pressure in a 1D cut along the $z$-axis of 
Figure~\ref{fig11}(b). In 1D models, the oscillation of the jet is a transient phenomena and terminated when the pressure inside 
the jet falls to the same level as that of the ambient medium (See section 2.2.3). On the other hand, in 2D case, the pressure 
inside the jet continues to vary and a pressure balance between the jet and the ambient medium is not achieved. The predicted size 
of the reconfinement region apparently seems to exist only near the injection point.

Figure~\ref{fig14} shows the time-distance diagram for the pressure along the $z$-axis. The reconfinement region near the 
injection point does not move with time while the reconfinement region formed behind the jet head propagates along the jet 
direction. Though the modulation properties of the jet change due to the ambient medium inhomogeneity, the typical size of the 
reconfinement region is expected to be maintained at a constant value at any time as shown in Figure~\ref{fig14}. 

The temporal evolution of the Fourier spectrum of the pressure measured along the jet axis is given in Figure~\ref{fig15}.
Here the Fourier transformation $P(k)$ is obtained from 
\begin{equation} 
P(k) = \frac{1}{L_{\rm jet}} \int_{0}^{L_{\rm jet}} P_{z \rm -axis}(z) \exp^{-ikz} dz \;, 
\end{equation} 
where $L_{\rm jet}$ is the length of the jet in the propagation direction.
The solid, dashed and dotted curves indicate the spectra at the different phases $t=600$, $1200$ and $2000$, respectively.
Despite the different evolutionary phase, all the spectra show a peak at around $k/2\pi \simeq 1/40$ which corresponds to 
the inverse of the typical size of the reconfinement region derived from equation~(\ref{eq: lambda1}) and we can find that
there is a second harmonic in each spectrum. This suggests that the typical size of the reconfinement region inside the jet 
is essentially determined by the modulation caused by the interaction of rarefaction waves even though the ambient medium 
inhomogeneity causes fluctuation in jet structure.

\subsubsection{Power-law Models}

There is one main important difference between the power law models with $\alpha \ne 0$ and 
the uniform ambient model ($\alpha = 0$). For the models with decreasing pressure profile with 
distance from the launching point of the jet, the jet expands more than in the uniform ambient 
model. The jet is then further accelerated because the pressure inside the reconfinement region 
decreases more drastically and a larger amount of thermal energy is converted into bulk kinetic energy. 

Figure~\ref{fig16} shows the temporal evolution of the injected relativistically hot jet for the power law 
ambient medium model with $\alpha = 0.8$. The meanings of the color contours and the columns are the same 
as those of Figure~\ref{fig10}. Note that the vertical and horizontal scale of Figure~\ref{fig16} is 2.5 times 
larger than that of Figure~\ref{fig10}. It is found that, though the inflow flux is the same for each model, 
the maximum Lorentz factor inside the reconfinement region reaches $\simeq 180$ for the power-law model 
with case $\alpha=0.8$, while it is $\simeq 45$ for the uniform ambient medium model (see Figure~\ref{fig10}(c) 
and Figure~\ref{fig12}(b)). 

The decreasing ambient pressure results in a longer oscillation timescale as suggested in equation~(\ref{eq: t_oscillation}). 
The size of the reconfinement region in the power law models thus becomes larger than that in the uniform ambient 
medium model. Additionally, the results shown in Figure~\ref{fig16} imply that the reconfinement region evolves 
self-similarly. 

In order to confirm the self-similarity, we investigate the time trajectory of the cusp of the reconfinement region 
on the jet axis, depicted by diamonds in Figure~\ref{fig17}, after the cusp-shaped region is formed. Overplotted 
on Figure~\ref{fig17} is the curve denoting the scaling relation~(\ref{eq: lambda2}) which is obtained from the 
results of the 1D models (see section 3.2). Panels (a) and (b) correspond to the cases with $\alpha = 0.4$ and $0.8$ 
respectively. These figures confirm that the reconfinement region evolves self-similarly once the cusp-shaped 
reconfinement region is formed.

\section{Discussion}
\citet{Aloy06} discussed, for the first time, a new mechanism in relativistic 
hydrodynamics that can act as a powerful booster in jets, that is rarefaction acceleration. This 
mechanism is purely hydrodynamical and operates when the jet-ambient system satisfies 
three physical conditions, that is, the jet should be hot and overpressured, and have a 
relativistic velocity to the jet-ambient medium interface. The hot jet stores an 
internal energy comparable to or larger than its rest mass energy. When the pressure of 
the jet is larger than that of the ambient medium, the rarefaction wave is excited at the jet-ambient 
medium interface and the relativistic thermal energy of the plasma can be converted into the bulk 
kinetic energy of the jet near the interface. An expansion of the relativistically hot jet to 
the underpressured ambient medium is responsible for the rarefaction acceleration of the jet. 

The rarefaction acceleration should be commonly operated in the jet which satisfies 
these conditions, and which have been observed in many previous simulations as 
expected \citep[e.g.,][]{Gomez97, Aloy00b, Scheck02, Zhang03}. The reconfinement 
region confined by the oblique shock has been also formed in these previous simulations 
as a natural outcome of the rarefaction acceleration in the multidimensional system. However, 
the rarefaction acceleration can not be observed in the cold jet where the internal energy of the fluid 
is less than the rest mass energy of the fluid even if it shows the reconfinement shock \citep[e.g.,][]{Komissarov97}. 

\citet{Daly88} explored the dynamics of relativistic, hot and overpressured jets 
on the basis of a simplified, quasilinear, hydrodynamic equations for adiabatic, steady, 
cylindrically symmetric, and irrotational flows. In that work, they have already found the 
variety of flow structures with oscillating cross section or standing shocks. 
In addition, they have discussed scaling laws which 
relate the intrinsic properties of the jet to the pressure of the ambient medium. 

In this sense, our finding is that there exists a close connection between the rarefaction 
acceleration mechanism discussed by Aloy \& Rezzola and the flow structure with the 
reconfinement region and oscillating cross section quantitatively by using 1D and 2D 
hydrodynamic simulations.
It is remarkable that simple 1D model based on \citet{Aloy06}'s mechanism 
can reproduce complicated 2D structure of jets not only in rarefaction acceleration region 
but also in the reconfinement region so well. Since 1D model is simple and easy to 
understand, these findings would be useful for more detailed study of jets.

It should be finally emphasized that the scaling law we obtained in this work 
(equation~(\ref{eq: lambda1})) is different from that found by \citet{Daly88}. This would 
be because the approximations used for deriving the scaling in their work is applicable 
only to the system with a small pressure difference between the jet and ambient medium. 
Our scaling law can capture the flow properties of the jet-ambient medium system 
in the wider parameter range than that covered by the scaling law of \citet{Daly88}. 

The simple scaling law for the self-similarity of the reconfinement region~(\ref{eq: lambda2}) would predict 
the evolution of the reconfinement region in the context of both AGN jets and GRBs. The reconfinement 
shocks due to the interacting rarefaction waves may be able to account for the confinement of the relativistic jet from 
AGN on a sub-parsec scale \citep{Junor99, Kovalev07, Acciari09}. Some authors propose that the radio knots in 
the jet may correspond to reconfinement regions \citep[e.g.,][]{Daly88, Gomez95, Bicknell96, Komissarov97, Stawarz06, Mimica09}.
The scaling law~(\ref{eq: lambda2}) predicts that radio knots in the jet from AGNs evolve self-similarly with time
depending on the power law index of the ambient pressure distribution when the 
relativistically hot and steady jet is formed near the central engine.  

Although the numerical work in this paper are based on simulations with a constant engine which powers the jet, 
the variability injected by the central engine also leads to inhomogeneity in relativistic jets \citep{Gomez97, Mimica09, Morsony10}.
This might be a promising mechanism for generating internal shocks in relativistic jets, accounting for the 
time-variable emission properties of AGNs and GRBs \citep{Takahashi00, Piran04, Mimica05, Meszaros06}. 
The relation for inhomogeneity in the jet between the variability injected by the central engine and the interaction of 
the relativistic jet with the ambient medium is also needed to be investigated in a more realistic situation. 
Actually, for the case of power-law ambient models, we need a larger calculation domain while retaining 
the same resolution to study the effect of the vortices developed around the jet head on the propagation 
dynamics of the jet. This is because the jet expands more rapidly and the modulation structure becomes 
larger as the power-law index $\alpha$ increases. However, these studies go beyond the scope of this paper, 
but will be studied in a separate paper. 

\section{Summary}
The nonlinear evolution of the interacting rarefaction waves excited at the cylindrical jet-ambient medium interface is studied 
through one-dimensional relativistic hydrodynamic simulations. It is found that an enhanced decrease in the relativistic pressure 
due to the interaction of rarefaction waves transiently yields a more powerful boost of the Lorentz factor of the bulk jet than 
that expected from a single rarefaction wave. The cyclic in-situ energy conversion between thermal energy and bulk kinetic 
energy is a natural relativistic outcome of the jet scenario studied and responsible for the radial oscillating motion of the jet. 
The oscillation timescale is characterized by the initial pressure ratio of the jet to the ambient medium, and follows a simple 
scaling relation $\tau_{\rm oscillation} \propto (P_{\rm jet,0} / P_{\rm amb,0})^{1/2}$. 

It is confirmed from extended two-dimensional simulations that 
repeated excitation and convergence of rarefaction waves result in the alignment of the interacting regions of rarefaction 
waves, confined by oblique shocks, along the propagation direction of the jet
when a relativistically hot jet propagates through an ambient medium.
The evolution of the reconfinement region in which the fluid of the jet is powerfully boosted due to the interacting rarefaction waves 
has a self-similar property when the relativistic jet propagates through the ambient medium which has a power law 
pressure distribution. The evolution of the size of the reconfinement region along the jet direction $\lambda$ follows a simple 
scaling law $\lambda \propto t^{\alpha/2}$ where $\alpha$ is the power law index of the pressure distribution. Especially, 
in the uniform ambient medium model ($\alpha=0$) the typical size of the reconfinement region inside the jet is essentially 
determined by the modulation caused by the interaction of rarefaction waves. 
\acknowledgments
We thank Hiroyuki R. Takahashi and Akira Mizuta for thier useful discussions.
We thank Andrew Hillier for his careful reading of the manuscript. 
This work was supported by the Grant-in-Aid for the global COE program 
``The Next Generation of Physics, Spun from Universality and Emergence''
from the Ministry of Education, Culture, Sports, Science and Technology (MEXT) of Japan. 
Numerical computations were carried out on Cray XT4 at Center for Computational Astrophysics, 
CfCA, of National Astronomical Observatory of Japan and on SR16000 at  YITP in Kyoto University.
%The page charge of this paper is supported by CfCA.
J.M. acknowledges support by the Research Fellowship of the Japan Society for the Promotion of Science (JSPS).
%%%%%%%%%%%%%%%%%%%%%%%%%%%%%%%%%%%%%%%%%%%%%%%%%%%%%%%%%%%%%%%%%%%%%%%%%%%%%%%%%%%%%%%%%%%%%%%
\appendix
\section{Convergence Tests}

\subsection{One-dimensional Riemann Problem with Transverse Velocity}

The 1D relativistic Riemann problem with large transverse velocity requires high resolution in order to numerically resolve 
the complicated structure \citep{Zhang06, Mizuta06, Mizuno08}. 
We have tested our code on a 1D (x-direction) Riemann problem with transverse velocity using uniform resolutions of $\Delta x = 10^{-3}$ 
and $2 \times 10^{-5}$ when the calculation domain spans $0 < x < 1$. We have used the initial conditions from \citet{Zhang06} with 
adiabatic index $\Gamma = 5/3$ as follows. 

Left state ($0 < x < 0.5$): $\rho_{\rm L}=1$, $P_{\rm L}=10^{3}$, $v_{x, \rm L}=0$ and $v_{z, \rm L}=0.99c$.

Right state ($0.5 < x < 1$): $\rho_{\rm R}=1$, $P_{\rm R}=10^{-2}$, $v_{x, \rm R}=0$ and $v_{z, \rm R}=0.99c$.

Figure~\ref{fig18} denotes numerical and analytic solutions of this problem at $t=0.6$. Analytic solutions in the appendix 
are calculated with the code of \citet{Giacomazzo06}. Upper and lower panels correspond to 
the case with $\Delta x = 10^{-3}$ and $2 \times 10^{-5}$, respectively. In the low-resolution run, we can find the diffusion 
of the contact discontinuity. The position of the right-going shock front is not correctly captured. On the contrary, the left-going 
rarefaction wave is resolved with good accuracy. In the high-resolution run, the numerical solution captures the analytic solution well, 
although there is some undershoot in the transverse velocity $v_{z}$ at the contact discontinuity.

\subsection{One-dimensional Hydrodynamic Relativistic Jet Model}

We have also performed another convergence test using our jet model parameters in 1D (x-direction) cartesian coordinate 
using a uniform high resolution of $\Delta x =2 \times 10^{-5}$ and the resolution of $\Delta x = 10^{-3}$ adopted in section 2. 
We set the initial conditions as follows.

Left state ($0 < x < 0.5$): $\rho_{\rm L}=0.1$, $P_{\rm L}=1$, $v_{x, \rm L}=0$ and $v_{z, \rm L}=0.99c$.

Right state ($0.5 < x < 1$): $\rho_{\rm R}=1$, $P_{\rm R}=0.1$, $v_{x, \rm R}=0$ and $v_{z, \rm R}=0$.

Figure~\ref{fig19} shows the numerical and analytic solutions of our jet model at $t=0.1$ in the case with 
$\Delta x = 2 \times 10^{-5}$. The numerical solution resolves the hydrodynamic structure of the analytic solution very well. 
In the case with $\Delta x = 10^{-3}$, the position of the right-propagating shock front is not correctly captured 
while the numerical solution is able to capture the  hydrodynamic structure of the rarefaction wave and the contact discontinuity 
as shown in Figure~\ref{fig20}. In this paper, we focus on the interaction of rarefaction waves excited at the jet-ambient 
medium interface. Since the shock wave propagates outward from the jet-ambient medium interface, the interaction of rarefaction 
waves inside the jet is not affected by the outward-going shock wave.

\begin{figure}[!htbp]
\begin{center}
\scalebox{0.7}{\rotatebox{0}{\includegraphics{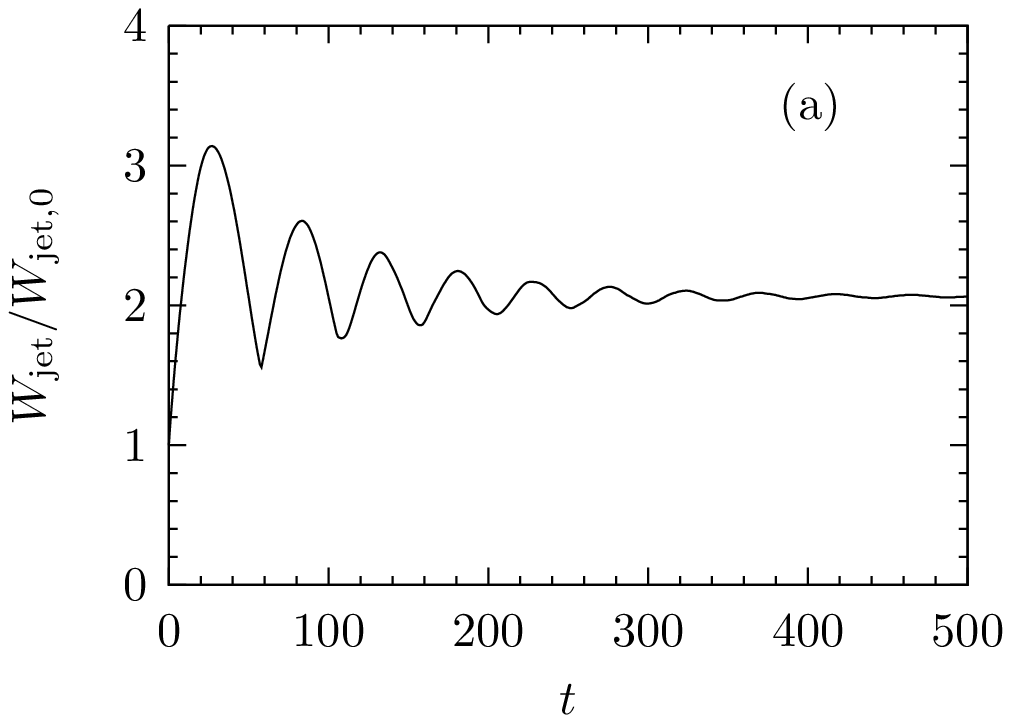}}}
\scalebox{0.7}{\rotatebox{0}{\includegraphics{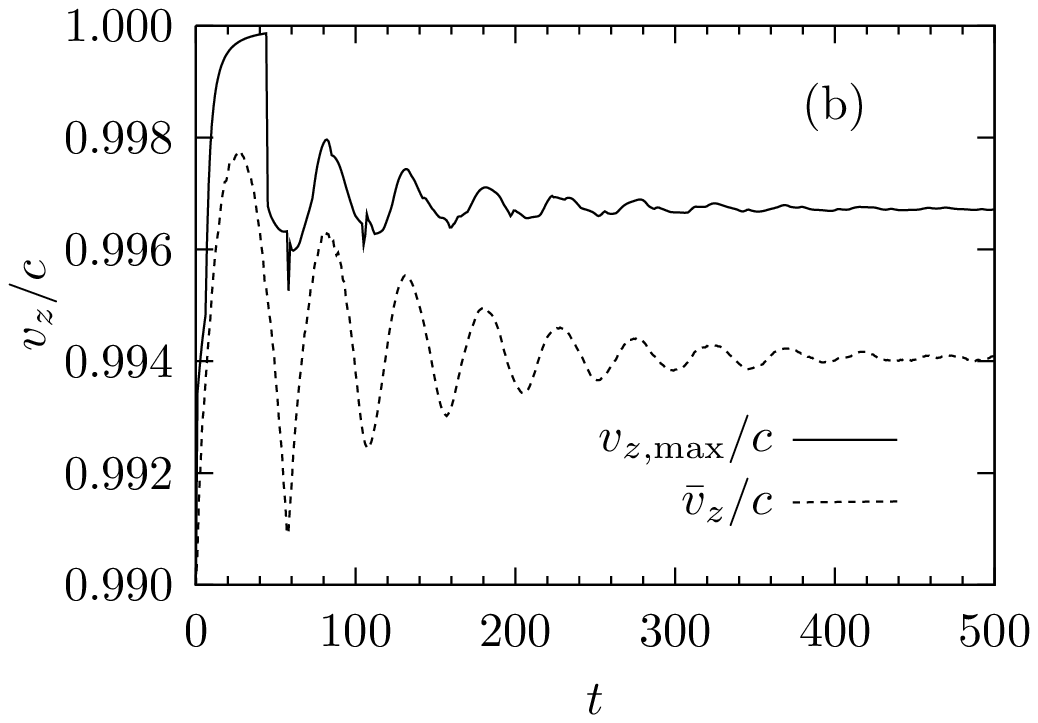}}}\\
\scalebox{0.7}{\rotatebox{0}{\includegraphics{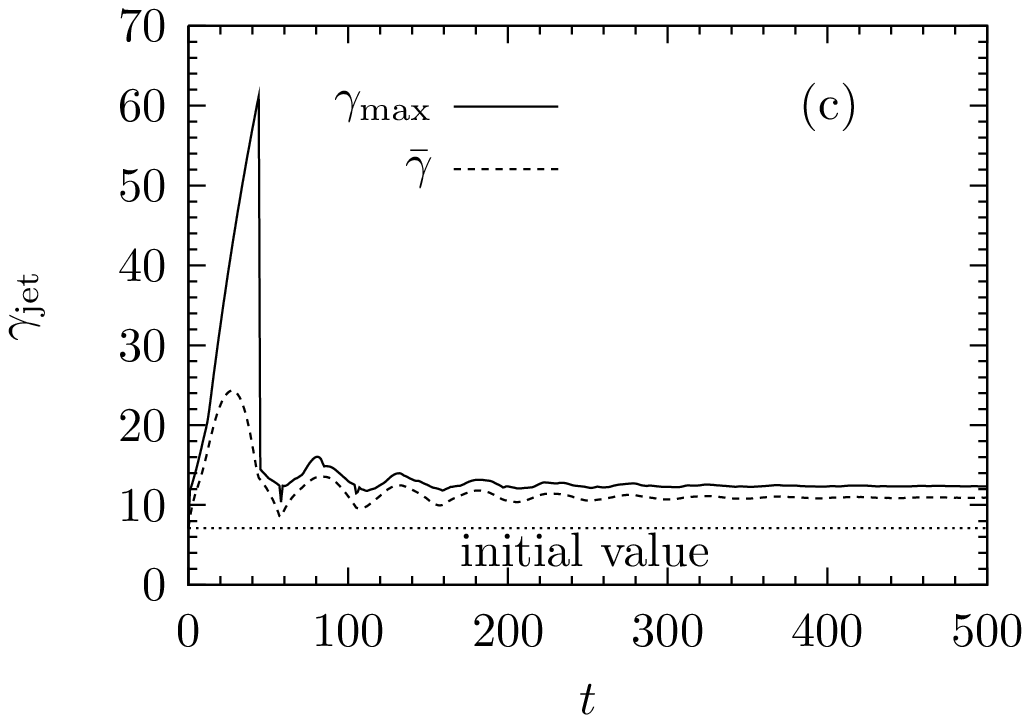}}}
\scalebox{0.7}{\rotatebox{0}{\includegraphics{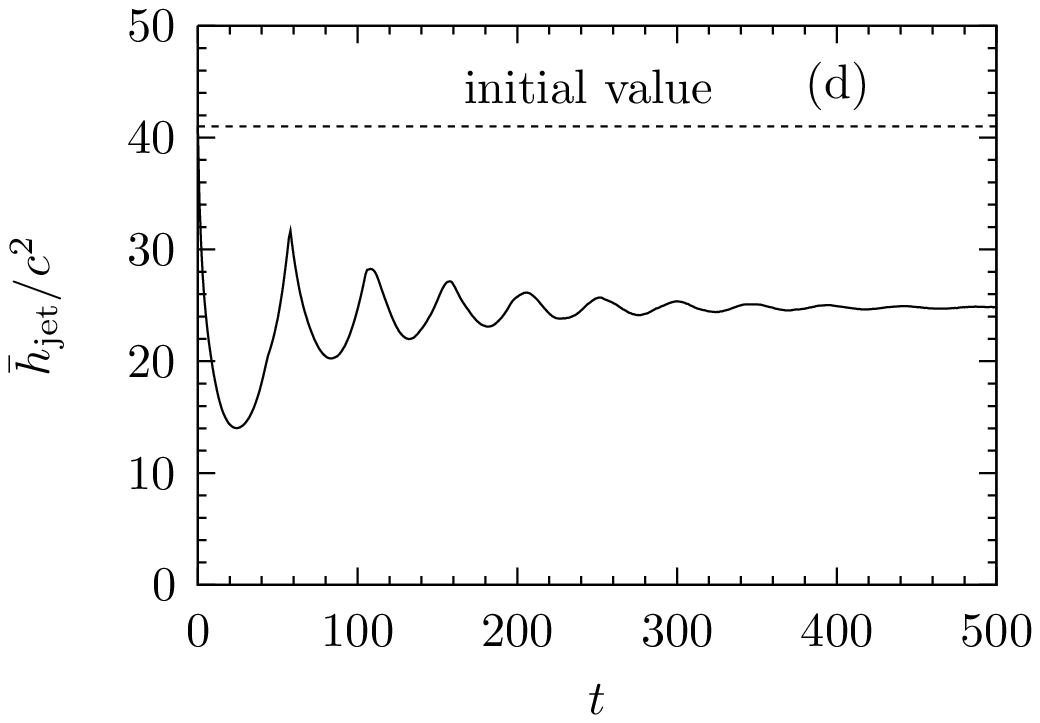}}}
\caption{The temporal evolution of (a) the jet width, (b) the maximum and average of the tangential velocity ($v_{z, \rm max}$ and $\bar{v}_{z}$), (c) the maximum and average of the Lorentz factor ($\gamma_{\rm max}$ and $\bar{\gamma}$), and (d) the average of the specific enthalpy in the jet. The initial Lorentz factor inside the jet is roughly 7 and shown by the dotted line in panel (c). However, the initial specific Lorentz factor of the jet is 41 and presented by a dashed line in panel (d).
}
\label{fig7}
\end{center}
\end{figure}

\begin{figure}[!htbp]
\begin{center}
\scalebox{0.8}{\rotatebox{0}{\includegraphics{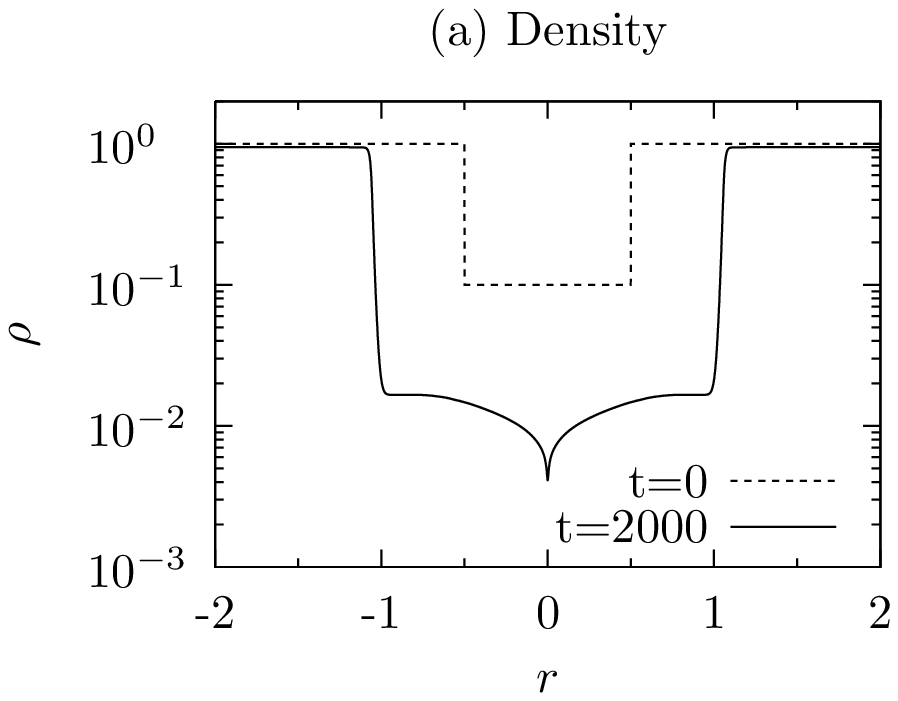}}}
\scalebox{0.8}{\rotatebox{0}{\includegraphics{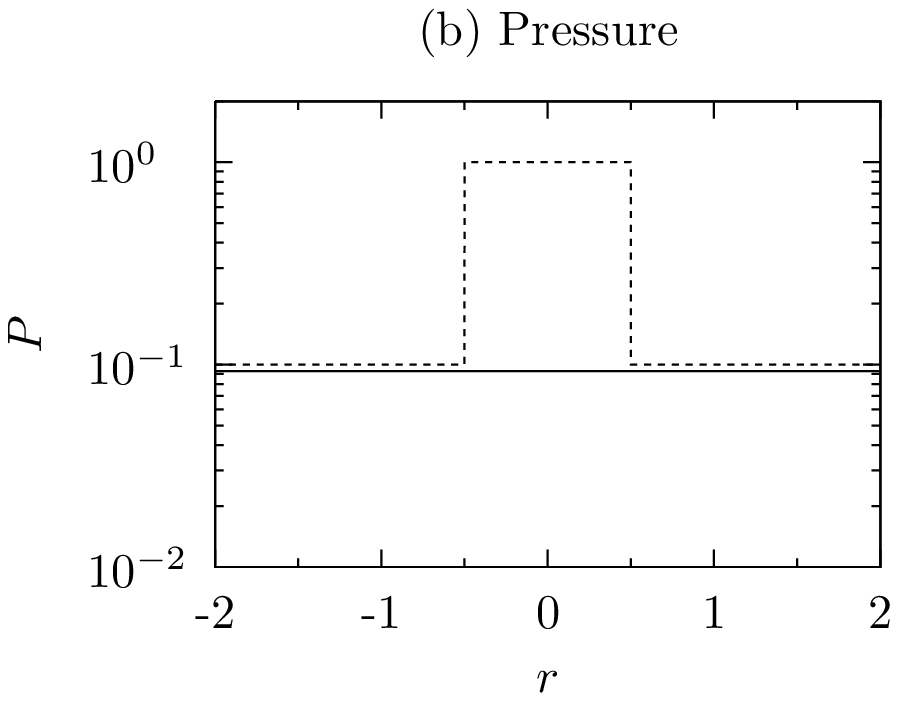}}}\\
\scalebox{0.8}{\rotatebox{0}{\includegraphics{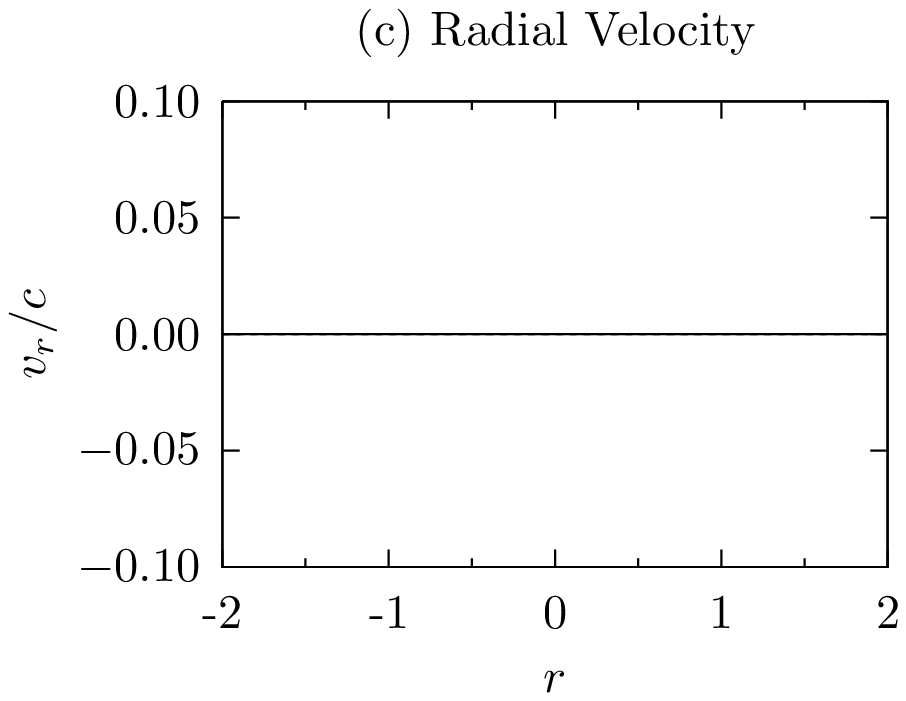}}}
\scalebox{0.8}{\rotatebox{0}{\includegraphics{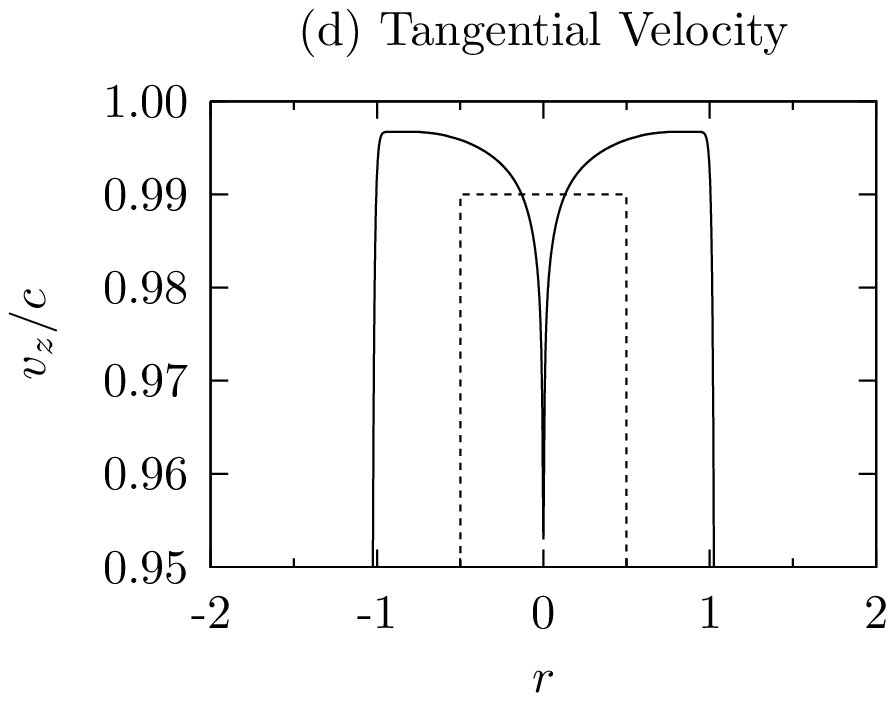}}}\\
\scalebox{0.8}{\rotatebox{0}{\includegraphics{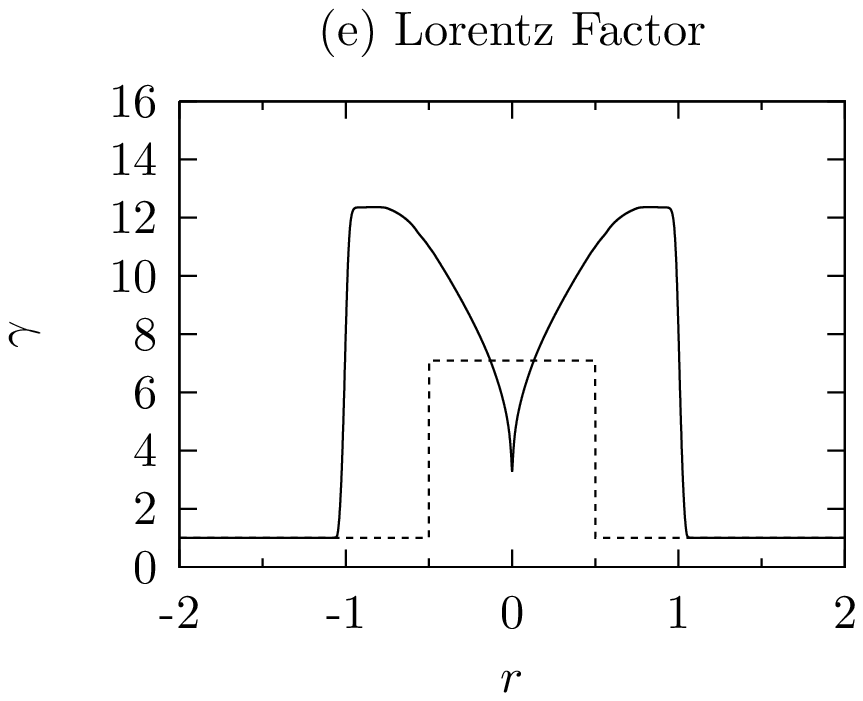}}}
% or
%\scalebox{0.58}{\rotatebox{0}{\includegraphics{f8a.eps}}}
%\scalebox{0.58}{\rotatebox{0}{\includegraphics{f8b.eps}}}
%\scalebox{0.58}{\rotatebox{0}{\includegraphics{f8c.eps}}}\\
%\scalebox{0.58}{\rotatebox{0}{\includegraphics{f8d.eps}}}
%\scalebox{0.58}{\rotatebox{0}{\includegraphics{f8e.eps}}}
\caption{The spatial distribution of the initial non-equilibrium and final quasi-steady states of the jet: (a) the density, (b) the pressure, (c) the radial velocity, (d) the tangential velocity and (e) the Lorentz factor. The dashed and solid lines represent the initial ($t=0$) and final ($t=2000$) state, respectively.}
\label{fig8}
\end{center}
\end{figure}

\begin{figure}[!htbp]
\begin{center}
\scalebox{1}{\rotatebox{0}{\includegraphics{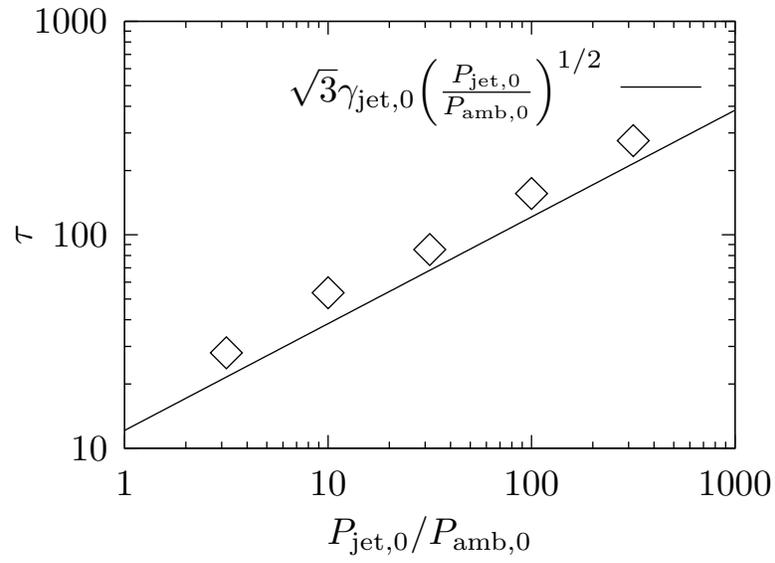}}}\\
\caption{Relation between the oscillation time scale of the system and initial pressure ratio of the jet to the ambient gas (Equation~(\ref{eq: t_oscillation})). Diamonds denote the oscillation time averaged over ten cycles for each parameter.}
\label{fig9}
\end{center}
\end{figure}

\begin{figure}[!htbp]
\begin{center}
\scalebox{0.45}{\rotatebox{0}{\includegraphics{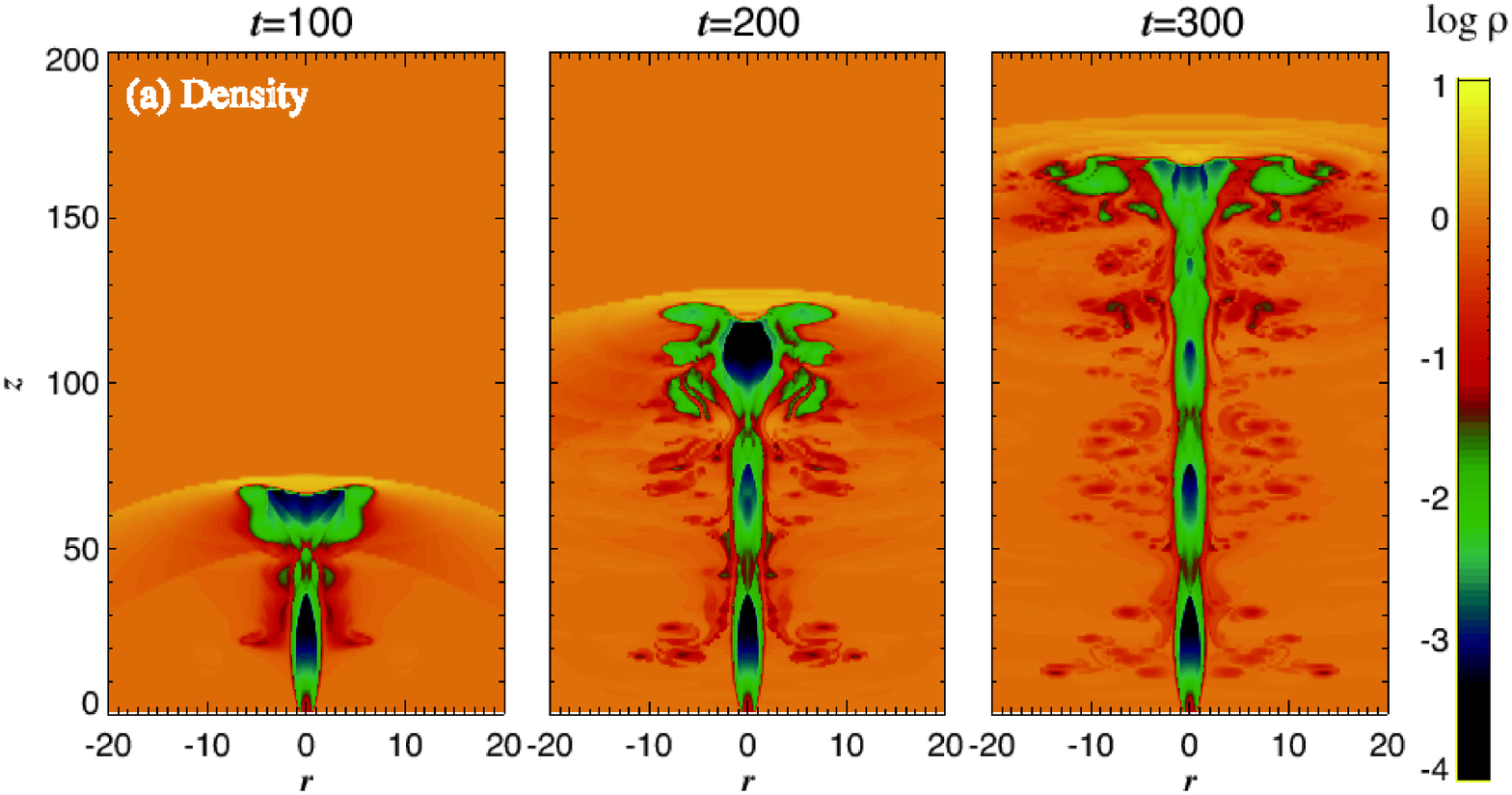}}}\\
\scalebox{0.45}{\rotatebox{0}{\includegraphics{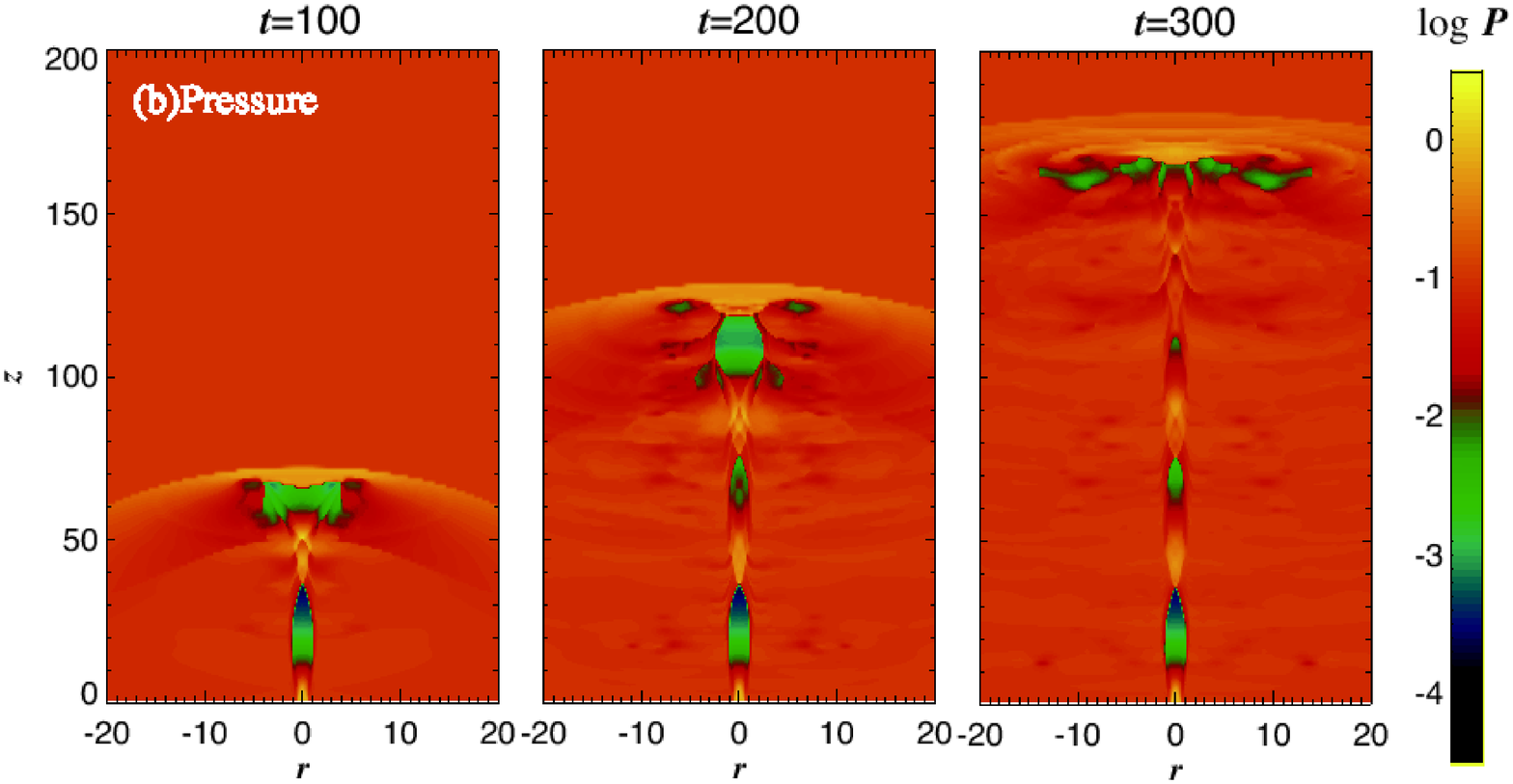}}}\\
\scalebox{0.45}{\rotatebox{0}{\includegraphics{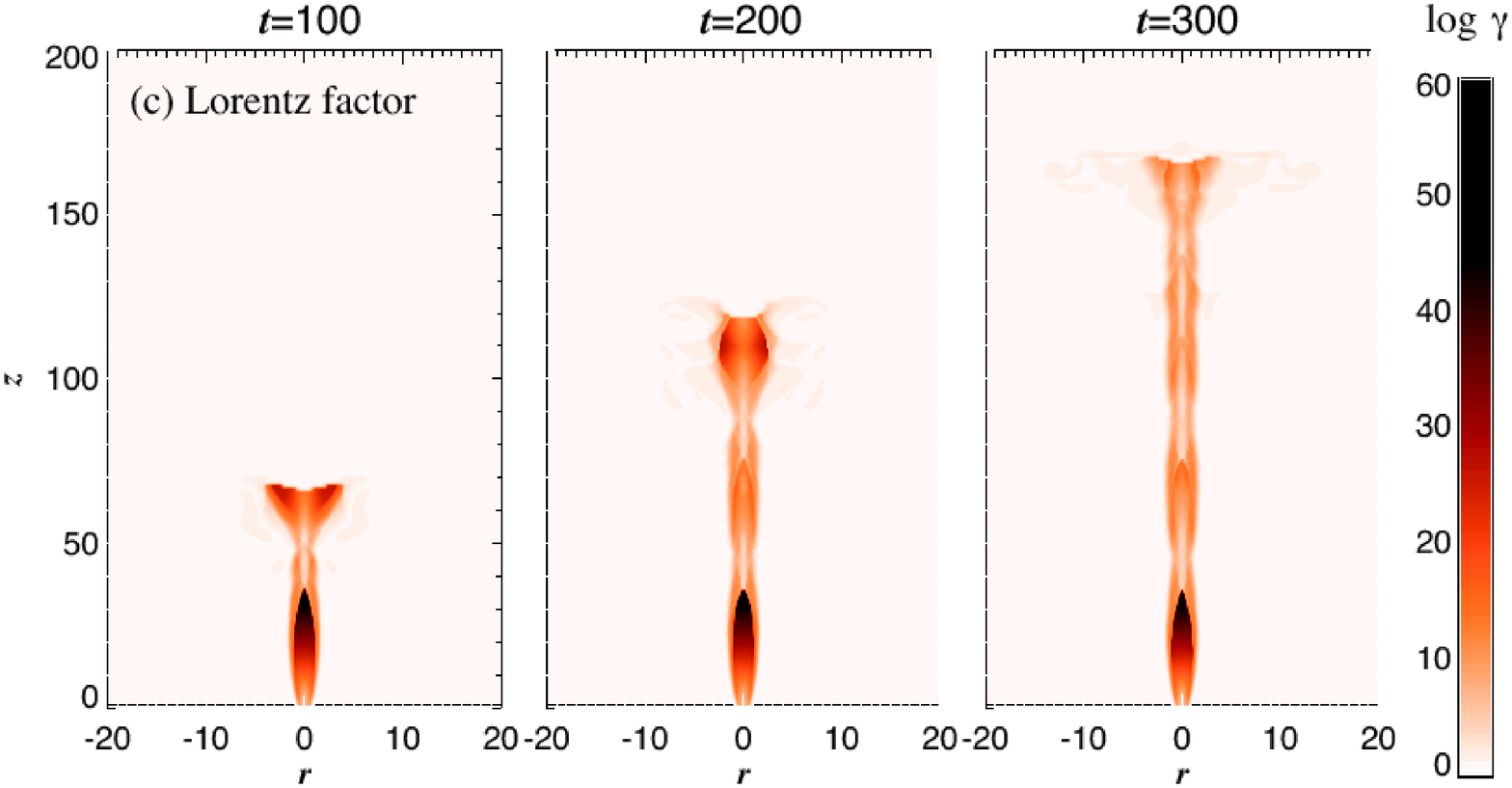}}}
\caption{
Time evolution of the injected relativistic jet into the uniform ambient medium: (a) the density, (b) the pressure and (c) the Lorentz factor. The left, middle, and right columns correspond to $t = 100$, $200$, and $300$, respectively. 
(A color version of this figure is available in the online journal.)
}
\label{fig10}
\end{center}
\end{figure}

\begin{figure}[!htbp]
\begin{center}
\scalebox{0.6}{\rotatebox{90}{\includegraphics{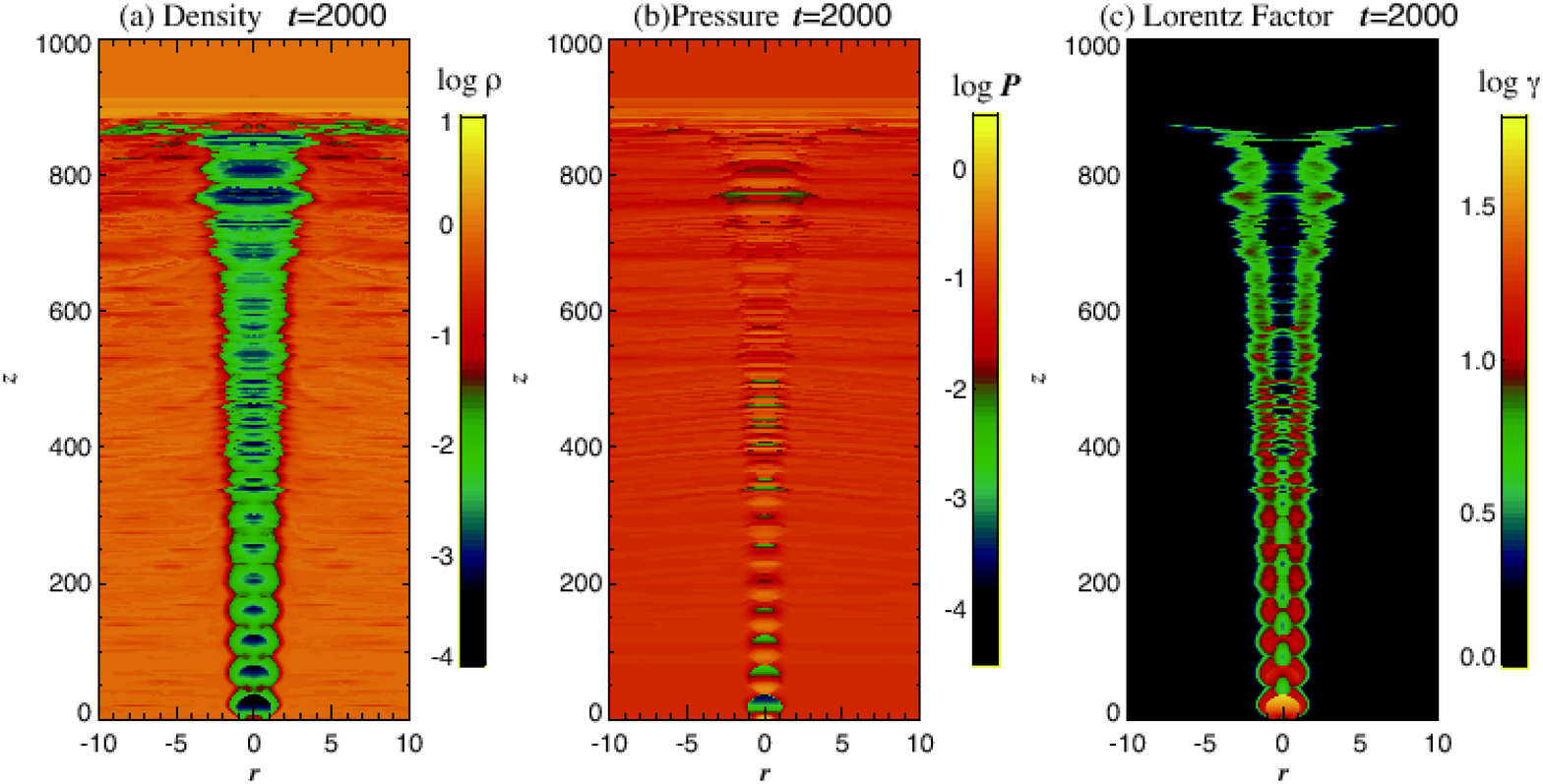}}}
\caption{
Snapshots of the spatial distribution of (a) the density, (b) the pressure and (c) the Lorentz factor in the uniform ambient medium model when $t=2000$. (A color version of this figure is available in the online journal.)
}
\label{fig11}
\end{center}
\end{figure}

\begin{figure}[!htbp]
\begin{center}
\scalebox{2}{\rotatebox{0}{\includegraphics{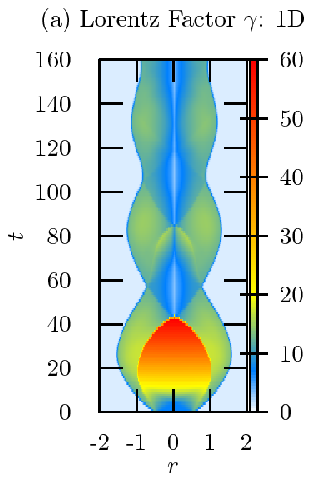}}}
\scalebox{2}{\rotatebox{0}{\includegraphics{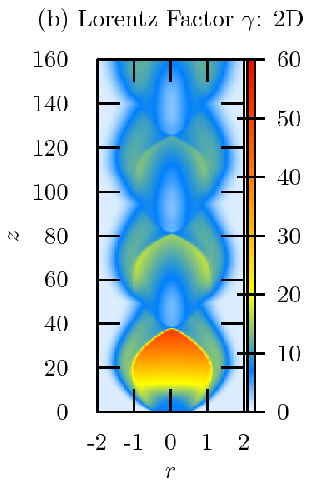}}}
\caption{Left panel represents the time evolution of the Lorentz factor in the jet-ambient medium system of 1D simulation. Right panel demonstrates the spatial distribution of the Lorentz factor at $t=2000$ in the uniform ambient model ($\alpha=0$) of 2D simulation. (A color version of this figure is available in the online journal.)
}
\label{fig12}
\end{center}
\end{figure}

\begin{figure}[!htbp]
\begin{center}
\scalebox{1.1}{\rotatebox{0}{\includegraphics{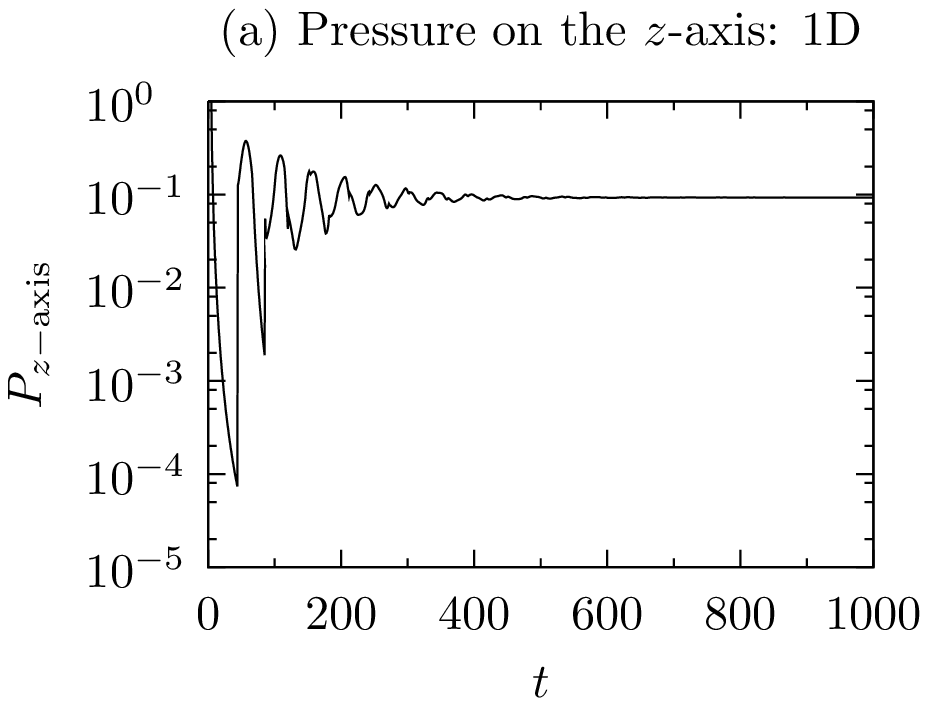}}}\\
\scalebox{1.1}{\rotatebox{0}{\includegraphics{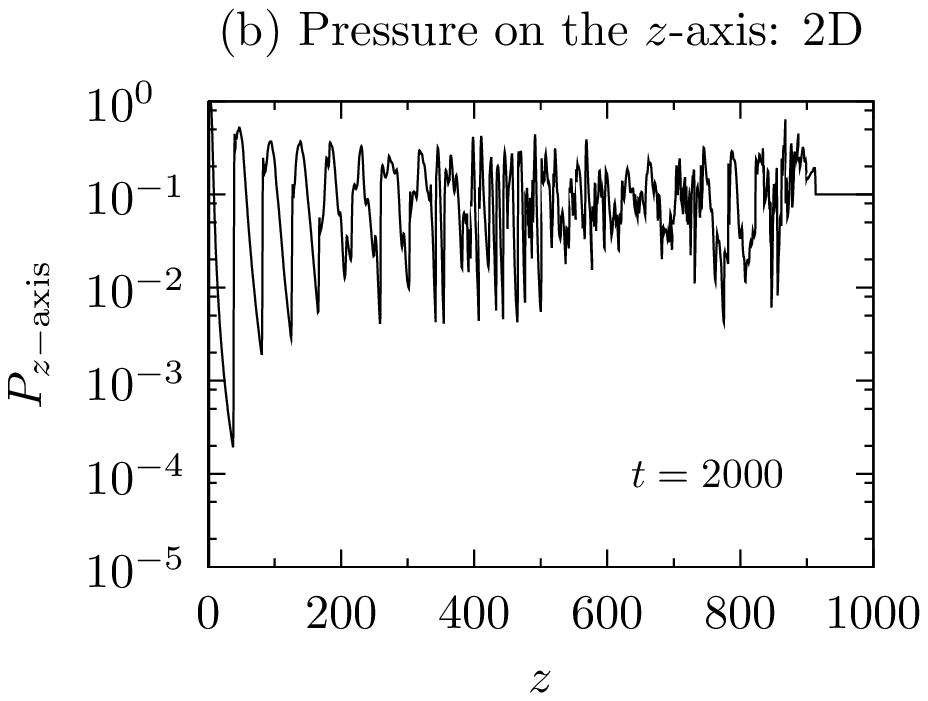}}}
\caption{Panel (a): temporal evolution of the pressure at the $z$-axis in 1D simulation. Panel (b): spatial distribution of the pressure along the z-axis at $t=2000$ in the uniform ambient model ($\alpha=0$) of 2D simulation.}
\label{fig13}
\end{center}
\end{figure}

\begin{figure}[!htbp]
\begin{center}
\scalebox{1.5}{\rotatebox{0}{\includegraphics{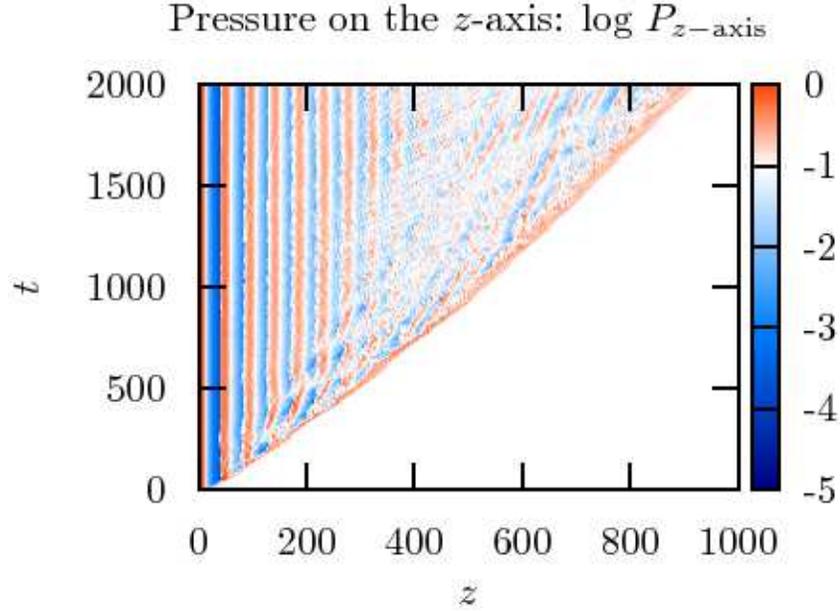}}}
\caption{The time-distance diagram for the pressure on the z-axis in the uniform ambient model ($\alpha=0$) of 2D simulation. (A color version of this figure is available in the online journal.)
}
\label{fig14}
\end{center}
\end{figure}

\begin{figure}[!htbp]
\begin{center}
\scalebox{1}{\rotatebox{0}{\includegraphics{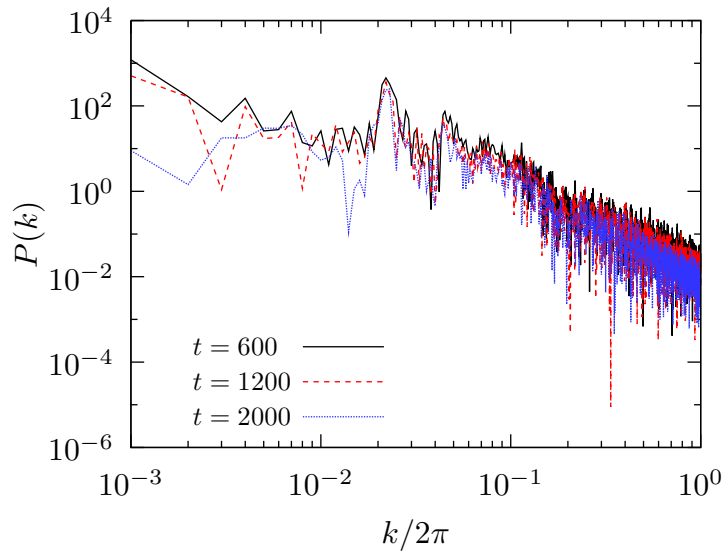}}}
\caption{
Temporal evolution of Fourier spectrum of the pressure measured along the jet axis $P (k) = \frac{1}{L_{\rm jet}} \int_{0}^{L_{\rm jet}} P_{z \rm -axis} (z) \exp^{-ikz} dz$, 
where $L_{\rm jet}$ is the length of the jet in the propagation direction. 
The solid, dashed and dotted curves correspond to the case $t=600$, $1200$ and $2000$, respectively. 
(A color version of this figure is available in the online journal.)
}
\label{fig15}
\end{center}
\end{figure}

\begin{figure}[!htbp]
\begin{center}
\scalebox{0.45}{\rotatebox{0}{\includegraphics{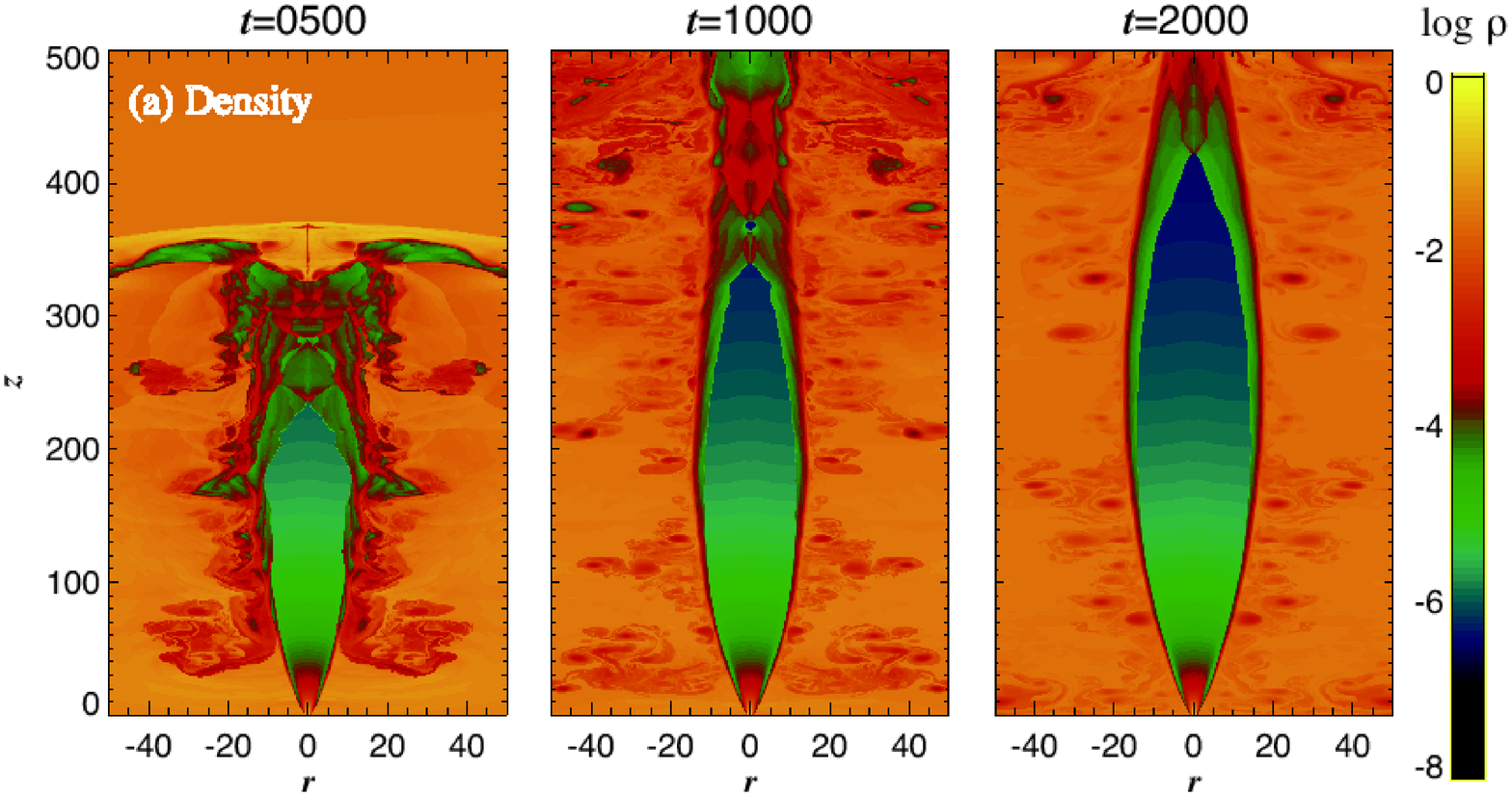}}}\\
\scalebox{0.45}{\rotatebox{0}{\includegraphics{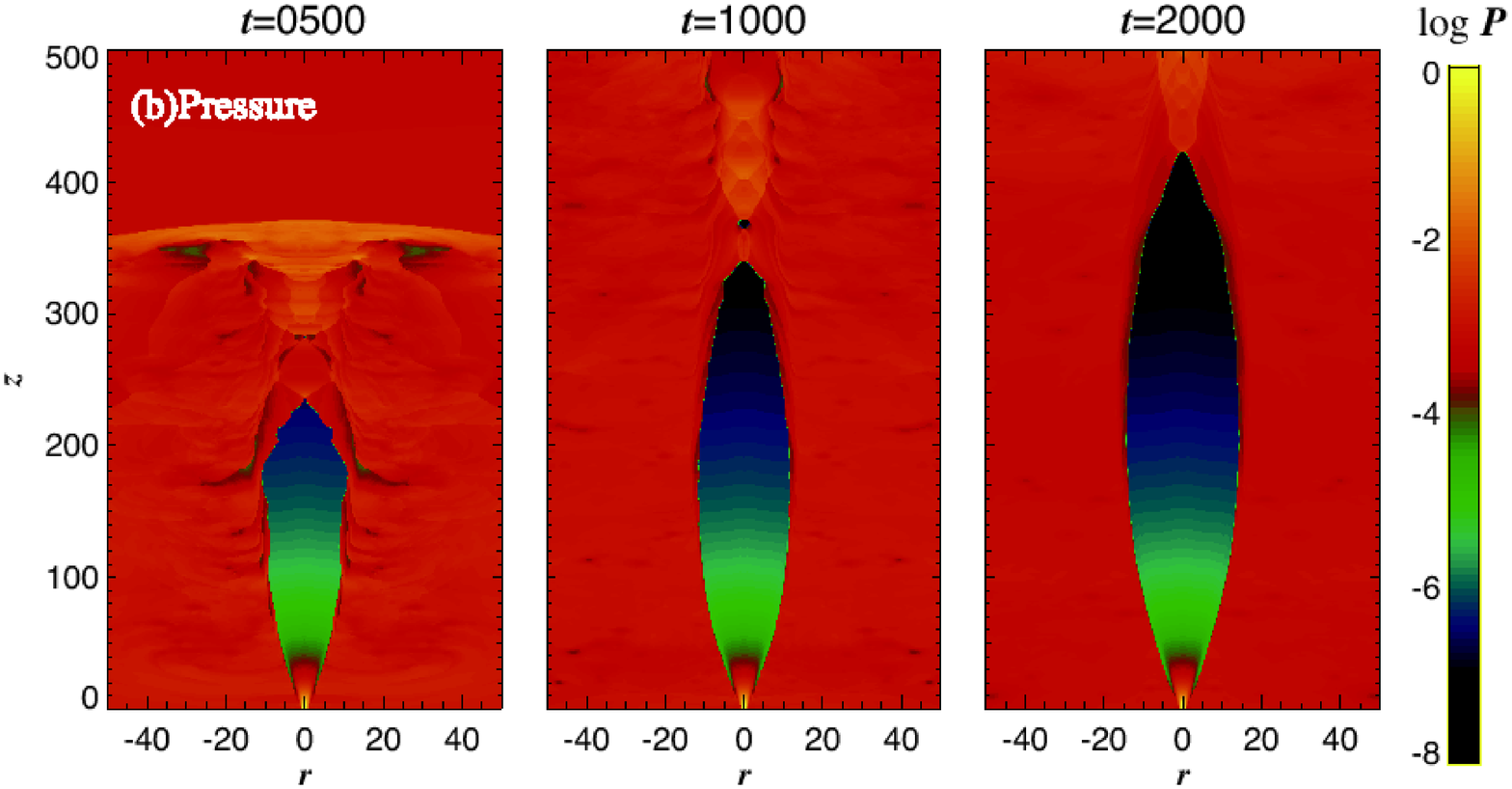}}}\\
\scalebox{0.45}{\rotatebox{0}{\includegraphics{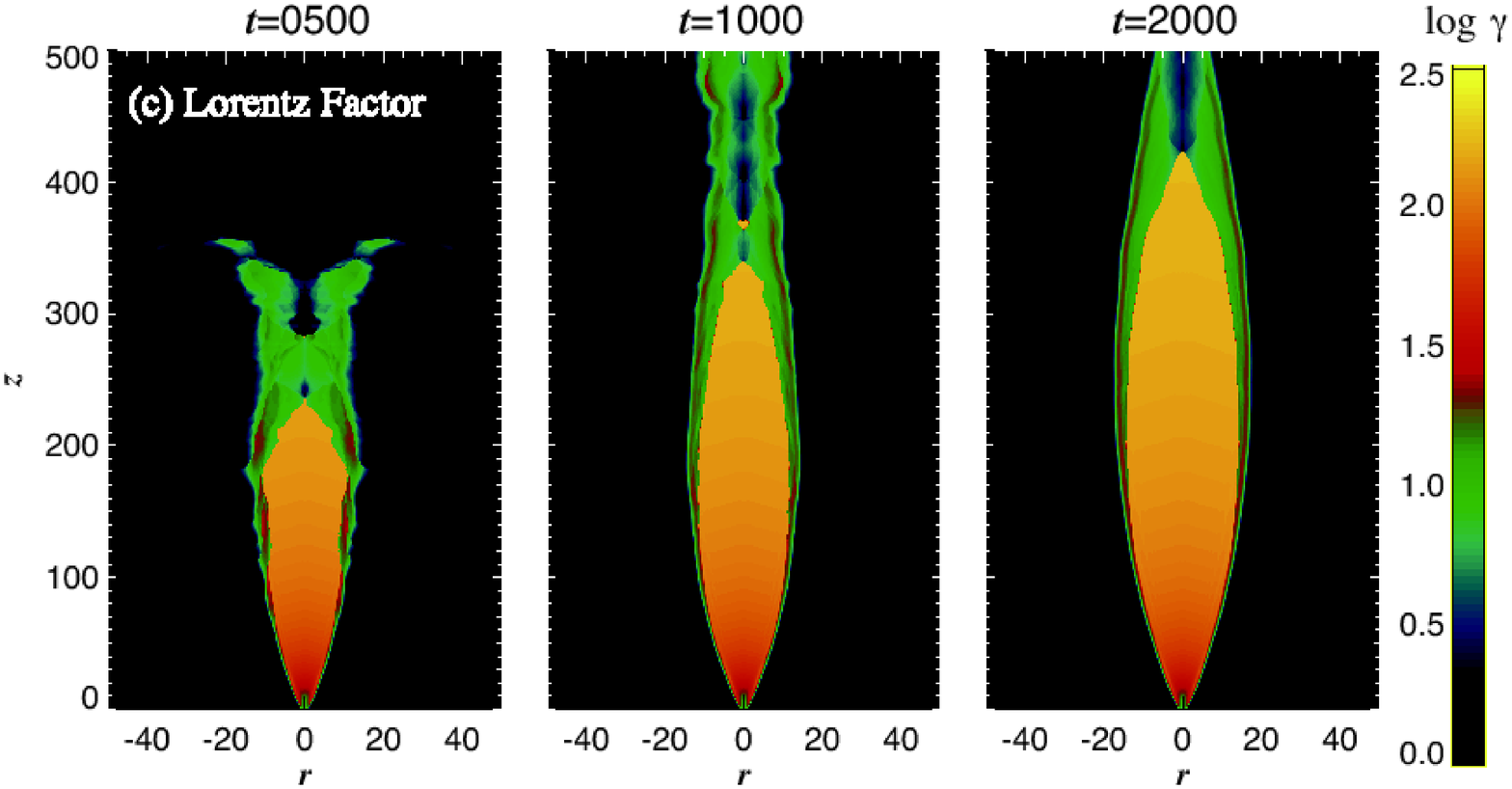}}}
\caption{
Time evolution of the injected relativistic jet in the power-law pressure distribution model where the power-law index $\alpha=0.8$: (a) the density, (b) the pressure and (c) the Lorentz factor. The left, middle, and right columns correspond to $t = 500$, $1000$, and $2000$, respectively. (A color version of this figure is available in the online journal.)
}
\label{fig16}
\end{center}
\end{figure}

\begin{figure}[!htbp]
\begin{center}
\scalebox{1}{\rotatebox{0}{\includegraphics{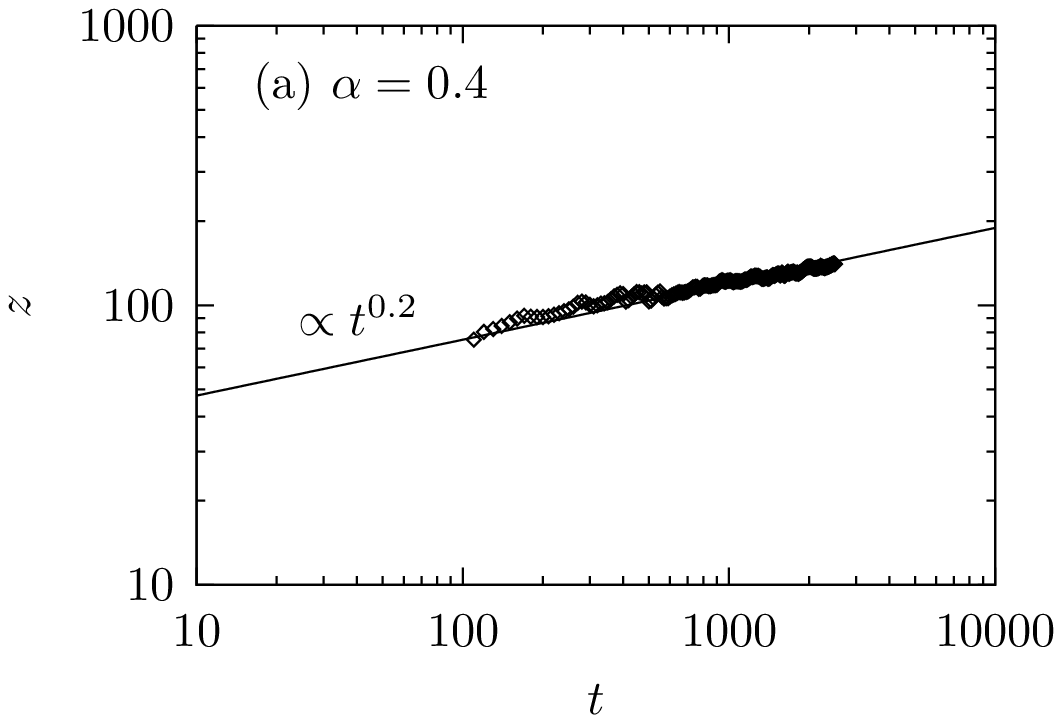}}}\\
\scalebox{1}{\rotatebox{0}{\includegraphics{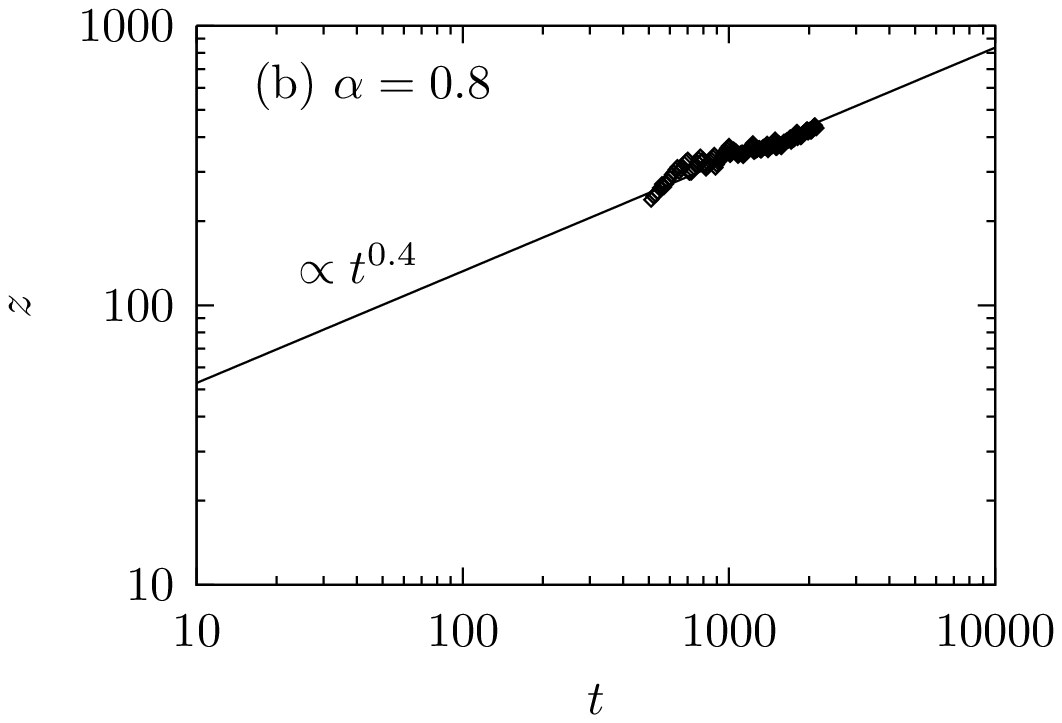}}}
\caption{
Time trajectory of the position of the cusp of the first reconfinement region on the z-axis is depicted by diamonds. Solid lines demonstrate the scaling law $\lambda \propto t^{\alpha/2}$ we derived in section~3.2 where $\lambda$ and $\alpha$ are the length of the cusp-shaped reconfinement region along the z-axis and the index of the power-law pressure  distribution of the ambient medium, respectively. Panel~(a) and (b) correspond to the case with $\alpha=0.4$ and $\alpha=0.8$.
}
\label{fig17}
\end{center}
\end{figure}

In our calculations in section 2, shock waves are excited inside the cylindrical jet. In order to investigate the effect of these shock 
waves, we have compared results of the calculation we performed in section 2 to those of the high-resolution calculation, 
in which a uniform grid with the grid size $\Delta r = 2 \times 10^{-5}$ is adopted in the cylindrical coordinate. 
Figure~\ref{fig21} illustrates the temporal evolution of (a) the jet width, (b) the maximum and (c) the average Lorentz factor in the jet 
in both numerical runs. Note that horizontal axis corresponds to the time for the duration $0 < t < 100$ demonstrated 
in Figure~\ref{fig7}. The jet width in the case with $\Delta r = 10^{-3}$ is up to a factor $1.07$ larger than in the high-resolution 
case because of numerical diffusion of the contact discontinuity. This leads to a 2 percent decrease in the average Lorentz factor inside 
the jet compared to the high-resolution case. We can find, however, only small differences of the maximum Lorentz factor in the 
jet between these numerical runs. Therefore, our choice of resolution does not impact significantly on our main results.

%%%%%%%%%%%%%%%%%%%%%%%%%%%%%%---------------------- BODY OF THE PAPER ------------------------%%%%%%%%%%%%%%%%%%%%%%%%%%%%%%%%%%%

%%%%%%%%%%%%%%%%%%%%%%%%%%%%%%%%%%%%%%---------------------- Table 1 ------------------------%%%%%%%%%%%%%%%%%%%%%%%%%%%%%%%%%%%%

%%%%%%%%%%%%%%%%%%%%%%%%%%%%%%%%%%%%%%---------------------- Figure 1 ------------------------%%%%%%%%%%%%%%%%%%%%%%%%%%%%%%%%%%%%
%%%%%%%%%%%%%%%%%%%%%%%%%%%%%%%%%%%%%%---------------------- Figure 1 ------------------------%%%%%%%%%%%%%%%%%%%%%%%%%%%%%%%%%%%%

%%%%%%%%%%%%%%%%%%%%%%%%%%%%%%%%%%%%%%---------------------- Figure 2 ------------------------%%%%%%%%%%%%%%%%%%%%%%%%%%%%%%%%%%%%
%%%%%%%%%%%%%%%%%%%%%%%%%%%%%%%%%%%%%%---------------------- Figure 2 ------------------------%%%%%%%%%%%%%%%%%%%%%%%%%%%%%%%%%%%%

%%%%%%%%%%%%%%%%%%%%%%%%%%%%%%%%%%%%%%---------------------- Figure 3 ------------------------%%%%%%%%%%%%%%%%%%%%%%%%%%%%%%%%%%%%
%%%%%%%%%%%%%%%%%%%%%%%%%%%%%%%%%%%%%%---------------------- Figure 3 ------------------------%%%%%%%%%%%%%%%%%%%%%%%%%%%%%%%%%%%%

%%%%%%%%%%%%%%%%%%%%%%%%%%%%%%%%%%%%%%---------------------- Figure 4 ------------------------%%%%%%%%%%%%%%%%%%%%%%%%%%%%%%%%%%%%
%%%%%%%%%%%%%%%%%%%%%%%%%%%%%%%%%%%%%%---------------------- Figure 4 ------------------------%%%%%%%%%%%%%%%%%%%%%%%%%%%%%%%%%%%%

%%%%%%%%%%%%%%%%%%%%%%%%%%%%%%%%%%%%%%---------------------- Figure 5 ------------------------%%%%%%%%%%%%%%%%%%%%%%%%%%%%%%%%%%%%
%%%%%%%%%%%%%%%%%%%%%%%%%%%%%%%%%%%%%%---------------------- Figure 5 ------------------------%%%%%%%%%%%%%%%%%%%%%%%%%%%%%%%%%%%%

%%%%%%%%%%%%%%%%%%%%%%%%%%%%%%%%%%%%%%---------------------- Figure 6 ------------------------%%%%%%%%%%%%%%%%%%%%%%%%%%%%%%%%%%%%
%%%%%%%%%%%%%%%%%%%%%%%%%%%%%%%%%%%%%%---------------------- Figure 6 ------------------------%%%%%%%%%%%%%%%%%%%%%%%%%%%%%%%%%%%%

%%%%%%%%%%%%%%%%%%%%%%%%%%%%%%%%%%%%%%---------------------- Figure 7 ------------------------%%%%%%%%%%%%%%%%%%%%%%%%%%%%%%%%%%%%
%%%%%%%%%%%%%%%%%%%%%%%%%%%%%%%%%%%%%%---------------------- Figure 7 ------------------------%%%%%%%%%%%%%%%%%%%%%%%%%%%%%%%%%%%%

%%%%%%%%%%%%%%%%%%%%%%%%%%%%%%%%%%%%%%---------------------- Figure 8 ------------------------%%%%%%%%%%%%%%%%%%%%%%%%%%%%%%%%%%%%
%%%%%%%%%%%%%%%%%%%%%%%%%%%%%%%%%%%%%%---------------------- Figure 8 ------------------------%%%%%%%%%%%%%%%%%%%%%%%%%%%%%%%%%%%%

%%%%%%%%%%%%%%%%%%%%%%%%%%%%%%%%%%%%%%---------------------- Figure 9 ------------------------%%%%%%%%%%%%%%%%%%%%%%%%%%%%%%%%%%%%
%%%%%%%%%%%%%%%%%%%%%%%%%%%%%%%%%%%%%%---------------------- Figure 9 ------------------------%%%%%%%%%%%%%%%%%%%%%%%%%%%%%%%%%%%%

%%%%%%%%%%%%%%%%%%%%%%%%%%%%%%%%%%%%%%---------------------- Figure 10  ------------------------%%%%%%%%%%%%%%%%%%%%%%%%%%%%%%%%%%%%
\clearpage
%%%%%%%%%%%%%%%%%%%%%%%%%%%%%%%%%%%%%%---------------------- Figure 10  ------------------------%%%%%%%%%%%%%%%%%%%%%%%%%%%%%%%%%%%%

%%%%%%%%%%%%%%%%%%%%%%%%%%%%%%%%%%%%%%---------------------- Figure 11 ------------------------%%%%%%%%%%%%%%%%%%%%%%%%%%%%%%%%%%%%
\clearpage
%%%%%%%%%%%%%%%%%%%%%%%%%%%%%%%%%%%%%%---------------------- Figure 11 ------------------------%%%%%%%%%%%%%%%%%%%%%%%%%%%%%%%%%%%%

%%%%%%%%%%%%%%%%%%%%%%%%%%%%%%%%%%%%%%---------------------- Figure  12 ------------------------%%%%%%%%%%%%%%%%%%%%%%%%%%%%%%%%%%%%
%%%%%%%%%%%%%%%%%%%%%%%%%%%%%%%%%%%%%%---------------------- Figure 12 ------------------------%%%%%%%%%%%%%%%%%%%%%%%%%%%%%%%%%%%%

%%%%%%%%%%%%%%%%%%%%%%%%%%%%%%%%%%%%%%---------------------- Figure  13 ------------------------%%%%%%%%%%%%%%%%%%%%%%%%%%%%%%%%%%%%
%%%%%%%%%%%%%%%%%%%%%%%%%%%%%%%%%%%%%%---------------------- Figure 13 ------------------------%%%%%%%%%%%%%%%%%%%%%%%%%%%%%%%%%%%%

%%%%%%%%%%%%%%%%%%%%%%%%%%%%%%%%%%%%%%---------------------- Figure  14 ------------------------%%%%%%%%%%%%%%%%%%%%%%%%%%%%%%%%%%%%
%%%%%%%%%%%%%%%%%%%%%%%%%%%%%%%%%%%%%%---------------------- Figure 14 ------------------------%%%%%%%%%%%%%%%%%%%%%%%%%%%%%%%%%%%%

%%%%%%%%%%%%%%%%%%%%%%%%%%%%%%%%%%%%%%---------------------- Figure 15 ------------------------%%%%%%%%%%%%%%%%%%%%%%%%%%%%%%%%%%%%
%%%%%%%%%%%%%%%%%%%%%%%%%%%%%%%%%%%%%%---------------------- Figure 15 ------------------------%%%%%%%%%%%%%%%%%%%%%%%%%%%%%%%%%%%%

%%%%%%%%%%%%%%%%%%%%%%%%%%%%%%%%%%%%%%---------------------- Figure 16 ------------------------%%%%%%%%%%%%%%%%%%%%%%%%%%%%%%%%%%%%
\clearpage
%%%%%%%%%%%%%%%%%%%%%%%%%%%%%%%%%%%%%%---------------------- Figure 16 ------------------------%%%%%%%%%%%%%%%%%%%%%%%%%%%%%%%%%%%%

%%%%%%%%%%%%%%%%%%%%%%%%%%%%%%%%%%%%%%---------------------- Figure 17 ------------------------%%%%%%%%%%%%%%%%%%%%%%%%%%%%%%%%%%%%
\clearpage
%%%%%%%%%%%%%%%%%%%%%%%%%%%%%%%%%%%%%%---------------------- Figure 17 ------------------------%%%%%%%%%%%%%%%%%%%%%%%%%%%%%%%%%%%%

%%%%%%%%%%%%%%%%%%%%%%%%%%%%%%%%%%%%%%---------------------- Figure 18 ------------------------%%%%%%%%%%%%%%%%%%%%%%%%%%%%%%%%%%%%
\begin{figure}[!htbp]
\begin{center}
\scalebox{0.52}{\rotatebox{0}{\includegraphics{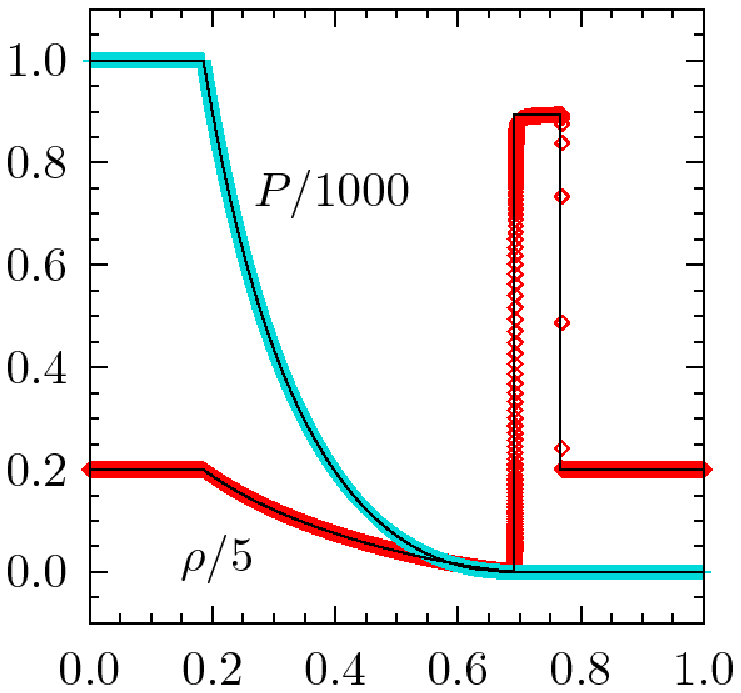}}}
\scalebox{0.52}{\rotatebox{0}{\includegraphics{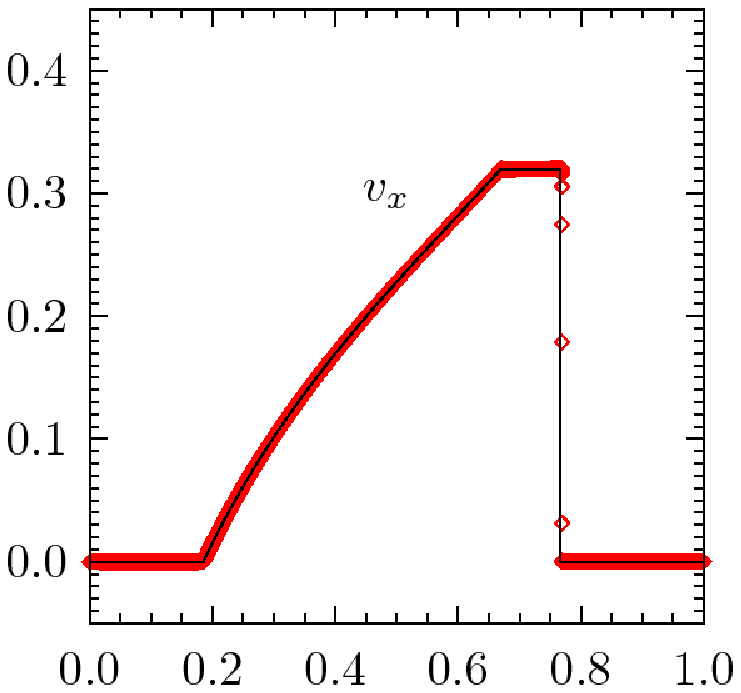}}}
\scalebox{0.52}{\rotatebox{0}{\includegraphics{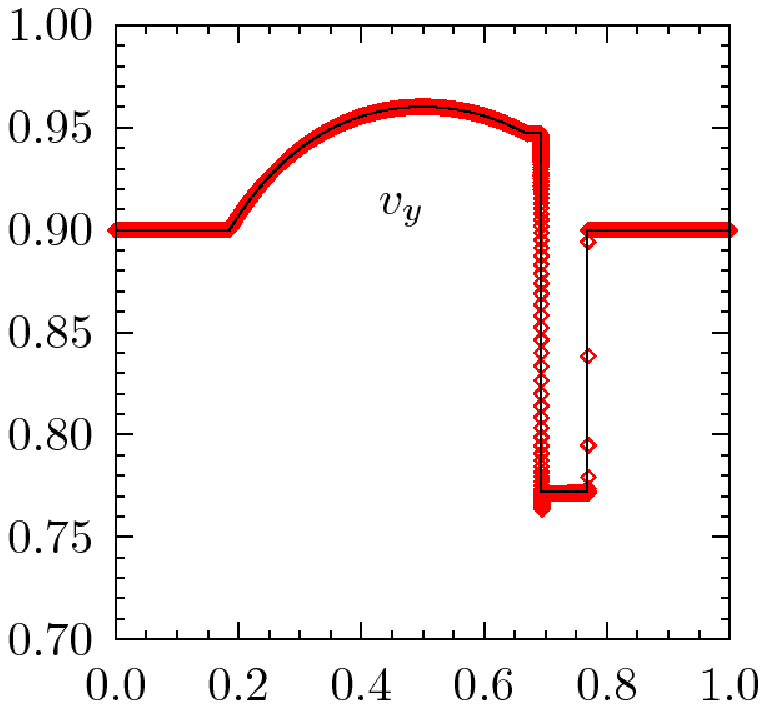}}}\\
\scalebox{0.52}{\rotatebox{0}{\includegraphics{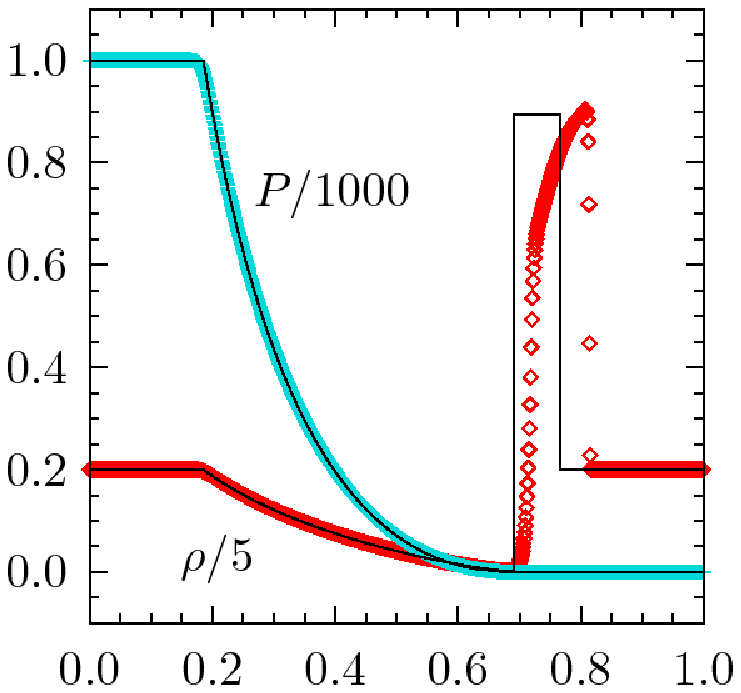}}}
\scalebox{0.52}{\rotatebox{0}{\includegraphics{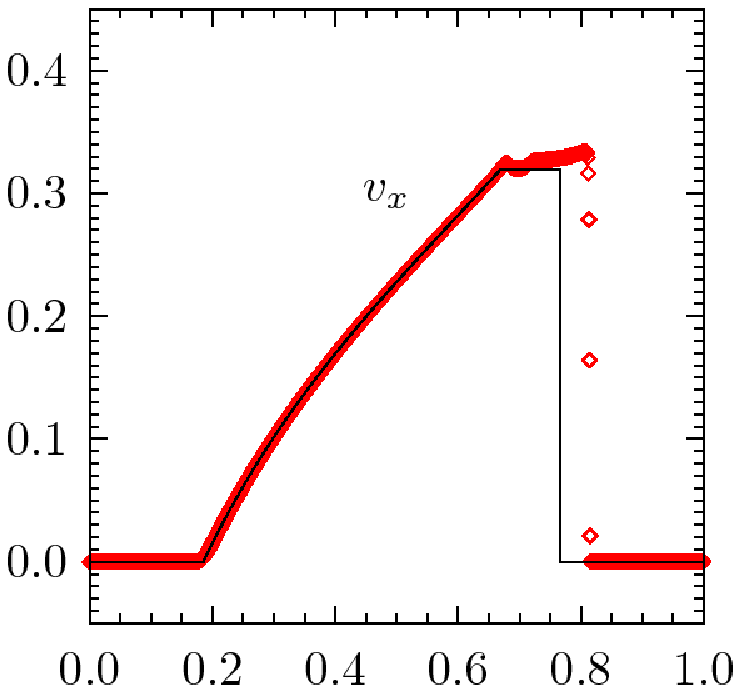}}}
\scalebox{0.52}{\rotatebox{0}{\includegraphics{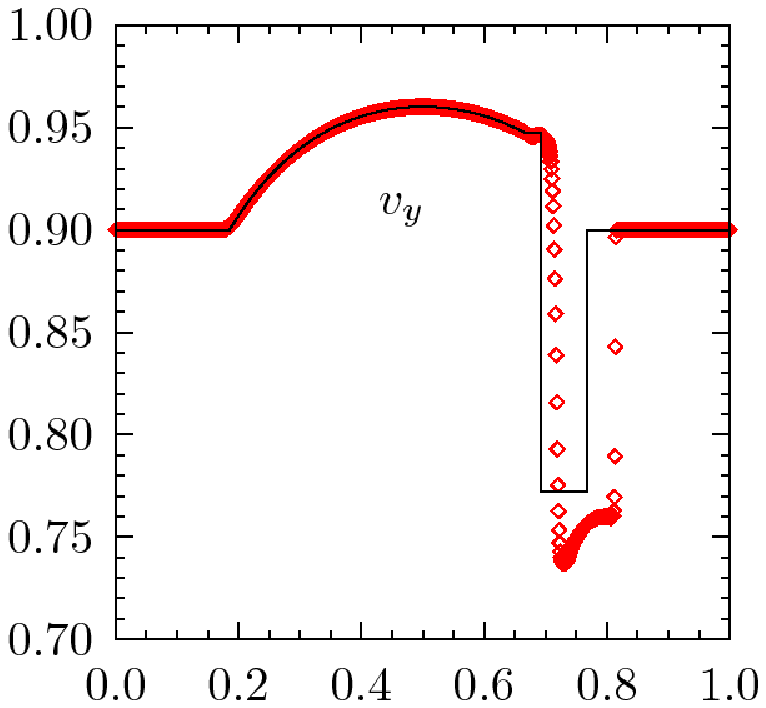}}}
\caption{
Solutions of 1D Riemann problem with large transverse velocity at $t=0.6$. Upper and lower panels correspond to the high-resolution ($\Delta x =2 \times 10^{-5}$) and low-resolution ($\Delta x =10^{-3}$) cases, respectively. Diamonds and crosses represent the numerical solution while solid lines are analytic solution. (A color version of this figure is available in the online journal.)
}
\label{fig18}
\end{center}
\end{figure}
\clearpage
%%%%%%%%%%%%%%%%%%%%%%%%%%%%%%%%%%%%%%---------------------- Figure 18 ------------------------%%%%%%%%%%%%%%%%%%%%%%%%%%%%%%%%%%%%

%%%%%%%%%%%%%%%%%%%%%%%%%%%%%%%%%%%%%%---------------------- Figure 19 ------------------------%%%%%%%%%%%%%%%%%%%%%%%%%%%%%%%%%%%%
\begin{figure}[!htbp]
\begin{center}
\scalebox{0.52}{\rotatebox{0}{\includegraphics{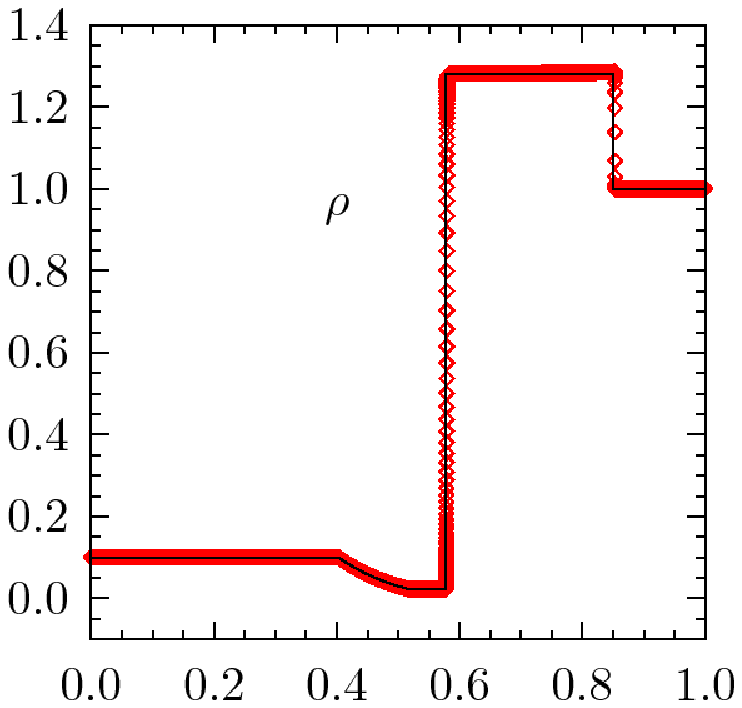}}}
\scalebox{0.52}{\rotatebox{0}{\includegraphics{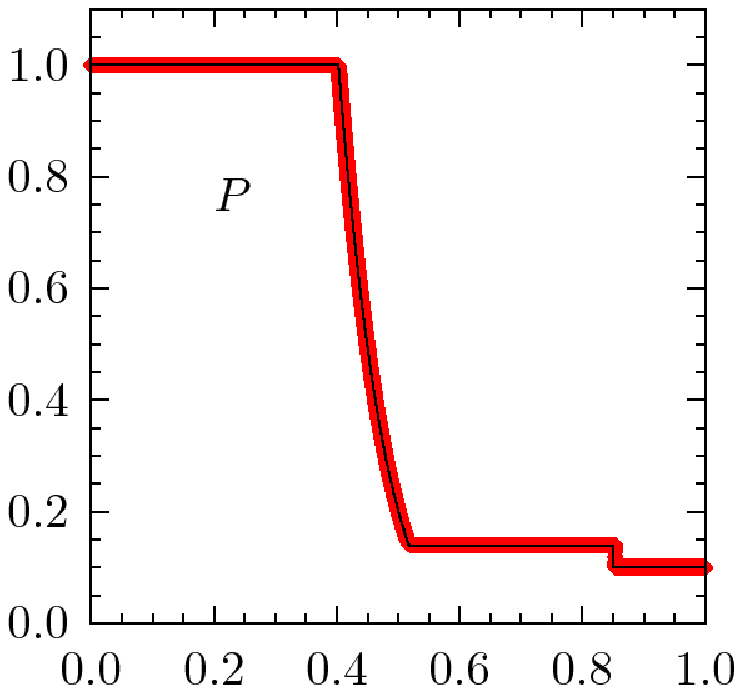}}}
\scalebox{0.52}{\rotatebox{0}{\includegraphics{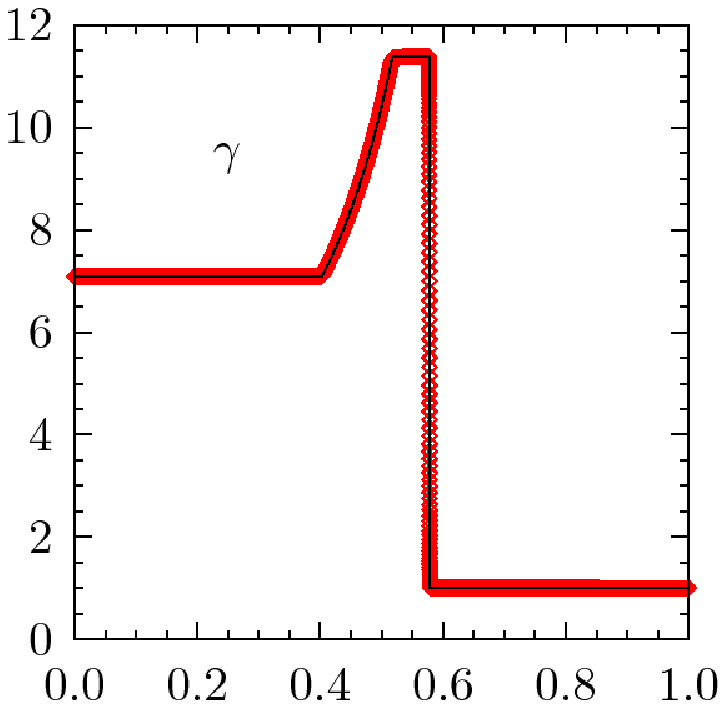}}}\\
\scalebox{0.52}{\rotatebox{0}{\includegraphics{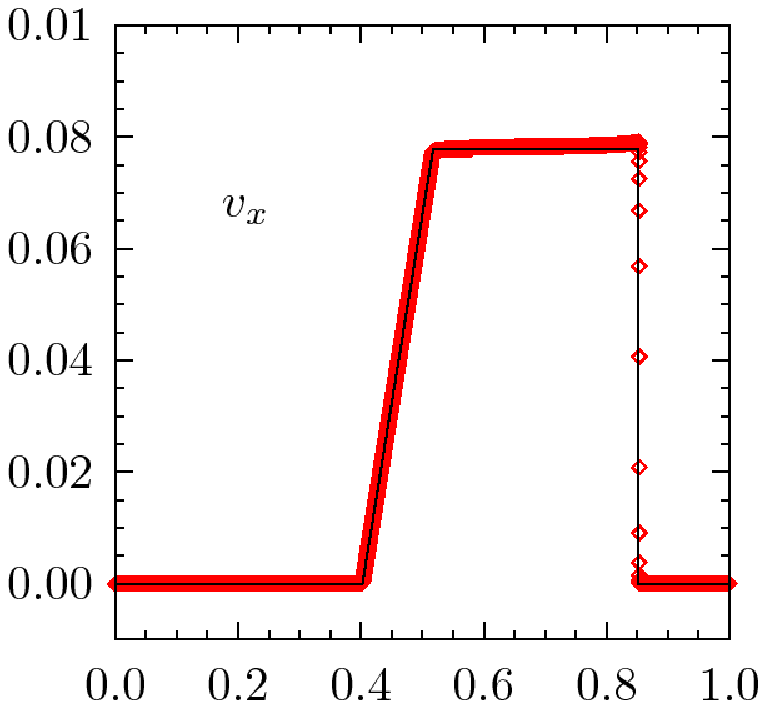}}}
\scalebox{0.52}{\rotatebox{0}{\includegraphics{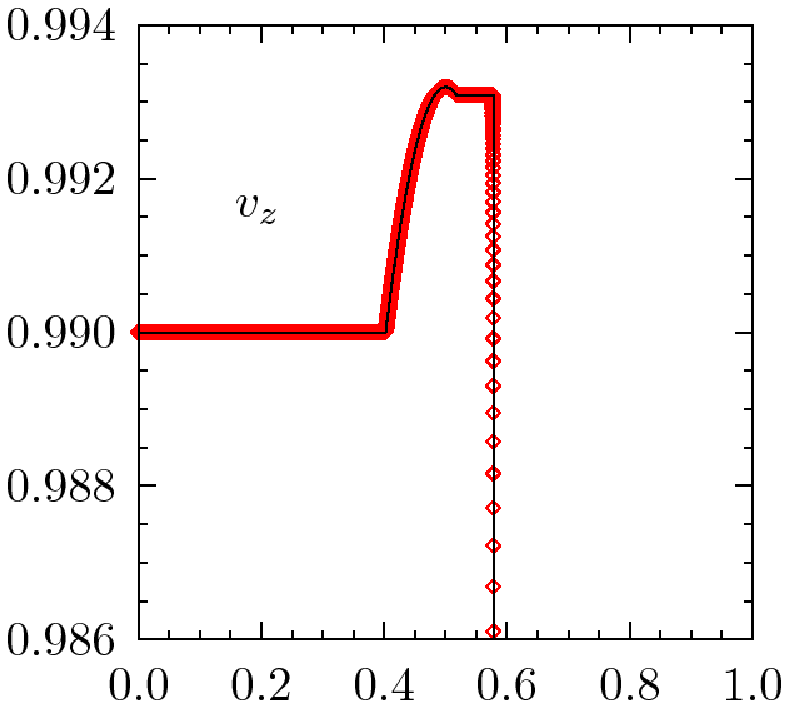}}}
\caption{
Solutions of Riemann problem of our jet model in 1D (x-direction) cartesian coordinate with the resolution of $\Delta x =2 \times 10^{-5}$. Diamonds and solid lines illustrate the numerical and analytic solutions, respectively. (A color version of this figure is available in the online journal.)
}
\label{fig19}
\end{center}
\end{figure}
%%%%%%%%%%%%%%%%%%%%%%%%%%%%%%%%%%%%%%---------------------- Figure 19 ------------------------%%%%%%%%%%%%%%%%%%%%%%%%%%%%%%%%%%%%

%%%%%%%%%%%%%%%%%%%%%%%%%%%%%%%%%%%%%%---------------------- Figure 20 ------------------------%%%%%%%%%%%%%%%%%%%%%%%%%%%%%%%%%%%%
\begin{figure}[!htbp]
\begin{center}
\scalebox{0.52}{\rotatebox{0}{\includegraphics{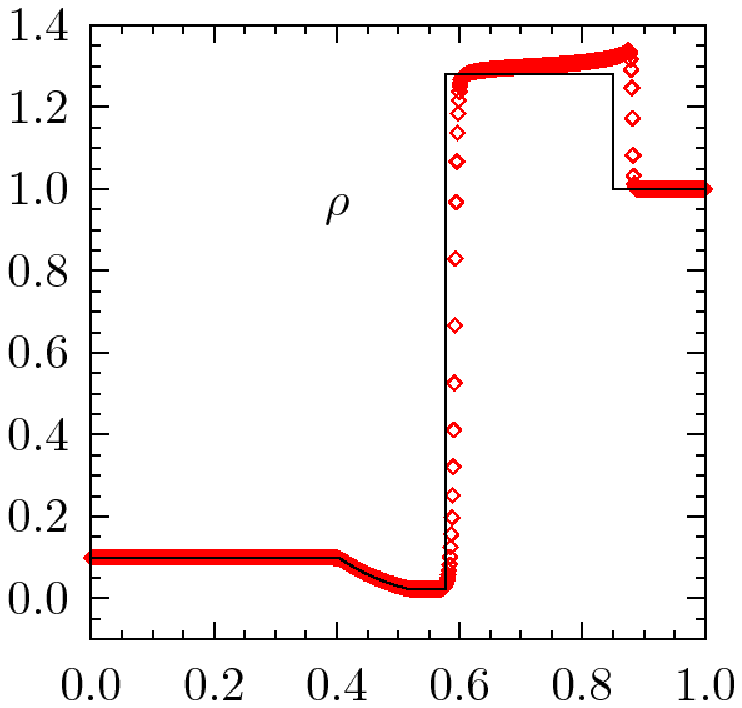}}}
\scalebox{0.52}{\rotatebox{0}{\includegraphics{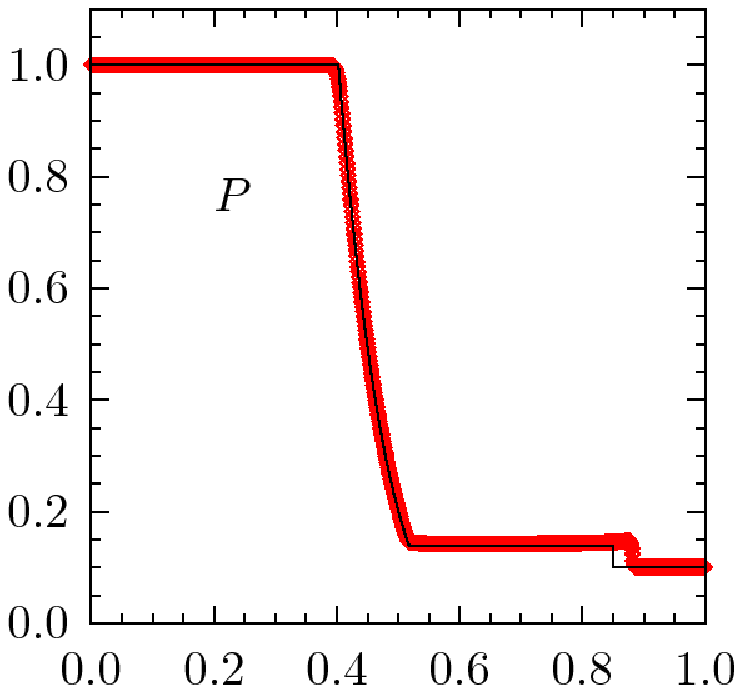}}}
\scalebox{0.52}{\rotatebox{0}{\includegraphics{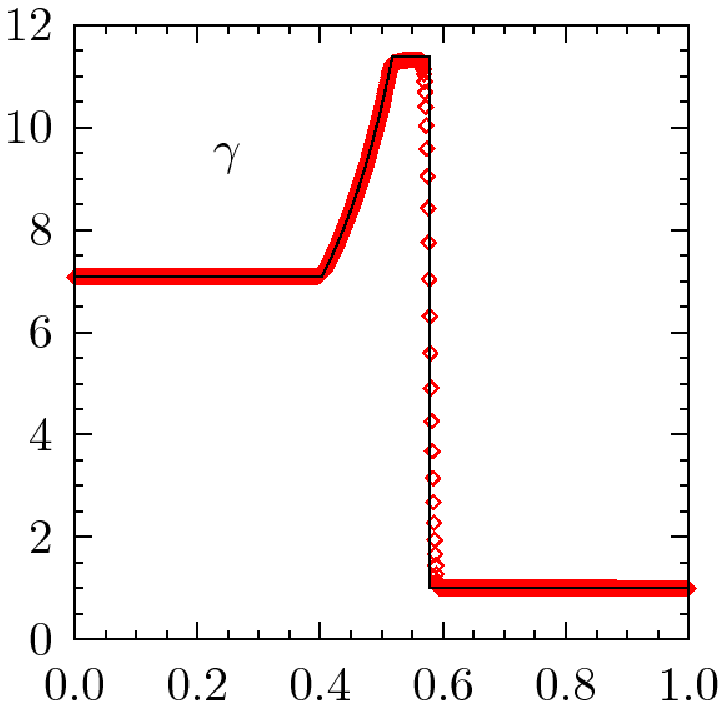}}}\\
\scalebox{0.52}{\rotatebox{0}{\includegraphics{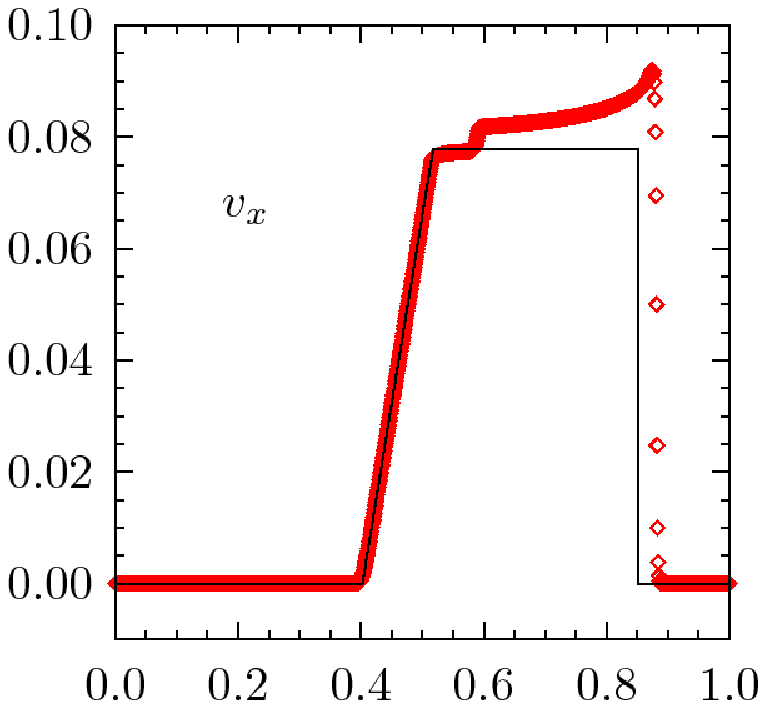}}}
\scalebox{0.52}{\rotatebox{0}{\includegraphics{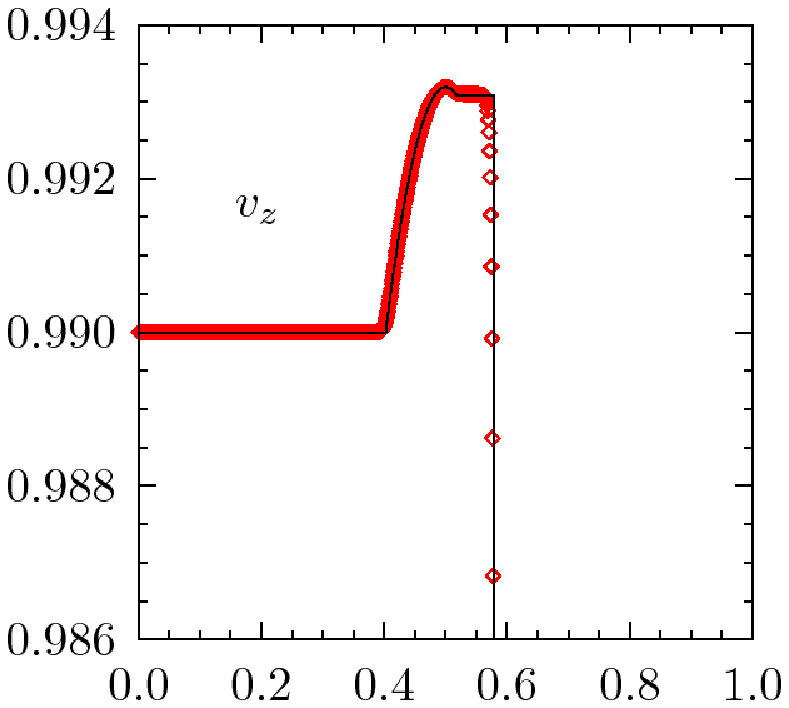}}}
\caption{
Same as Figure~\ref{fig19} but the case with $\Delta x =10^{-3}$. (A color version of this figure is available in the online journal.)
}
\label{fig20}
\end{center}
\end{figure}
\clearpage
%%%%%%%%%%%%%%%%%%%%%%%%%%%%%%%%%%%%%%---------------------- Figure 20 ------------------------%%%%%%%%%%%%%%%%%%%%%%%%%%%%%%%%%%%%

%%%%%%%%%%%%%%%%%%%%%%%%%%%%%%%%%%%%%%---------------------- Figure 21 ------------------------%%%%%%%%%%%%%%%%%%%%%%%%%%%%%%%%%%%%
\begin{figure}[!htbp]
\begin{center}
\scalebox{0.7}{\rotatebox{0}{\includegraphics{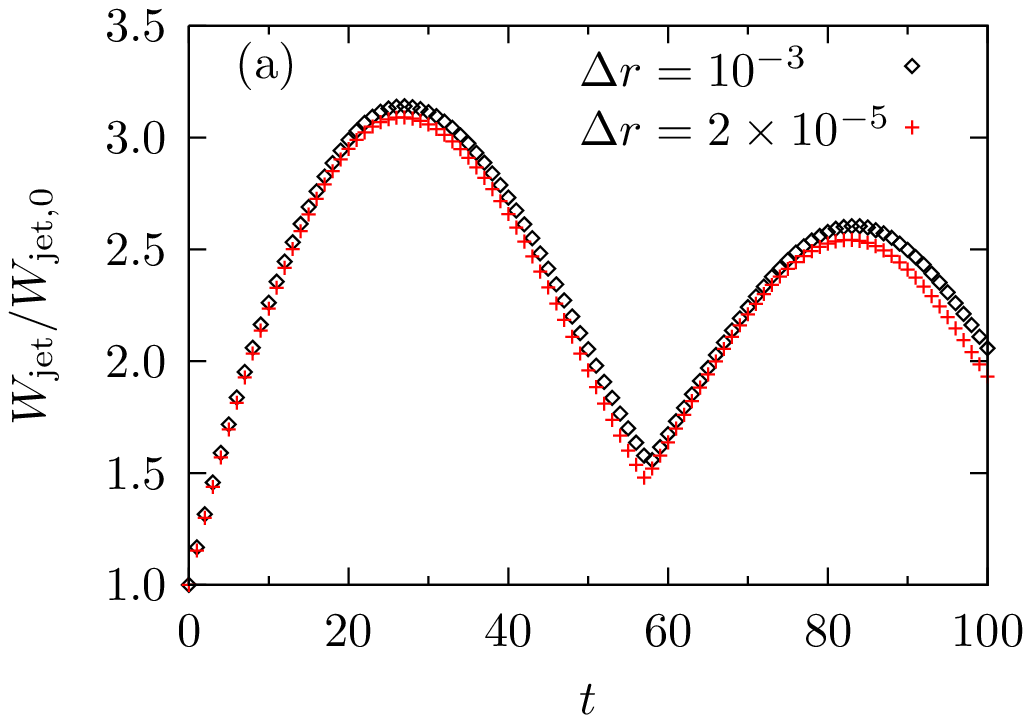}}}\\
\scalebox{0.7}{\rotatebox{0}{\includegraphics{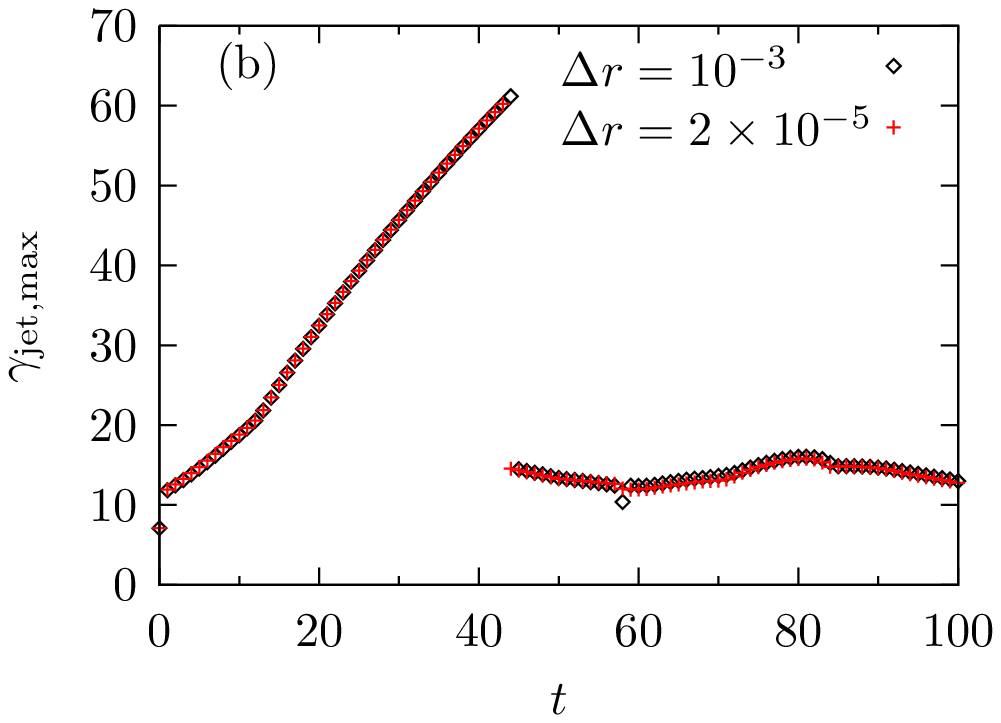}}}\\
\scalebox{0.7}{\rotatebox{0}{\includegraphics{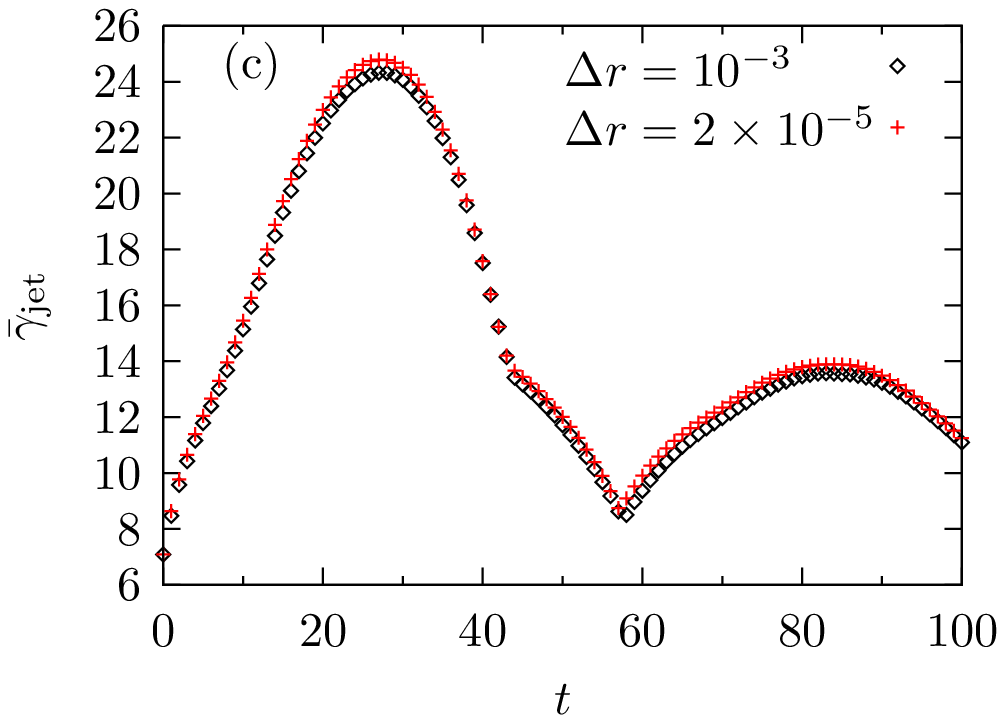}}}
\caption{
Temporal evolution of (a) the jet width, (b) the maximum and (c) the average Lorentz factor in the jet. Diamonds and crosses correspond to the resolution of $\Delta x =10^{-3}$ and $\Delta x =2 \times 10^{-5}$, respectively. (A color version of this figure is available in the online journal.)
}
\label{fig21}
\end{center}
\end{figure}
%%%%%%%%%%%%%%%%%%%%%%%%%%%%%%%%%%%%%%---------------------- Figure 21 ------------------------%%%%%%%%%%%%%%%%%%%%%%%%%%%%%%%%%%%%

%%%%%%%%%%%%%%%%%%%%%%%%%%%%%%%%%%%---------------------- END OF PAPER ------------------------%%%%%%%%%%%%%%%%%%%%%%%%%%%%%%%%%%%
\end{document}